\documentclass[11pt]{article}
\usepackage{latexsym,color,amsmath,amssymb}
\usepackage{amsthm,amscd,amsfonts}

\usepackage{thm-restate}

\usepackage{scalefnt}
\usepackage[font=small,labelfont=bf]{caption}
\usepackage{float}
\usepackage{graphicx}
\usepackage{relsize}
\usepackage{mathrsfs}
\usepackage{upgreek}

\usepackage{fullpage}

\newtheorem{thm}{Theorem}[section]

\newtheorem{cor}[thm]{Corollary}
\newtheorem{lem}[thm]{Lemma}
\newtheorem{lemma}[thm]{Lemma}

\newtheorem{definition}[thm]{Definition}

\newtheorem{rem}[thm]{Remark}

\def\xx{{x}}

\def\zz{{z}}

\newcommand{\R}{\mathbb R}
\newcommand{\RR}{\mathbb R}

\newcommand{\sph}{\mathbb S}

\newcommand{\iprod}[2]{\langle #1,#2 \rangle} 

\def\grad{{\nabla}}

\def\dom{{\rm dom}\,}

\def\eps{{\varepsilon}}
\def\dom{{\rm dom}}
\def\conv{{\rm conv}}

\begin{document}
	
\title{On Radial Isotropic Position: Theory and Algorithms\thanks{
Work on this paper by Shiri Artstein was supported in part by the European Research
Council (ERC) under the European Union's Horizon 2020 research and
innovation programme (grant agreement No.~770127), and in part by ISF grant 665/15.
Work by Haim Kaplan was supported in part by
by ISF grant 1595/19 and grant 1367/2016 from the German-Israeli Science Foundation (GIF).
Work by Micha Sharir was supported in part by
Grant 260/18 from the Israel Science Foundation,
Grant G-1367-407.6/2016 from the German-Israeli Foundation for Scientific Research and Development,
by the Blavatnik Research Fund in Computer Science at Tel Aviv University,
and by the Israeli Centers of Research
Excellence (I-CORE) program (Center No.~4/11).
}}

	\author{Shiri Artstein-Avidan\thanks{
			School of Mathematics, Tel Aviv University, Tel Aviv, Israel; {\tt shiri@tau.ac.il}}
		\and
		Haim Kaplan\thanks{
			School of Computer Science, Tel Aviv University, Tel Aviv, Israel and Google; {\tt haimk@tau.ac.il}}
		\and
		Micha Sharir\thanks{
			School of Computer Science, Tel Aviv University, Tel Aviv, Israel; {\tt michas@tau.ac.il}}}
	
	\maketitle

\begin{abstract}
	We review the theory of, and develop algorithms for transforming a finite point set
	in $\R^d$ into a set in \emph{radial isotropic position} by a nonsingular linear 
	transformation followed by rescaling each image point to the unit sphere.
	This problem, defined in detail in the introduction, arises in a wide spectrum of applications, ranging from 
	communication complexity, robust subspace recovery, 
	aspects of Brascamp-Lieb inequalities, and locally correctable codes,
	up to a recent application of Kane, Lovett and Moran  to point location in arrangements
	of hyperplanes in high dimensions.
	
        Our algorithms use gradient descent methods for a particular
	convex function $f$ whose minimum defines the transformation,
        and our main focus is on analyzing their performance.       
        Although the minimum can be computed exactly, by (rather expensive)
        symbolic techniques from computational real algebraic geometry,
        gradient descent only approximates the desired minimum (and
        corresponding transformation), to any desired level of accuracy.
	We show that computing the gradient of $f$ amounts to
	computing the Singular Value Decomposition (SVD) of a certain matrix
	associated with the input set, making gradient descent simple to implement. 
	We believe that gradient descent is superior to 
	other (also approximate) algorithmic techniques (mainly the ellipsoid algorithm) 
        previously used in the literature to find this
	transformation, and it should run much faster in practice.
	
	We prove that $f$ is smooth, and therefore, by applying gradient descent with an 
        appropriate stepsize, we get convergence rate proportional to $1/\eps$, where 
        $\eps$ is the desired approximation accuracy.
	To complete the analysis, we provide upper bounds on the norm of the 
        optimal solution which depend on new parameters measuring ``the degeneracy'' in our input. 
        We believe that our parameters capture degeneracy better than other, seemingly weaker, parameters 
        introduced in previous works, in particular in that they are stable with respect 
        to perturbation of the input, and may be useful in other contexts.
	
	We next analyze the strong convexity of our function $f$, and present two 
        worst-case lower bounds on the smallest eigenvalue of its Hessian. 
	This gives another worst-case bound on the convergence rate of another variant of
        gradient decent that depends only logarithmically on $1/\eps$.
	
	\end{abstract}

\pagestyle{plain}

\section{Introduction} \label{sec:intro}

A set $X = \{ x_i\}_{i=1}^n$ of $n\ge d$ (column) vectors in $\RR^d$
is in \emph{isotropic position}
if one (and thus all) of the following equivalent conditions holds.
\begin{align*}
& (1)\;\; u = \sum_{i=1}^n \frac{d}{n}  \iprod{x_i}{u}x_i, \;\; \forall u\in\R^d , \\
& (2)\;\; I_d = \sum_{i=1}^n \frac{d}{n}  x_i\otimes x_i , \\
& (3)\;\; |u|^2  = \sum_{i=1}^n \frac{d}{n}  \iprod{x_i}{u}^2, \;\; \forall u\in\R^d\ .
\end{align*}
Here we denote by $x\otimes x$ the rank-one operator $(x\otimes x)y = \iprod{x}{y}x$ 
(which is $xx^T$ in matrix form), and $I_d$ is the $d\times d$ identity matrix.
In words, the third condition asserts that the sum of the squares of the projections of our vectors
in any direction has the same value $n/d$. If the $x_i$'s are unit vectors then the third
condition also says that the sum of the squares of the projections of any unit vector on the $x_i$'s is $n/d$.

As an easy example, the standard basis $\{e_i\}_{i=1}^d$ is a set in isotropic position.
As a slightly less obvious example, the
vertices of a regular simplex, so that its center is at the origin and the vertices lie on $\sph^{d-1}$,
form a set of $d+1$ points in isotropic position. In more generality,
consider a subspace $E$ of $\RR^n$ of dimension $d$.
Let $\{e_i\}_{i=1}^n$ be an orthonormal basis of $\RR^n$,
and let $y_i = P_{E} e_i$ be their orthogonal projections onto $E$.
We have $\sum_{i=1}^n e_i \otimes e_i = I_n$, and therefore
$\sum_{i=1}^n y_i \otimes y_i = I_d$, as is easily verified
(note that $P_{E} P_{E}^T = I_d$). This means that by rewriting
$x_i = y_i /|y_i|$ (assuming that all the $y_i$'s are non-zero),
and letting $c_i = |y_i|^2\ge 0$, we get that
$I_d = \sum_{i=1}^n c_i x_i\otimes x_i.$
In the case of a simplex, one can present the vertices of the regular simplex
as the projections of an orthonormal basis in $\RR^{d+1}$ such that
all projections have equal length, which is why all the
$c_i$'s are equal (to $d/(d+1)$) and we get isotropic position.
This naturally brings us to the following generalizations of isotropy.

\begin{definition} \label{def1}
	{\bf $c$-isotropy.}
	Let $c = (c_1,\ldots,c_n)$ be an $n$-sequence of positive real weights.
	A set $X = \{ x_i\}_{i=1}^n$ of $n\ge d$ (column) vectors in $\RR^d$
	is in \emph{$c$-isotropic position} if
	the following equivalent properties hold.
	\begin{align*}
	& (1)\;\; u  = \sum_{i=1}^n c_i  \iprod{x_i}{u}x_i, \;\; \forall u\in\R^d , \\
	& (2)\;\; I_d  = \sum_{i=1}^n c_i x_i\otimes x_i , \\
	& (3)\;\; |u|^2  = \sum_{i=1}^n c_i  \iprod{x_i}{u}^2, \;\; \forall u\in\R^d\ .
	\end{align*}
\end{definition}
Isotropy is the special case of $c$-isotropy when $c_i = d/n$ for all $i$.

\medskip

It is a classical fact that \emph{any} set of $n$ (not necessarily unit) vectors $\{y_i\}_{i=1}^n$ 
that span $\RR^d$ can be put in $c$-isotropic position, for any positive weight sequence $c$, using the
linear transformation $A= \Bigl( \sum_i c_i y_i \otimes y_i \Bigr)^{-1/2}$ (which is the unique
positive definite matrix $A= A^T$ for which $A^2 = \Bigl(\sum_i c_i y_i \otimes y_i\Bigr)^{-1}$,
and which exists as $\sum c_i y_i \otimes y_i$ is positive definite), because
\[
\sum c_i Ay_i \otimes Ay_i  = A \left( \sum c_i y_i \otimes y_i \right) A^T = I_d.
\]

In this paper we are specifically interested in the notion of \emph{radial isotropy}, defined as follows.
\begin{definition} \label{def2}
	{\bf Radial $c$-isotropy.}
	A set $X = \{ x_i\}_{i=1}^n$ of $n\ge d$ nonzero (column) vectors in $\RR^d$
	is in \emph{radial $c$-isotropic position} (for a vector $c$ as in Definition~\ref{def1})
	if the normalized vectors $x_i/|x_i|$, for $i=1,\ldots,n$,
	are in  $c$-isotropic position. We say that the vectors are in (standard, or uniform)
        radial isotropic position when the above property holds with $c_i = d/n$ for all $i$.
\end{definition}

It is easily checked that any weights $c_i$ that admit a point set in
radial $c$-isotropic position must satisfy $\sum_{i=1}^n c_i = d$.

We are interested in the existence and the computation of a linear transformation that maps
a given set of vectors into radial $c$-isotropic position, for some prescribed sequence $c$
of positive weights. Formally,
\begin{definition} \label{def:T}
	We say that a nonsingular linear transformation $T :\R^d\to\R^d$
	puts a set $X$ of $n$ nonzero vectors $x_1,\dots,x_n\in\R^d$ in radial $c$-isotropic position,
	for a coefficient vector $c\in(\R^+)^n$ of $\ell_1$-norm $d$, if
	$$
	\sum_{i=1}^n c_i \frac{Tx_i}{|Tx_i|}
	\otimes\frac{Tx_i}{|Tx_i|} = I_d.
	$$
\end{definition}

The existence of such a linear transformation, even if we consider only the simpler form of 
radial isotropy ($c_i = d/n$ for each $i$),
is more intricate than the standard, non-radial setup (and in general it may fail to exist).
Forster \cite{For} proved that when the vectors of $X$ are in \emph{general position}
(i.e., every $d$ vectors among them are linearly independent),
such a transformation exists. In fact, earlier, Barthe \cite{Bar} proved 
that the vectors $x_i$, $i=1,\ldots,n$, can be put in a radial $c$-isotropic position if and only if 
$c$ is in the relative interior of the so called {\em basis polytope} (see below) defined by the $x_i$'s.
We give a complete, and somewhat enhanced version of this theory in Section \ref{sec:shiri}.
Since this theory is well known,\footnote{%
  It is well known to the experts. The details that we provide in the appendix are spelled
  out for the convenience of the non-expert reader.}
except for the enhancements that we derive and add to it, Section~\ref{sec:shiri} contains
only the highlights of the theory, and most details, including proofs, are presented in
Appendix \ref{sec:shiri-full}.
In a later study, Carlen, Lieb and Loss~\cite{CLL} gave an equivalent characterization 
of the basis polytope, which we use extensively in our study (we prove it too, for completeness,
in Appendix \ref{sec:shiri-full}).

Radial isotropy arises when we have a set of $n$ subspaces, say lines, and we want to linearly transform these
subspaces so that the sum of the squared projections of any unit vector $x$ on their transformed copies is $n/d$.
A similar interpretation can be given for general weights $c_i$.\footnote{%
  Radial isotropy corresponds to the case of lines; higher-dimensional subspaces require further extension of the notion.}
Furthermore, as suggested by Hardt and Moitra \cite{HM},
radial isotropic position can also be thought of as a stable analogue of isotropic position.
That is, while isotropic position has important applications both in algorithms and in exploratory data analysis,
it is rather sensitive to even a small number of outliers.
Radial isotropic position is more robust in the presence of outliers. 


\medskip
\noindent{\bf Specific Applications.}
The problem arises in several, rather diverse applications, and several other studies
address problems of very similar or more general nature. Among these applications and related work we mention

\medskip
\noindent{\bf (i)} 
the derivation of a linear lower bound on the unbounded error probabilistic communication complexity~\cite{For}, 

\medskip
\noindent{\bf (ii)}
robust subspace recovery in the presence of outliers in machine learning~\cite{HM}, 

\medskip
\noindent{\bf (iii)}
algorithmic and optimization aspects of Brascamp-Lieb inequalities~\cite{Garg18},

\medskip
\noindent{\bf (iv)}
a superquadratic lower bound for 3-query locally correctable codes over the reals~\cite{DSW},
and (somewhat more remote but still related) 

\medskip
\noindent{\bf (v)}
a deterministic polynomial-time algorithm for 
approximating mixed discriminants and mixed volumes~\cite{GS00}. 

See below for some additional details.

Another context in which radial isotropy arises (often as a special case) is in
entropy maximization; see the works of Carlen et al.~\cite{CLL}, Singh and Vishnoi~\cite{vishnoi1}
and Straszak and Vishnoi~\cite{vishnoi2}, and also Lee's blog~\cite{Lee} and 
see below for an additional discussion. It also related to so-called finite tight frames 
(see~\cite{waldron} for a recent monograph on this topic).

\paragraph{Point location in high dimensions.}
Last, but not least, is
a recent application in Kane et al.~\cite{KLM2}, who use radial isotropic position for
comparison-based algorithms for point location in a high-dimensional arrangement of hyperplanes.
This work, which served as a starting motivation for our work on this paper,
followed an earlier breakthrough result of Kane et al.~\cite{KLM},
in which they showed that one can solve the 3-SUM problem (decide whether any three
out of $n$ given real numbers sum to $0$) using only $O(n\log^2n)$ simple linear queries on the input.
As a matter of fact, Kane et al.~established in \cite{KLM} a more general result,
namely that one can answer point location queries
in an arrangement of a set $H$ of hyperplanes in $\RR^d$, each of which has
integer coefficients with a small $\ell_1$-norm, using only
$O(d \log d \log |H|)$ simple linear comparisons involving the query point $x$.
Here $d$ is the dimension of the ambient space. (In the 3-SUM application we have
$d=n$ and $|H| = {n\choose 3}$.)

In the follow-up study~\cite{KLM2}, Kane et al.~extended their technique
to sets of general hyperplanes, with arbitrary coefficients, allowing only
two types of comparisons involving the input $x$: \emph{sign tests} 
(determining on which side of a hyperplane $x$ lies), and 
\emph{generalized comparison queries}, in which one asks 
for the sign (positive, negative, or zero) of expressions
of the form $\alpha h_1(x) + \beta h_2(x)$, where $\alpha$ and $\beta$
are arbitrary real parameters. Kane et al.~showed that $O(d^3 \log d \log|H|)$
sign tests and generalized comparisons suffice for point location.\footnote{%
  This performance is worse than the best known recent bound of
  Ezra and Sharir~\cite{ES}, but the queries that their algorithm
  performs, and the space decomposition that is induced by the algorithm, are
  much simpler than those in \cite{ES}.}

A crucial step that the algorithm of \cite{KLM2} performs is transforming the
normals of the input hyperplanes into a set in radial (uniform)
isotropic position by a linear transformation $T$ followed by a normalization.
They then locate $T^{-1}x$ in the arrangement of the transformed hyperplanes
(via standard comparisons, which are equivalent to generalized comparisons in
the original space). After the transformation they have the property that
the sum of the suitably defined squared `scalar products' of the hyperplanes
with any normalized query point $x$ is the same, and sufficiently large,
a fact which is crucial for their analysis. Since the latter property is
all that they need, a relaxed, approximate notion of radial isotropy certainly
suffices for their needs. In their work, since they only measure the number of
sign and (generalized) comparison queries, they do not care about
the cost of computing the transformation that brings the hyperplanes
into (approximate) radial isotropic position. In another recent study
by Ezra et al.~\cite{EHKS}, a full implementation of the first technique
of Kane et al.~\cite{KLM} is presented. Having an efficient procedure
for transforming the input hyperplanes into approximate radial isotropic
position, like the one presented in this paper, facilitates a
straightforward adaptaion of the machinery in \cite{EHKS} to
obtain a full implementation of the second, isotropy-based technique
of Kane et al.~\cite{KLM2} for point location in arrangements of arbitrary
hyperplanes.


\paragraph{Robust subspace recovery.}
A second interesting application, due to
Hardt and Moitra \cite{HM}, studies the problem of \emph{robust subspace recovery}.
In this problem we are given a set $X$ of $n$ vectors in $\RR^d$ and we
want to determine whether there exists some subspace of some dimension
$\ell$ that contains more than $(\ell/d)n$ of these vectors.
The motivation is to detect whether a
dataset $X$ in $\RR^d$ does in fact reside in a lower-dimensional
space if we remove a relatively small subset of \emph{outliers}. Hardt
and Moitra gave a Las Vegas algorithm that makes $O(d^2n)$ iterations
on average, where in each iteration it draws $d$ vectors from $X$,
and when it finds $d$ linearly dependent vectors, it identifies
the subspace that they span as a candidate rich subspace.

Hardt and Moitra also argue that when such a subspace exists, the
vector $\frac{d}{n}{\bf 1}$ must be outside the basis polytope of $X$,
and use this fact, together with a polynomial-time algorithm for
detecting membership in the basis polytope, to derandomize their algorithm.

As follows from the theory of radial isotropic position (that we will review
in Section~\ref{sec:shiri} and Appendix~\ref{sec:shiri-full}),
the condition that $\frac{d}{n}{\bf 1}$ is not in the basis polytope of $X$
is equivalent to the condition that we cannot put $X$ into radial isotropic
position with respect to the vector $\frac{d}{n}{\bf 1}$. Saying it the
other way around, it follows that putting $X$ into radial isotropic position
with respect to $\frac{d}{n}{\bf 1}$ is a proof that there is no
$\ell$-dimensional subspace that contains more than $(\ell/d)n$ vectors of $X$,
for any $\ell$.

Motivated by this observation, Hardt and Moitra applied Barthe's
characterization of attaining radial $c$-isotropic position, in order
to derive the aforementioned polynomial-time ellipsoid-based algorithm that finds such a
certificate (i.e., puts the vectors into radial isotropic position).

\paragraph{Additional related work.}
As mentioned, our problem can be viewed as a special case of a more general problem considered 
by Singh and Vishnoi \cite{vishnoi1} and Straszak and Vishnoi \cite{vishnoi2} on 
entropy maximization. They consider the problem of finding a distribution $q$ of 
maximum entropy over a (possibly large) collection ${\cal F}$ of subsets $F$ of 
$[n] =\{1,2,\ldots,n\}$, among all distributions  with a given vector $c$ of marginals.

A somewhat more distantly related, and more general, line of research is on 
operator scaling and its connection to Brascamp-Lieb constants (see for example 
\cite{Garg18,GS00}). The most recent and closely related among these works is 
by Garg et al.\ \cite{Garg18}, who give an algorithm to compute the 
Brascamp-Lieb constant for a particular Brascamp-Lieb ``datum'' 
(a set of linear transformations and a vector of exponents). 
The algorithm uses an alternative minimization technique (as the problem is not convex) to 
bring the instance into a so called ``geometric position'',
from which it can deduce the desired constant. In this terminology our problem 
is to find a transformation that brings a ``Brascamp-Lieb {\em rank-one} datum'' 
into isotropic position. For rank one this problem is convex and thereby easier. 
It was treated before by Hardt and Moitra \cite{HM} and in a somewhat different 
settings also by Gurvitz and Samorodnitsky \cite{GS00} who applied the ellipsoid 
algorithm to solve it.

\paragraph{Our results.}
See the end of the introduction for a summary of highlights of the novel contributions
of our work.

Given a set $X=\{x_1,\ldots,x_n\}$ of $n$ nonzero vectors in $\RR^d$, and a weight vector 
$c=(c_1,\ldots,c_n)\in \left(\RR^+\right)^n$,
such that there exists a linear transformation sending the $x_i$'s into radial $c$-isotropic position, we develop
algorithms to find such a transformation.

We first consider algorithms that find an exact solution. 
We show how to do it either by solving a system of $n$ polynomial equations, 
in $n$ variables, each of degree $d$, or a different system of $d^2$ equations 
in $d^2$ variables, each of degree $2n$. The running times are $d^{O(n)}$ or 
$n^{O(d^2)}$, respectively, using well known techniques from symbolic algebra 
(or computational real algebraic geometry~\cite{BPR}).

Our main focus, however, is on computing such a transformation approximately and 
considerably more efficiently. That is, we want to compute a linear transformation
that puts the $x_i$'s into radial $c_{\rm apx}$-isotropic position, for some vector $c_{\rm apx}$ that
satisfies $|c_{\rm apx}-c| \le \eps$, say in the $\ell_2$-norm, for some prespecified accuracy parameter $\eps$.

We assume that $X \subset \sph^{d-1}$; this is a natural assumption for radial isotropy,
it occurs in the applications we are aware of, and it involves no loss of generality.

As follows from Barthe's results, detecting whether the $x_i$'s can be put in radial $c$-isotropic position
is equivalent to testing whether the vector $c$ is in the relative interior of
the {\em basis polytope} associated with $X$, given by 
\begin{equation}\label{eq:basic-pol} 
K_X =\conv \Bigl\{ {\bf 1}_S \mid S\subseteq[n], \; |S| = d =
\dim\big(\mathrm{span}\{x_i\mid i\in S\}\big) \Bigr\} ,
\end{equation}
where ${\bf 1}_S$ is the $n$-dimensional indicator vector of the set $S$,
or, equivalently (as proved in~\cite{CLL}), by
\begin{equation} \label{eq:basis-cll}
K_X = \Bigl\{ c\in [0,1]^n \mid \sum_{j=1}^n c_j = d, \text{ and }
	\forall J\subset [n]\, ,\, \sum_{i\in J} c_i \le \dim ({\rm span~} \{x_i\}_{i\in J}) \Bigr\} .
\end{equation}
These two expressions are dual, in a sense: In (\ref{eq:basic-pol}), $K_X$ is defined
as the convex hull of a set of point, whereas in (\ref{eq:basis-cll}) it is defined as the intersection of halfspaces.

The basis polytope plays a central role in matroid theory and submodular optimization;
see, e.g., \cite{Edm}, and there are efficient algorithms for detecting membership in
the basis polytope \cite{Cun,Edm,GLS,IWF01,Sch}. Furthermore, in some common cases, membership
of $c$ in the basis polytope is obvious, as, for example, in the case studied by Forster \cite{For},
where $c_i = d/n$ for each $i$ and the $x_i$'s are in general position.

Barthe's characterization reduces the problem of finding a transformation that puts
the input vectors in radial $c$-isotropic position to a problem of finding a point
$t^*\in\RR^n$ that attains the minimum of a specific convex function $f(t)$, 
defined in terms of $X$ and $c$. Concretely, 
$$
f(t) = \log \det \Bigl( \sum_{i=1}^n e^{t_i} x_i\otimes x_i \Bigr) - \iprod{c}{t} .
$$
It follows that a most natural (and simple) approach, which is the one proposed
in this paper, to finding the transformation that
puts the vectors in radial $c$-isotropic position is to find the minimizing vector $t^*$ using
an appropriate variant of the \emph{gradient descent} technique; see Bubeck~\cite{Bub} for details.
Our main set of results is an analysis of gradient descent applied to $f$, and of various parameters
that affect its efficiency.

We introduce a new concept promising that the vector $c$ lies ``deeply inside'' $K_X$, which is 
related to the representation \eqref{eq:basis-cll}. This notion is different and more robust than  the ones 
in previous works, and we discuss it in more detail and compare it to other parameters 
considered in the literature, in Section \ref{sec:gd}. 
\begin{definition}\label{def:etadelta}
	Let $X = \{ x_i\}_{i=1}^n \subset S^{d-1}\subset \RR^d$ and let $K_X$ be given 
        by \eqref{eq:basic-pol} or \eqref{eq:basis-cll}. We say, for $\eta, \delta >0$, that
	a vector $c$ lies {\em ($\eta, \delta$)-deep inside $K_X$} if,
	for any subspace $E$ with $\dim(E)= k \in \{ 1, \ldots, d-1\}$, we have that
	\begin{equation}\label{eq:definition-of-asterix-condition}
	\sum_{x_j \in E_\delta} c_j \le k(1 - \eta), \qquad\text{where}\qquad
	E_\delta = \{ x\in \sph^{d-1} \mid d(x, E) \le \delta\} \ ,
	\end{equation}	
        and the distance $d(x,E)$ is the Euclidean distance.
\end{definition}

Note that for $\eta=\delta=0$ this condition just says that $c$ is in the basis polytope, by \eqref{eq:basis-cll}.
In general this is a stronger constraint, as we also include in $E_\delta$ vectors that lie close
to $E$, and impose a stricter inequality on the corresponding $c_j$'s.
Our algorithm does not need to compute or know $\eta$ and $\delta$. These parameters are used only for analysis.

\paragraph{The first technique, using smoothness.}
We establish, in Section \ref{sec:gd}, the following results, after discussing, in some detail,
a few variants of the gradient descent method, and investigating in depth the associated parameters
that control their efficiency.

The first set of variants that we use are the \emph{projected gradient descent for smooth functions} \cite[Section 3.2]{Bub}
and \emph{Nesterov's accelerated version} \cite[Section 3.7]{Bub}; see Section~\ref{sec:gd} for more details.

\begin{restatable}{thm}{smooththeorem}
	\label{thm:smoothintro}
	\label{thm:smooth}
	For a vector $c$ that is $(\eta, \delta)$-deep inside $K_X$, we can construct
	a transformation that brings $X$ into radial $c_{\rm apx}$-isotropic position,
        for some vector $c_{\rm apx}$ that satisfies $|c_{\rm apx} - c| \le \sqrt{\eps}$,
	in $O(|t^*|^2/\eps) = O(n|t^*|_\infty^2/\eps)$ iterations of gradient descent 
	(or $O(\sqrt{n}|t^*|_\infty/\sqrt{\eps})$ iterations of accelerated gradient descent),
        where $t^*$ is the (unique) extremizing vector of $f$ with $\min_i t^*_i = 0$.
	Each iteration takes
	$O\left(n d^2 \log\left(\log n + |t^*|_\infty + \log \frac{1}{\eps}\right)\right)$
	arithmetic operations on words of $\log n + |t^*|_\infty + \log \frac{1}{\eps}$ bits.
        Moreover, putting $c_{\rm min} := \min_i c_i$, we have
	\begin{equation} \label{tinf1}
	|t^*|_\infty \le \log \frac{1}{c_{\rm min}} + (d-1) 
	\log \left(\frac{8}{\eta\delta^2}\right) .
	\end{equation}
\end{restatable}

Theorem \ref{thm:smoothintro} implies that, for 
moderate values of $\eps$ (such as $O(d/n)$, which suffices for the applications 
we are aware of), gradient descent should be reasonably fast, especially when 
$c_{\rm min}$ is not too close to $0$. Verifying this experimentally, though, and 
comparing its performance in practice with the other approaches (such as
ellipsoid-based techniques), is left for future research.

\medskip
The two main steps in our proof are as follows. 

\medskip
\noindent
{\bf (1)} In Section \ref{sec:grad-bound} we show that the $\ell_1$-norm of the 
gradient of $f$ is bounded by $2d$ (and the $\ell_2$-norm by $\sqrt{2}d$), for all $t$,
and the computation of $\nabla f$ amounts to computing the singular value decomposition (SVD)
of the vector set $\{e^{t_i/2}x_i\}_{i=1}^n$, 
where $t=(t_1,\ldots,t_n)$ is the current approximation maintained by the
gradient descent. See Section~\ref{sec:iso-svd} for details.
Since SVD is used in numerous applications and is available
in many scientific and statistical packages, the application  of gradient descent
to $f$ is particularly simple to implement.

\smallskip
\noindent
{\bf (2)}
In Section \ref{sec:beta} we show that the largest eigenvalue of the Hessian of 
$f$ is $\le 1/2$ (for all $t$). This justifies using a variant of gradient descent 
for smooth functions (given in Bubeck~\cite[Section 3.2]{Bub}).
Such a variant finds a point $t'$ such that $|f(t')-f(t^*)| \le \eps$ 
in $O(|t^*|^2/\eps)$ steps (or $O(|t^*|/\sqrt{\eps})$ steps if we 
use Nesterov's accelerated gradient descent~\cite[Section 3.7]{Bub}),
when we start the descent from the origin $t=0$. 
We also show (Section \ref{sec:bypass}), using again our bound
on the eigenvalues of the Hessian, that $|\grad f(t') - \grad f(t^*)| \le \sqrt{\eps}$,
which implies in our setting that $t'$ yields a transformation that maps $X$
to radial $c_{\rm apx}$-isotropic position, for some $c_{\rm apx}$
satisfying $|c_{\rm apx} - c|_2 \le \sqrt{\eps}$.

To complete this part of the analysis, we establish an upper bound on $|t^*|$, 
in Section \ref{sec:R}. 
Our bound, stated in Theorem~\ref{thm:smooth},
is logarithmic in the paranmeters $\eta$ and $\delta$ of Definition \ref{def:etadelta} (see (\ref{tinf1})).
We note that Hardt and Moitra \cite{HM}, as well as Singh and Vishnoi \cite{vishnoi1}, use
a different set of parameters for bounding $|t^*|_\infty$ (Singh and Vishnoi 
do this in their more general setting). We use our technique to significantly 
strengthen the bound of Hardt and Moitra \cite{HM}, and finally to deduce Theorem \ref{thm:smoothintro}. 

\paragraph{The second technique, using strong convexity.}

We say that a function $f$ is \emph{$\alpha$-strongly convex} if it satisfies
$$
f(y) - f(x) \ge \nabla f(x)^T (y-x) + \frac{\alpha}{2} |y-x|^2 ,
$$
for any $x$ and $y$.

For smooth and strongly convex functions,
gradient descent converges faster, in $O\left(\kappa\log\frac{|t^*|}{\eps}\right)$ steps
(or in $O\left(\sqrt{\kappa}\log\frac{|t^*|}{\eps}\right)$ steps with Nesterov's
acceleration), where $\kappa$ is the ratio between the largest and smallest positive
eigenvalues of the Hessian; see Section~\ref{sec:gd} for more details.
Unfortunately, our function $f$ is not strongly convex, as there are directions,
such as the all-1 vector ${\bf 1}$, in which $f$ is constant.
Nevertheless, we show that, under certain irreducibility assumptions
(which automatically hold if $c$ is deep inside the basis polytope, in the sense of Definition \ref{def:etadelta}),
and under a suitable definition of the optimization domain, $f$ is strongly convex in that domain.
This gives the  bound stated in the following theorem, which depends only logarithmically 
on $\eps$, but with considerably worse dependence on the other parameters of the problem. 
\begin{restatable}{thm}{strongtheorem}
\label{thm:strongintro}
\label{thm:strong}
For a vector $c$ that is $(\eta, \delta)$-deep inside $K_X$, we can construct
the transformation that brings $X$ into radial $c_{\rm apx}$-isotropic position, 
for some $c_{\rm apx}$ satisfying $|c_{\rm apx} - c|_2 \le \eps$,
in $O\left(\kappa \log \frac{|t^*|}{\eps} \right)$ iterations of gradient descent 
(or $O\left(\sqrt{\kappa} \log \frac{|t^*|}{\eps} \right)$ iterations of accelerated gradient descent).
Each iteration takes $O\left(n d^2 \log\left(\log n + |t^*|_\infty + \log \frac{1}{\eps}\right)\right)$
arithmetic operations on words of $\log n + |t^*|_\infty + \log \frac{1}{\eps}$ bits, with
the same upper bound on $|t^*|_\infty$ as in Theorem~\ref{thm:smooth} and with
\[
\kappa \le
\frac {dn^4e^{2|t^*|_{\infty}} \left( 1 + \frac{\sqrt{n}e^{|t^*|_{\infty}} }{\delta} \right)^{2d}}{2\delta^2}\ .		
\]
If $X$ is in general position then we also have
$$
\kappa \le \frac{e^{4d|t^*|_\infty}} { 2(\Delta_S^{\rm min})^2 } \frac{n(n-1)}{d(n-d)} .  
$$
Here $\Delta_S^{min}$ is the minimal square determinant of a $d$-tuple of vectors from $X$. 
\end{restatable}

The latter result is obtained by a careful analysis of the Hessian of the function $f$, 
given in two completely different forms. The first form works for general $X$ and a 
vector $c$ that is deeply inside the basic polytope, whereas the second bound, 
with better dependence on $n$ and $d$, depends on $X$ being in general position, with a bound that depends  
on the minimum square determinant of any $d$-tuple $S$ from $X$ (as do earlier studies, such as \cite{HM}). 
Finally, since for implementing gradient descent one
pretends to have access to an exact gradient, whereas in practice we compute the gradient only
approximately using SVD, we show in Section \ref{App-Micha-approx-grad}, that we may account 
for these errors within the same asymptotic bounds as in the theorems above. 

\medskip
\noindent{\bf Other algorithmic approaches.}
The aforementioned related work uses two other algorithmic approaches to our problem.
The first approach is already implicit in the proof of Forster~\cite{For}.
As mentioned, Forster considers radial isotropy 
(with $c_i=d/n$ for each $i$ and for vectors in general position) and proposes to
transform the vectors by the mapping $\zz_i = B\xx_i/ | B\xx_i |$,
where $B=\Sigma^{-1}V^T$, and where $V$ and $\Sigma$ are
two of the matrices that are produced as part of the SVD
$X^T = U\Sigma V^T$ of $X^T$ (see Section~\ref{sec:iso-svd} for details).
Forster proves that the smallest
eigenvalue of the linear operator $\sum_{i=1}^n  z_i\otimes z_i$ is
either greater than the smallest eigenvalue of $\sum_{i=1}^n  x_i\otimes x_i$,
or is the same but with a strictly smaller multiplicity.
Using this fact, Forster shows that if we iterate this step it converges
to a set of vectors in radial isotropic position.
However, no guarantee about the rate of convergence is given in \cite{For}.
The transformation that brings $X$ to (approximate)
radial isotropic position is obtained by composing the transformations
used in each iterative step.

Garg et al.~\cite{Garg18} also use a similar algorithm for the more
general problem of bringing a higher-rank Brascamp-Lieb datum into
geometric position. They bound the running time of this approach 
by a polynomial in the bit length of the input, the common denominator of the
entries in the vector $c$ (which they assume are rational whose common denominator
is not too large), and in $1/\eps$, where $\eps$ is the approximation parameter.

The second approach, due to Hardt and Moitra~\cite{HM} (and also used by 
Gurvits and Samorodnitsky \cite{GS00}, Singh and Vishnoi \cite{vishnoi1}, 
and Straszak and Vishnoi \cite{vishnoi2}, in other related settings),
applies an ellipsoid-based procedure for (roughly) halving the region 
containing the minimizing vector $t^*$.
For this, they bound the region in which $t^*$ lies, and quantify the strong convexity of $f$.
The resulting algorithm is polynomial in $\log 1/\eps$, in $1/\gamma$
(where $\gamma$ is another parameter that also measures (roughly) how 
deep is $c$ inside the basis polytope), in $L$ (the bit complexity
of the input vectors and of $c$), and inversely in the minimum 
square determinant $\Delta_S^{\rm min}$ of any $d$-tuple $S$ of linearly 
independent $x_i$'s (as in the second bound in Theorem~\ref{thm:strong}).
Our results here, when plugged into Hardt and Moitra's analysis,
improve their bounds considerably. Hardt and Moitra's algorithm is 
certainly harder to implement than those that use SVD-based gradient descent,
like our algorithms.

Straszak and Vishnoi \cite{vishnoi2} prove (in their more general setting,
of which ours is a special case) that there always exists 
a vector $t'$ such that $|f(t')-f(t^*)| \le \eps$ 
and $|t'|_2$ is polynomial in $\log(1/\eps)$. This independence of
other parameters allows us to actually strengthen our 
Theorems \ref{thm:smooth} and \ref{thm:strong}, by replacing
the upper bound on $|t^*|_\infty$ by the minimum between the actual upper bound 
that we derive here and ${\rm poly}(\log(1/\eps))$ (and
modify the optimization region accordingly; see Equation (\ref{eq:K})).

\paragraph{Summary of the highlights of our contribution.}

\smallskip
\noindent
{\bf (1)} We use gradient descent for computing the minimizer $t^*$,
instead of the other techniques proposed so far in the literature
(and reviewed above). We believe
it to be a superior technique, which is easy to implement and which
should run much faster in practice than the other approaches. \\

\smallskip
\noindent
{\bf (2)} The connection between radial isotropy and SVD, although
already noted by Forster~\cite{For}, is explored here in a deeper and more extended
context, and is shown to be very beneficial both for the algorithms themselves and
for their analysis. \\

\smallskip
\noindent
{\bf (3)} We offer detailed (and fairly nontrivial) analysis of several important 
parameters of the problem, such as $|t^*|_\infty$, the smoothness parameter
$\beta$, the parameter $\alpha$ of strong convexity, and more. \\

\smallskip
\noindent
{\bf (4)} 
We introduce a new notion of being ``deep inside'' the basis polytope, using the parameters $\eta$ and $\delta$
of Definition \ref{def:etadelta}. We use these parameters as an alternative to
the minimum square determinant
$\Delta_S^{\rm min}$ (which was used in previous studies, and which is very
sensitive to even a single `nearly dependent' $d$-tuple in $X$),
which makes our approach more stable with respect to any small perturbation
of the input, and allows us to obtain better bounds for the parameters 
mentioned in (3), and thereby for the performance of gradient descent.\\

\smallskip
\noindent
{\bf (5)} Last, but perhaps not least, we offer a comprehensive
treatment of this fascinating topic, which we believe to be helpful, given the scattered
nature of the existing relevant literature, where the problem is discussed in widely
different contexts, using different styles of terminology, often addressed only as 
a subproblem of other problems, and often receiving rather sketchy treatments,
and suboptimal analysis of its parameters.

\section{Putting a set in radial isotropic position:\\
	Characterization and properties (brief essential summary)} \label{sec:shiri-brief}
\label{sec:shiri}

In this section we  survey the theory of radial isotropy. Several ways to develop this theory
have been proposed in the literature \cite{For,Klartag,Lee,vishnoi1,vishnoi2},
and we follow the approach of Barthe \cite{Bar}, with our interpretation, clarifications, 
and some enhancements, including some features from the analysis of Carlen et al.~\cite{CLL}.
We give a brief summary here that is sufficient to present, develop and analyze our algorithms. 
For a complete presentation, including the proofs, we refer the reader to Appendix \ref{sec:shiri-full}.

To state the various equivalent conditions for attaining radial isotropy, we introduce the basis polytope
associated with a set $X$ of vectors.

\begin{restatable}{definition}{defbasicpoly}
\label{def:basic-poly}
The ``basis polytope'' associated with a set $X=\{x_i\}_{i=1}^n\subset \RR^d$ is the $\{0,1\}$-polytope
\[
K_X =\conv \Bigl\{ {\bf 1}_S \mid S\subseteq[n], \; |S| = d =
\dim\big(\mathrm{span}\{x_i\mid i\in S\}\big) \Bigr\} ,
\]
where ${\bf 1}_S$ is the $n$-dimensional indicator vector of the set $S$.
\end{restatable}

We also use the equivalent representation of $K_X$ given by Carlen et al.~\cite{CLL}.
%
\begin{restatable}{prop}{propcllversion}{\rm [Carlen, Lieb and Loss~\cite{CLL}]}
\label{cll-version}
    Given a set $X=\{x_i\}_{i=1}^n\subset \RR^d$, we have
    \[
    {K}_X = \Bigl\{ c\in [0,1]^n \mid \sum_{j=1}^n c_j = d, \text{ and }
    \forall J\subset [n]\, ,\, \sum_{i\in J} c_i \le \dim ({\rm span~} \{x_i\}_{i\in J}) \Bigr\} .
    \]
\end{restatable}

The theory is developed via the
\emph{Legendre transform} of the function $\Phi: \R^n\to \RR$, defined by
\[
\Phi(t) =  \log \det (Q(t))
\]
where $t=(t_1,\ldots,t_n)$ and
$
Q(t) = \sum_{i=1}^n e^{t_i} x_i \otimes x_i .
$

As long as $X=\{x_i\}_{i=1}^n$ spans $\RR^d$, an assumption that we will make
throughout this paper, the determinant of $Q(t)$ is positive, $Q(t)$ is
positive definite and invertible, and $\Phi$ is everywhere defined and finite.

For a set $S\subseteq [n]$, $|S|=d$, define (as in the introduction)
${\displaystyle \Delta_S = \det( (x_i)_{i \in S})^2 = \det\left(\sum_{i\in S} x_i\otimes x_i\right)}$.
Then $\Delta_S > 0$ if and only if $d =
\dim\big(\mathrm{span}\{x_i\mid i\in S\}\big)$.


\begin{restatable}{lem}{lemcbrestated}
\label{lem:cb}
    The function $\Phi$ is convex, everywhere differentiable,  monotonically increasing in each coordinate, and admits
    the representation
    ${\displaystyle \Phi (t)  = \log \left( \sum_{|S|=d} e^{\sum_{i\in S} t_i} \Delta_S \right)}$,
    where the sum is over all $d$-element subsets $S$ of $[n]$.
    In particular, for $j=1,\ldots,n$,
    \[
    \frac {\partial \Phi}{\partial t_j} (t) =
    \frac{\sum_{\{S\mid j\in S, |S|=d\}} e^{\sum_{i\in S} t_i} \Delta_S }{\sum_{|S|=d} e^{\sum_{i\in S} t_i} \Delta_S },
    \qquad \frac {\partial \Phi}{\partial t_j} (t)>0, \qquad {\rm and} \qquad \sum_{j=1}^n \frac {\partial \Phi}{\partial t_j} (t)=d.
    \]
\end{restatable}

We also have the following alternative form of the gradient of $\Phi$.

\begin{restatable}{lem}{lemdiffoflogdetsum}
    \label{lem:diff-of-logdetsum}
    For every $t\in\RR^n$ and $j=1,\ldots,n$,
    ${\displaystyle \frac{\partial\Phi }{\partial t_j}(t)=  e^{t_j}|Q^{-1/2}(t)x_j|^2}$.
\end{restatable}
(As discussed in the introduction, $Q^{-1/2}(t)$ exists for every $t$.)
The \emph{Legendre dual} of $\Phi$, denoted by $\Phi^*$, is a map from $\RR^n$ to $\RR\cup\{+\infty\}$,
given at $\xi\in \RR^n$ by
\[
\Phi^*(\xi) = \sup_{t\in \RR^n} \Bigl\{ \iprod{t}{\xi} - \Phi(t) \Bigr\} .
\]

It is not hard to see that the domain of
$\Phi^*$ (i.e., the region where it is finite) is contained in 
$\{\xi \in \RR^n \mid \sum \xi_i = d,\;\xi_i\ge 0\;\text{for each $i$} \}$. 
A complete characterization of the domain is given by 

\begin{restatable}{lem}{propPhiSfiniterestated}
    \label{prop:PhiSfinite-iff-attained-iff-cinK}
    For a set $X$ of $n$ vectors $x_1,\dots,x_n\in\sph^{d-1} \subset \R^d$ and a vector $c\in (\R^+)^n$,
    $\Phi^*(c)$ is finite if and only if $c \in K_X$.
\end{restatable}

Radial $c$-isotropy is linked with not only the finiteness of $\Phi^*(c)$, but also with it actually
being attained, namely with the property that there exists $t^*\in \RR^n$ such that
$
\Phi^*(c) = \iprod{t^*}{c}-\Phi(t^*) .
$

\begin{restatable}{lem}{lemattainedrestated}
    \label{lem:attained}
    The supremum in the definition of $\Phi^*(c)$ is (finite and) attained if and only if 
    $c$ is in the relative interior of $K_X$.
\end{restatable}

Finally, Barthe's celebrated characterization of when $X$ can be put in
radial $c$-isotropic position is as follows.

\begin{restatable}{prop}{propPhiSfiniteiffradiposexistsrestated}
    \label{prop:PhiSfinite-iff-rad-i-pos-exists}
    For a set $X$ of $n$ vectors $x_1,\dots,x_n\in\sph^{d-1} \subset \R^d$ and a vector $c\in (\R^+)^n$,
    $\sum_i c_i = d$,
    $\Phi^*(c)$ is finite and attained at some point $t^*$ if and only if $X$ can be put in radial $c$-isotropic
    position. Moreover, $\Phi^*(c)$ is attained at the point $t^*\in \RR^n$ if and only if
     the matrix $Q^{-1/2}(t^*)$ puts $X$ in radial $c$-isotropic position,
    where, as above, $Q(t^*) = \sum_{i=1}^n e^{t^*_i} x_i\otimes x_i$.
\end{restatable}

Clearly, the vector $t^*$ that attains $\Phi^*(c)$ has to satisfy 
$\nabla \Phi(t^*) = c$. We will show, in Lemma~\ref{lem:unique:rip}, that, under a suitable
irreducibility assumption (described shortly), the radial $c$-isotropic position is unique up to rotation.
\smallskip

The function $\Phi$ is linear in the all-1 direction ${\bf 1} = {\bf 1}_{[n]}$. 
The existence of other directions of linearity turns out to depend on the dimension of the polytope
$K_X$ and is equivalent to a notion of irreducibility that we describe next.
Given a set $X = \{x_i\}_{i=1}^n$ of vectors in $\RR^d$,
we define an equivalence relation  $\sim$ on $[n]$, considered in Barthe~\cite{Bar}, as follows: 
Two indices $i,j$ satisfy $i\sim j$ if there exists a subset $S\subset [n]$, 
$|S| = d-1$, such that $S_1 = S\cup\{i\}$ and $S_2 = S\cup\{j\}$ both satisfy $\Delta_{S_i}>0$ 
(we show in Appendix \ref{sec:transitivity}, that this relation is  
an equivalence relation\footnote{%
    In earlier treatments (as in \cite{Bar}), transitivity was not established, and the relation
    was made transitive by taking its transitive completion.}).
The equivalence classes of $\sim$ form a partition  $[n] = \sigma_1 \cup \cdots \cup \sigma_k$.
Letting $E_j = \mathrm{span}\{ x_i\mid i\in \sigma_j\}$, for $j=1,\ldots,k$, it was shown in Barthe~\cite{Bar} that 
$\RR^d = \bigoplus_{j=1}^k E_j$, and this is the maximal splitting of
$\RR^n$ into a direct sum of subspaces that collectively contain all the vectors of $X$.

\begin{restatable}{lem}{lemequivrelrestated}
    \label{lem:equivrel}
    The dimension of the affine hull of $K_X$ is equal to $n-k$, where $k$
    is the number of equivalence classes of the equivalence relation $\sim$.
\end{restatable}

When there is only {\em one} equivalence class, we say that $X$ is {\em irreducible};
in this case the only affine subspace that contains $K_X$ is $\sum z_i = d$.
In the reducible case we have additional affine subspaces that
contain $K_X$, given by $\sum_{i\in\sigma_j} z_i = \dim (E_j)$, for each $j=1,\ldots,k$, 
and additional linearity directions, as shown in the following lemma. 

\begin{restatable}{lem}{lemwhenlinrestated}
     \label{lem:whenlin}
    Assume that the set $X$ admits a maximal splitting into subsets $X_j$ contained in
    respective components $E_j$ of a direct sum $\RR^d = \bigoplus_{j=1}^k E_j$,
    with a corresponding maximal index splitting $[n] = \sigma_1\cup \cdots \cup \sigma_k$,
    so that $E_j = {\rm span} \{ x_i \mid i\in \sigma_j\}$ for each $j$. Then
    $\Phi ((1-\lambda) t + \lambda s) = (1-\lambda)\Phi ( t) + \lambda \Phi(s)$, for $\lambda\in \RR$,
    if and only if $t = s + \sum_{i=1}^k \alpha_j {\bf 1}_{\sigma_j}$, for any choice of scalars $\alpha_j$.
\end{restatable}

\smallskip

When $\Phi$ is reducible, we get a partition of $\RR^d$ into a nontrivial direct sum
$\bigoplus_{j=1}^k E_j$.
If $E_i$ and $E_j$ are orthogonal for every pair $i\ne j$ then one can check that
$\Phi(t) = \sum_{j=1}^k \Phi_j((t_i)_{i\in \sigma_j})$,
where $\Phi_j$ is the restriction of $\Phi$ to $E_j$.
Consequently, everything else factorizes too, so 
$\Phi^*(c) = \sum_{j=1}^k \Phi_j^*((c_i)_{i\in \sigma_j})$,
and the domain of $\Phi^*$ is the intersection of the domains of the $\Phi^*_j$'s.
Furthermore, the optimization problem of finding a vector $t$ that maximizes $\iprod{c}{t} - \Phi(t)$,
which we handle using gradient descent (in Section \ref{sec:gd}), decomposes too into 
 irreducible independent subproblems.

In case $\Phi$ is reducible but some pairs of the subspaces in the decomposition $\RR^d=\bigoplus_{j=1}^m E_j$
are not orthogonal, we first transform $X$ into isotropic position (not necessarily radial).
After the transformation, the decomposition becomes orthogonal and the problem decomposes,
as the following lemma shows.

\begin{restatable}{lem}{orthinisorestated}
    \label{orth:iniso}
    Let $X$ be a set of $n$ vectors in $\RR^d$ in  $c$-isotropic position, and assume that
    $\RR^d$ is the direct sum $F_1 \oplus F_2$ such that each vector of $X$ lies in $F_1\cup F_2$.
    Then $F_1$ and $F_2$ are orthogonal.
\end{restatable}

\section{Isotropy and SVD} \label{sec:iso-svd}
The singular value decomposition (SVD) plays a crucial role in
our implementation and analysis of gradient descent  for computing
a transformation that brings $X$ into radial $c$-isotropic position. 
We review in this section the minimal essential background on SVD (see Blum et al.~\cite{BHK} for more details). 

The SVD of an arbitrary $n\times d$ matrix $A$ is a
decomposition of $A$ as  $A = U\Sigma V^T$, where $U$ is an $n\times d$ matrix,
whose $d$ columns, the so-called \emph{left-singular vectors} of $A$, are mutually orthonormal,
$\Sigma$ is a $d\times d$ diagonal matrix whose entries are the (nonnegative)
\emph{singular values} of $A$, denoted as $\sigma_1,\ldots,\sigma_d$, and $V$
is an orthonormal $d\times d$ matrix, whose columns are called the
\emph{right singular vectors} of $A$.

The $d\times d$ matrix $B=\Sigma^{-1} V^T$ will play a crucial role in our analysis. Note that
$B^T B = V\Sigma^{-2} V^T = (A^TA)^{-1}$. Also note that $BA^T = (\Sigma^{-1} V^T) A^T =U^T$.

In view of Lemma~\ref{lem:diff-of-logdetsum}, the computation of the gradient of $\Phi$ can be performed by computing
the SVD decomposition  $U\Sigma V^T$ of $X^T(t)$ where $X(t)$ has column vectors $\{e^{t_i/2}x_i\}_{i=1}^n$. Indeed, in this case 
$Q(t) = X(t)X^T(t) =V(t)\Sigma^2(t) V^T(t)$ and thus  
$Q^{-1/2}(t) = V(t)\Sigma^{-1}(t)V^T(t)$, allowing us to conclude that 
$\nabla \Phi(t) = \Bigl( e^{t_j}|\Sigma^{-1}V^T x_j|^2 \Bigr)_{j=1}^n$. 
Here, for convenience, we prefer to work with the non-symmetric version
$Q^{-1/2}(t) = \Sigma^{-1}(t)V^T(t)$; the only difference is that
the former version rotates space (by $V(t)$) after applying the latter version, 
which is irrelevant for the analysis.
Furthermore, with the suitable assumptions on $c$, there is a $t^*$ 
(as specified in Lemma \ref{prop:PhiSfinite-iff-rad-i-pos-exists})
such that $\nabla \Phi(t^*) = c$, and the same $t^*$ is such that
$Q^{-1/2}(t^*)$ (again, obtained from the SVD of $X(t^*)$)
brings $X$ into radial $c$-isotropic position.

The SVD also plays a crucial role in the analysis of the convergence rates
of our algorithms, and shows up, in particular, in the representation of the 
Hessian matrix $\nabla^2 \Phi$ (Section \ref{sec:beta}), and the analysis of 
the range of its singular values (Section \ref{sec:alpha}). This is, in our opinion,
a fascinating connection between SVD and the theory of radial $c$-isotropy in general,
and gradient descent in particular. This connection (to a lesser extent) was 
implicit in the work of Forster~\cite{For}.

\smallskip
\noindent
{\bf Computing the SVD.}
There are several well established methods for computing (a numerical approximation of)
the SVD of a given $n\times d$ matrix $A$ of real numbers. The fastest methods
\cite{Golub2013,Stewart2001} take $O\left(n d^2 \log\log \frac{1}{\eps}\right)$
arithmetic operations (on words of $\log \frac{1}{\eps}$ bits) to
compute the singular values and the singular vectors, up to an
additive  error of $\eps |A|$, where $|A|$ is the Frobenius norm of $A$.

\section{Computing approximate radial isotropy via gradient descent} \label{sec:gd}

\subsection{An exact solution} \label{sec:exact}

Before embarking upon a systematic study of the use of gradient descent to
compute (an approximation of) the extremizing vector $t^*$, we first consider briefly
the issue of exact computation of $t^*$. We recall that the problem is to find the
exact vector $t^*$ such that $Q^{-1/2}(t^*)$ puts $X$ into radial isotropic position.
We recall that, by Lemma \ref{prop:PhiSfinite-iff-rad-i-pos-exists},
for $t^*$ to exist, $c$ must be in the relative interior of
$K_X$. So for every $S\subseteq [n]$,
such that $|S|=d$ and $\Delta_S > 0$, there exists a coefficient $\lambda_S \ge 0$ such that
$\sum_{|S|=d,\;\Delta_S > 0} \lambda_S = 1$, and
\[
\sum_{|S|=d,\; \Delta_S > 0,\; i\in S} \lambda_S = c_i ,
\quad\text{for $i=1,\ldots,n$} \ .
\]
By the proof of Lemma \ref{lem:attained} we get that the parameters $\lambda_S$ are related to
the extremizing vector $t^*=(t_1,\ldots,t_n)$ as follows.
\[
\lambda_S   = \frac{ e^{\sum_{i\in S}t_i}\Delta_S}{\sum_{|S'|=d  }e^{\sum_{i\in S'}t_i}\Delta_{S'}} ,
\]
for each $|S|=d$ with $\Delta_S > 0$.

This implies that we can find $t$ by solving
the following system of $n$ equations in the $n$ variables $t_1,\ldots,t_n$.

$$
c_i \sum_{|S| = d,\; \Delta_S>0} \Delta_Se^{\sum_{j\in S} t_j} =
\sum_{|S|=d,\; \Delta_S > 0,\; i\in S} \Delta_Se^{\sum_{j\in S} t_j} ,
\quad\text{for $i=1,\ldots,n$} .
$$
Write $\zeta_i = e^{t_i}$ for $i=1,\ldots,n$. Then, for any $d$-tuple $S$,
$$
e^{\sum_{i\in S} t_i} = \zeta^S := \prod_{i\in S} \zeta_i ,
$$
and we get the following polynomial system
$$
c_i \sum_{|S| = d,\; \Delta_S>0} \Delta_S \zeta^S =
\sum_{|S|=d,\; \Delta_S > 0,\; i\in S} \Delta_S \zeta^S ,
\quad\text{for $i=1,\ldots,n$} .
$$
This is a system of $n$ polynomial equations in the $n$ variables $\zeta_1,\ldots,\zeta_n$,
where each equation is a polynomial of degree $d$. Using the extensive theory of
computational real algebraic geometry, as presented, e.g., in Basu et al.~\cite{BPR}, we can solve this
system exactly, in the sense that the solution is represented symbolically and implicitly,
and any polynomial equation or inequality in the $\zeta_j$'s can be settled correctly
by a discrete procedure. For example, one can use the algorithms 12.16 or 12.17~in \cite{BPR},
whose running time is $d^{O(n)}$. (the degree of the polynomials raised to an exponent proportional to the
number of variables.) Resolving (exactly) any polynomial equality or inequality of
constant degree in the $\zeta_i$'s also takes $d^{O(n)}$ time.

Another exact (algebraic) solution can be obtained by solving explicitly
for the entries $T_{ij}$ of $T = Q^{-1/2}(t^*)$. Specifically, there are $d^2$ such entries,
and the equations that they need to satisfy are given by Definition \ref{def:T}; that is,
$$
\sum_{i=1}^n c_i \frac{Tx_i}{|Tx_i|}
\otimes\frac{Tx_i}{|Tx_i|} = I_d.
$$
This yields a system of $d^2$ equations, one for each combination of a row and
a column, in the $d^2$ entries of $T$.
The left-hand side of each equation is the sum of $n$ fractions, where
both the numerator and denominator of each fraction are quadratic expressions
in the entries of $T$. In other words, after canceling denominators,
each equation involves a polynomial in $d^2$ variables of degree $2n$.
Solving this system can be done using the same techniques from computational
real algebraic geometry mentioned above. The running time is bounded by $n^{O(d^2)}$.
This method is faster than the preceding solution when $n > cd^2$, for a suitable
constant $c$; otherwise the former approach is faster. Regardless of which
approach is faster, they are both highly inefficient. In the rest of this section
we will consider substantially faster approximating solutions.

\subsection{Approximating $t^*$ via gradient descent: An overview}

Let $c$ be a given vector in the relative interior of $K_X$.
By the analysis of Section \ref{sec:shiri}, the problem of finding 
a linear transformation that brings $X$ to radial $c$-isotropic position
amounts (see Lemma \ref{prop:PhiSfinite-iff-rad-i-pos-exists}) to finding a vector $t^*$
at which the supremum $\Phi^*(c)$ is (finite and) attained.
To this end, we proceed to present and analyze algorithms that
compute an approximation to a minimizing vector $t^*$ for the function
\begin{equation} \label{eq:ourf}
f(t) := \Phi(t)- \iprod{c}{t}
= \log\det\left(\sum_{i=1}^ne^{t_i}x_i\otimes x_i\right) - \iprod{c}{t} ,
\end{equation}
which is the negation of the function used to define $\Phi^*(c)$ in
Section~\ref{sec:shiri}. As noted,
at any point $t^*$ where $f$ is minimized we have $\nabla\Phi (t^*) = c$.

\medskip
\noindent
{\bf Gradient descent.}
Our algorithms are based on the classical \emph{gradient descent} technique, as presented,
e.g., in Bubeck~\cite[Chapters 3 and 4]{Bub}.

Since gradient descent only converges to the optimum $t^*$ (without attaining
it exactly), we make do with seeking a vector $t$ (call it $t_{\rm apx}$) 
for which the vector $\nabla f(t_{\rm apx})= \nabla\Phi (t_{\rm apx}) - c$
is sufficiently small, in which case we put $c_{\rm apx} := \nabla \Phi (t_{\rm apx})$
and use the matrix $Q^{-1/2}(t_{\rm apx})$ to map $X$ to radial $c_{\rm apx}$-isotropic
position, which, provided the vectors $c$ and $c_{\rm apx}$ are sufficiently close,
is good enough for the applications at hand (e.g., the one in \cite{KLM2}, and also in \cite{HM}).
The previously developed approaches, as reviewed in the introduction, are also approximate, in a similar sense.

The approximation provided by gradient descent can come in two flavors:
When the convex function $f$ that we minimize can be shown to be strongly convex 
(as defined in the introduction; see also below for the precise definition) we get a point $t_{\rm apx}$ such that 
$|t_{\rm apx}-t^*| \le \eps$, for our prespecified approximation parameter $\eps$.
When strong convexity cannot be guaranteed (or when we do not wish to exploit it), 
we get a point $t_{\rm apx}$ such that $|f(t_{\rm apx})-f(t^*)| \le \eps$.

We prove (in Lemma~\ref{lem:hnorm}) that our function $f(t)$, 
as defined in Equation (\ref{eq:ourf}), is smooth. Specifically, we show that
the Hessian of $\Phi$ (and thus of $f$) is bounded from above by $\frac12 I_n$
(as an inequality between positive definite symmetric matrices).
This property immediately implies that if we get a point $t_{\rm apx}$ such that
$|t_{\rm apx}-t^*| \le \eps$ then $|c_{\rm apx} := \nabla\Phi (t_{\rm apx}) - c|\le \eps/2$, as desired.

For the other kind of approximation, we show in Section \ref{sec:bypass} that for any smooth convex function $g$, a
point $t_{\rm apx}$ such that $|g(t_{\rm apx})-g(t^*)| \le \eps$ (where now $t^*$
is the minimizer of $g$) also
satisfies that $|\nabla g(t_{\rm apx})| \le b \sqrt{\eps}$ for some constant $b$ 
that depends on the smoothness of $g$ (Lemma~\ref{smallf:smallgrad}).
It then follows that if we find a point $t_{\rm apx}$ such that 
$|f(t_{\rm apx})-f(t^*)| \le \eps$ then $|\nabla\Phi (t_{\rm apx}) - c|\le b\sqrt{\eps}$,
for a suitable constamt $b$. Therefore, we do not necessarily need
to find a point $t$ which is close to the minimizer $t^*$ of $f$,
and it suffices to find a point $t$ for which $f(t)$ is close to its minimal 
value $f(t^*)$ (with a suitable adjustment of the error parameter $\eps$), 
to get the desired approximate minimizer of the gradient of $f$.

We continue with a brief review of the variants of the general gradient descent technique
that we  use, and of our results.

\medskip
\noindent
{\bf Gradient descent for smooth functions.}
This is the first version of gradient descent that we use. It
applies to cases where the gradient is Lipschitz continuous, satisfying the condition
\begin{equation} \label{eq:smooth}
|\nabla f(x) - \nabla f(y)| \le \beta |x-y| ,
\end{equation}
for a suitable $\beta>0$ and for any $x,y\in \RR^n$ (ours is such a case, with $\beta=1/2$,
as follows from the aforementioned bounds on the Hessian of $\Phi$).

The method then starts at some $t^{(1)}$ (say, $t^{(1)} = {\bf 0}$),
and iterates the step
\[
t^{(i+1)} = t^{(i)} - \frac{1}{\beta}\nabla f(t^{(i)}) .
\]
As shown, e.g., in \cite[Theorem 3.3]{Bub}, we then have, after $s+1$ iterations,
\begin{equation} \label{eq:sgd}
| f(t^{(s+1)}) - f(t^*) | \le \frac{2\beta |t^{(1)}-t^*|^2}{s} = \frac{2\beta |t^*|^2}{s} .
\end{equation}

An accelerated form of this method, proposed by Nesterov (see \cite[Section 3.7]{Bub}) goes
as follows. Define the sequences
\[
\lambda_0 = 0 ,\qquad
\lambda_m = \frac12 \left( 1 + \sqrt{1+4\lambda_{m-1}^2} \right) ,\qquad\text{and}\qquad
\gamma_m = \frac{1-\lambda_m}{\lambda_{m+1}} 
\]
(we have $\gamma_m < 0$).
Then, starting at an arbitrary $t^{(1)} = \zeta^{(1)}$, we repeatedly compute
\begin{align*}
\zeta^{(i+1)} & = t^{(i)} - \frac{1}{\beta}\nabla f(t^{(i)}) \\
t^{(i+1)} & = (1-\gamma_i) \zeta^{(i+1)} + \gamma_i \zeta^{(i)} .
\end{align*}
Using gradient descent with Nesterov's acceleration, we get faster convergence rate 
with $s^2$ in the denominator of (\ref{eq:sgd}) instead of $s$ (see \cite[Theorem 3.19]{Bub}).
It follows that to get a point $t_{\rm apx}$ such that $|f(t_{\rm apx})-f(t^*)| \le \eps$
we have to run $O(\beta |t^*|^2/\eps)$ steps of gradient descent (or $O(\sqrt{\beta} |t^*|/\sqrt{\eps})$ 
steps of accelerated gradient descent). By our discussion above, this guarantees that
$|\nabla\Phi (t_{\rm apx}) - c|\le b\sqrt{\eps}$, for a suitable constant $b$.

As will follow from our analysis, it is desirable (actually essential) 
to constrain gradient descent so as to keep the $t^{(i)}$'s in 
some convex set $K$ (say a ball of certain radius with respect to some norm)
that is guaranteed to contain $t^*$ and is sufficiently small,
so as to ensure that various parameters that control the convergence 
of gradient descent do not become too large. To do so,
we change the gradient descent step to 
\begin{equation} \label{proj:grad:step}
t^{(i+1)} = P_K \Bigl(t^{(i)} - \frac{1}{\beta}\nabla f(t^{(i)}) \Bigr),
\end{equation}
where $P_K$ is the projection operator on $K$. This variant, called 
\emph{projected gradient descent}, has the same convergence properties as  
unconstrained gradient descent (with a somewhat worse constant).

To bound the performance of the technique, we establish, in
Section \ref{sec:R}, a new bound on $|t^*|$, which we will use to
define the set $K$ within which we want to operate. Motivated by
the characterization of the basis polytope $K_X$ in Proposition \ref{cll-version},
we recall Definition \ref{def:etadelta}, in which we introduced 
parameters $\eta, \delta >0$, and defined a vector $c$ to lie
{\em ($\eta, \delta$)-deep inside $K_X$} if,
for any subspace $E$ with $\dim(E)= k \in \{ 1, \ldots, d-1\}$, we have that
\[
  \sum_{x_j \in E_\delta} c_j \le k(1 - \eta), \qquad\text{where}\qquad
  E_\delta = \{ x\in \sph^{d-1} \mid d(x, E) \le \delta\} \ .
\]
The case $\eta= \delta=0$ is precisely the alternative condition that $c\in K_X$
(see Proposition~\ref{cll-version}). We also argue that for any $c$ to be $(\eta,\delta)$-deep 
inside $K_X$, $X$ must be irreducible (which we may assume to be the case if we first
decompose the problem into its irreducible parts). 

Then we prove (in Lemma \ref{R-cll}) that the extremizing vector $t^*$, with 
$\min_i t^*_i=0$, satisfies\footnote{%
  $t^*$ is unique under this constraint, assuming irreducibility.}
\begin{equation} \label{our-tinf1}
|t^*|_\infty \le \log \frac{1}{c_{\rm min}} + (d-1) 
\log \left(\frac{8}{\eta\delta^2}\right) ,
\end{equation}
where $c_{\rm min}$ is the smallest coordinate of $c$.

In each step of gradient descent we have to compute $\nabla f (t)$
for the current value $t = t^{(i)}$.
As we will show (and as discussed in Section~\ref{sec:iso-svd}), 
we do this by computing the SVD of the matrix $X(t)^T=\{e^{t_i/2} x_i \}^T$.
As mentioned in Section \ref{sec:iso-svd},
we can compute this SVD up to an additive error of $\eps'|X(t)|$,
in $O\left(n d^2 \log\log \frac{1}{\eps'}\right)$
arithmetic operations on words of $\log \frac{1}{\eps'}$ bits, for any
prespecified $\eps'>0$. In our case, we only have to run gradient 
descent as long as $|\nabla f(t^{(i)})| > \eps$, and therefore, as 
we argue in Section \ref{sec:apx:gd}, it suffices to compute the 
gradient up to an (additive) accuracy of $\Theta(\eps^3)$, say
(which will guarantee a relative error of at most $\eps^2$, in terms of
the Euclidean norm, between the real and the approximate gradients).
This means that we have to compute the SVD using the parameter
$\eps' = \eps^3/|X(t)|$. If the lengths of all the $t^{(i)}$'s that we get, while 
running gradient descent, are $O(|t^*|)$,
then $|X(t)| = O(ne^{|t^*|_\infty/2})$.
It follows that we can compute the gradient to an accuracy of $\eps$
in $O\left(n d^2 \log(\log n + |t^*|_\infty + \log \frac{1}{\eps})\right)$
arithmetic operations on words of $\log n + |t^*|_\infty + \log \frac{1}{\eps}$ bits. 
To guarantee that gradient descent does not work with $t$'s of norm much larger 
than $|t^*|_\infty$, we need to use projected gradient descent (see (\ref{proj:grad:step}))
within the region $K$ defined by Equation (\ref{eq:K}) below. 
We show that the above approximation (in computing the gradient)
also suffices for the projected gradient descent technique.
This gives us the following main result, already stated in the introduction,
and reproduced here for convenience.

\smooththeorem*


Section \ref{sec:R} also discusses earlier approaches for measuring how deep inside $K_X$ is the vector $c$.
We show that our bound is stronger than previous bounds (which were stated in terms of different parameters).

\smallskip
\noindent
{\bf Remark.} Straszak and Vishnoi \cite{vishnoi2} prove (in their more general setting,
of which ours is a special case) that there always exists 
a vector $t'$ such that $|f(t')-f(t^*)| \le \eps$ 
and $|t'|_2$ is polynomial in $\log(1/\eps)$. This independence of
other parameters allows us to actually strengthen our 
Theorems \ref{thm:smooth} and \ref{thm:strong} (below), by replacing the
upper bound on $|t^*|_\infty$ by the minimum between the actual upper bound 
that we derive here and ${\rm poly}(\log(1/\eps))$ (we also have to 
modify the optimization region $K$ accordingly).

\medskip
\noindent
{\bf Gradient descent for strongly convex functions.}
We continue to assume that $c$ is $(\eta,\delta)$-deep inside $K_X$, 
for suitable positive parameters $\eta$, $\delta$.
As in Bubeck~\cite[Section 3.4]{Bub}, $f$ is said to be
\emph{$\alpha$-strongly convex} over a convex domain $K$ if, 
for any pair of points $x$, $y\in K$, we have
\begin{equation} \label{eq:strong}
f(y) - f(x) \ge \nabla f(x)^T (y-x) + \frac{\alpha}{2} |y-x|^2 .
\end{equation}
Let $\alpha$ and $\beta$ be, respectively, the smallest and largest eigenvalues
of the Hessian of $f$, extremized over $t\in K$ (that is, maximizing $\beta$ and minimizing $\alpha$).
Then $f$ is both $\beta$-smooth and $\alpha$-strongly convex over $K$ (see, e.g., \cite{Bub}).
Set $\kappa := \beta/\alpha$, and apply the projected gradient descent technique,
for the region $K$, as defined in (\ref{proj:grad:step}),
for $s+1$ steps. Then one has (see \cite[Theorem 3.10]{Bub})
\begin{equation} \label{eq:sconvx}
|t^{(s+1)}-t^*| \le e^{-s/(2\kappa)} |t^{(1)}-t^*| = e^{-s/(2\kappa)} |t^*| ,
\end{equation}
if we start the process with $t^{(1)}={\bf 0}\in K$.


For smooth and strongly convex functions, Nesterov's accelerated gradient descent
proceeds as follows. With $\alpha$, $\beta$, and $\kappa = \beta/\alpha$, as above,
we start at an arbitrary $t^{(1)} = \zeta^{(1)}$, and then repeatedly compute
\begin{align*}
\zeta^{(i+1)} & = t^{(i)} - \frac{1}{\beta}\nabla f(t^{(i)}) \\
t^{(i+1)} & = \left(1 + \frac{\sqrt{\kappa}-1}{\sqrt{\kappa}+1} \right) \zeta^{(i+1)} -
\frac{\sqrt{\kappa}-1}{\sqrt{\kappa}+1} \zeta^{(i)} .
\end{align*}
Using Nesterov's acceleration, we get a faster convergence rate 
with $\sqrt{\kappa}$ in the denominator of the exponent, instead of $\kappa$, 
in Equation (\ref{eq:sconvx}) (see \cite[Theorem 3.18]{Bub}).
It follows that 
to reach a point $t_{\rm apx}$ satisfying $|t_{\rm apx}-t^*| \le \eps$, we have 
to run $O\left(\kappa \log \frac{|t^*|}{\eps} \right)$ steps of the gradient 
descent (or $O\left(\sqrt{\kappa} \log \frac{|t^*|}{\eps} \right)$ steps of
the accelerated gradient descent).

This version has an exponential rate of convergence, so, in principle, it is certainly
the method of choice for functions that are both smooth and strongly convex.
Another advantage is that the technique approximates $t^*$ rather than $f(t^*)$, 
although, as already noted, this is not an essential feature in our case
(except for a better dependence on $\eps$ when $t^*$ is the parameter being approximated).

However, if we want to use the more efficient variant of gradient descent, 
for strongly convex functions, we need to address the issue, noted above, that
there are directions in which our $f$ is not strongly convex---it is actually
constant in these directions. 
Moreover, $f$ cannot satisfy (\ref{eq:strong}), for any $\alpha$, 
when $|x|$ and $|y|$ are very large (this is a consequence of the fact 
that $\nabla f$ is bounded---see Lemma~\ref{lem:gradbounded}).

To deal with the latter difficulty, we take $K$ to be the
intersection of the $L_\infty$-ball
\begin{equation} \label{eq:K}
  |t|_\infty \le \log \frac{1}{c_{\rm min}} 
  + (d-1) \log \left(\frac{8}{\eta\delta^2}\right) 
\end{equation}
with the positive orthant (we can do that since the upper bound on $|t^*|_\infty$ 
is under the assumption that $t^*_{\rm \min}=0$). Then $K$ contains $t^*$,
by the upper bound in Equation (\ref{our-tinf1}). 

Consider now the directions of linearity of $\Phi$. 
We showed in Section~\ref{sec:shiri} (see also Appendix~\ref{sec:shiri-full}) that when $X$ is reducible,
the problem fully decomposes into separate ``irreducible'' subproblems, each of which can be handled
individually. We may therefore assume that $X$ is irreducible,\footnote{%
  The assumption that $c$ is $(\eta,\delta)$-deep inside $K_X$, for $\eta>0$ 
  and $\delta>0$, holds only when $X$ is irreducible.} 
and then the only direction of linearity of $\Phi$ (and constancy of $f$) is ${\bf 1}$.
This means that $\nabla f(t)$ is in the orthogonal complement of ${\bf 1}$, which we denote by $E_0$. 
Hence, if we start with $t^{(1)}\in E_0$ (that is, $\sum_{j=1}^n t^{(1)}_j = 0$), 
and use the unconstrained gradient descent method, we stay in $E_0$ throughout the iterative process.
For projected gradient descent this is not true, since the projection into $K$ is outside of $E_0$
(we have $K\cap E_0 = \{{\bf 0}\}$). Therefore we run our projected gradient descent over $K_0 :=P_{E_0}K$.
Since $t^* \in K$, and $f$ is constant over the fibers $t + \RR{\bf 1}$, it follows that $P_{E_0}(t^*) \in K_0$ 
is also a minimizer of $f$ (albeit not with the minimum coordinate equal to $0$), 
and we can use the version of gradient descent for 
strongly convex functions to find (i.e., approximate) the minimum of $f$ over $K_0$. 
We note that projection into $E_0$ can increase the $\ell_\infty$ norm of a vector by at most a factor of $2$, so that our bounds on $|t^*|_\infty$ change only by a factor of $2$. 

The performance of this version of gradient descent
depends on how strong is the convexity of $f$ over $K_0$.
That is, we seek a lower bound on $\alpha = \alpha(t)$, as large as possible, 
for which Equation (\ref{eq:strong}) holds for every $x,y\in K_0$.
We give two different lower bounds on $\alpha(t)$.
First, in Section \ref{sec:alpha}, we show that the Hessian $H(t)$,
at any given point $t\in E_0$, is a strictly positive operator,
and we can bound from below its smallest eigenvalue, $\alpha=\alpha(t)$
(which can also be used in the definition of strong convexity (\ref{eq:strong})), by
\begin{equation} \label{uiuj-lb}
\alpha(t) \ge
\frac{\delta^2} {dn^4e^{2|t|_{\infty}} \left( 1 + \frac{\sqrt{n}e^{|t|_{\infty}} }{\delta} \right)^{2d}}\ .
\end{equation}
Second, in Section~\ref{App-more-on-strong-conv}, we give a different lower bound, which
so far we only have for $X$ in general position, and which does depend on
$\Delta_S^{\rm min}$. It is
\begin{equation} \label{alpha:shiri}
\alpha(t) \ge
\frac{ (\Delta_S^{\rm min})^2 }{e^{4d|t|_\infty}}\frac{d(n-d)}{n(n-1) } .  
\end{equation}
Substituting our upper bound on $|t|_\infty$ in $K$, given in Equation (\ref{eq:K}), 
we get two uniform lower bounds on $\alpha(t)$ for all $t\in K_0$.
Combining these bounds, we get the following theorem, already stated in the introduction. 
\strongtheorem*

The bounds in Theorem \ref{thm:strong} have logarithmic dependence on $1/\eps$, but,
unfortunately, their dependence on all other parameters is worse than
in Theorem \ref{thm:smooth}; in particular, both our bounds on $\kappa$ are super-exponential in $d$.
We expect the practical performance of the technique to be much better if $\alpha(t)$ 
is much larger along the trajectory followed by the gradient descent. We leave the 
question of whether our bound on $\kappa$ can be improved, and the project of 
evaluating experimentally the performance of this algorithm, for future work.

\subsection{Bounding and computing the gradient}
\label{sec:grad-bound}

To run gradient descent we need to compute the gradient of
$f$ at each step, which, by the definition of $f$, is equal to
$\nabla f (t) = \nabla\Phi(t) -c$.

One form of $\nabla\Phi(t)$ is given by Lemma \ref{lem:cb}:
\[
\frac{\partial \Phi}{\partial t_j}(t) =
\frac{\sum_{\{S\mid j\in S, |S|=d\}} e^{\sum_{i\in S} t_i} \Delta_S }{\sum_{|S|=d} e^{\sum_{i\in S} t_i} \Delta_S }\ ,
\qquad j=1,\ldots,n ,
\]
which, as noted in Lemma~\ref{lem:cb}, implies that $|\nabla\Phi (t)|_1 = d$. Thus we get:
\begin{lem} \label{lem:gradbounded}
    For all $t$, $|{\nabla f}(t) |_1 \le |c|_1 + |\nabla\Phi|_1 = 2d$.
\end{lem}

To compute $\nabla\Phi(t)$, though, we use its alternative form given by
Lemma \ref{lem:diff-of-logdetsum}:
\begin{equation} \label{eq:gradient-f}
\frac{\partial \Phi}{\partial t_j}(t) = e^{t_j}|Q^{-1/2}(t)x_j|^2
= (e^{t_j/2} x_j)^T Q^{-1}(t) (e^{t_j/2} x_j) ,
\end{equation}
for $j=1,\ldots,n$. This can be written as
${\displaystyle \frac{\partial \Phi}{\partial t_j}(t) = |u_j|^2}$,
where the vectors $u_j := Q^{-1/2}(t)e^{t_j/2} x_j$ are the column vectors of $U^T$
in the SVD decomposition $U\Sigma V^T$ of $X(t)^T = \left\{e^{t_j/2} x_j\right\}^T$.
We can therefore compute the gradient by computing the SVD of $X(t)^T$, as discussed in
Section~\ref{sec:iso-svd}. Note that when the values of the $t_i$'s are large, the
norm of $X(t)$ can be much larger than the norm of the gradient given by Lemma \ref{lem:gradbounded}. 
This requires computing the SVD to a relatively high accuracy, as discussed above.

\subsection{Bounding the region $K$} \label{sec:R}


In this section we bound $|t^*|_\infty$, assuming that $\min_i t^*_i = 0$.\footnote{%
  Recall that, for gradient descent for strongly convex functions, we need
  to run the gradient descent procedure within the subspace $E_0$ orthogonal
  to ${\bf 1}$, which requires that we modify $t^*$. We ignore this technical
  (and rather insignificant) issue in the present analysis.}
This bound is essential for determining the required accuracy of the computation 
of the SVD, and for determining the region in which we have to run the projected 
gradient descent procedure. Our bounds on the running time of gradient descent in 
Theorems \ref{thm:smooth} and \ref{thm:strong} depend on the upper bound on 
$|t^*|_\infty$ that we develop here. We also compare our bound to previous upper 
bounds on $|t^*|_\infty$, such as the one in \cite{HM}.

\subsubsection{The Hardt-Moitra approach}

We start by restating (and slightly correcting) a previous bound by Hardt and Moitra,
who claim in~\cite[Lemma 30]{HM} that
${\displaystyle |t^*|_\infty \le \frac{2}{\gamma} \log \frac{1}{\Delta_S^{{\rm min}}}}$,
where $\Delta_S^{{\rm min}}$ is the smallest positive value of the determinants $\Delta_S$,
and where $\gamma$ (called $\alpha$ in \cite{HM}) is such that
for any vector $t$ whose minimum coordinate is $0$ we have:
\begin{equation} \label{hm:inside}
\iprod{c}{t} \le (1-\gamma) \max_{S\in \cal S} \iprod{{\bf 1}_S}{t} .
\end{equation}

The analysis in \cite{HM} does not handle properly
one of the inequalities along the way. Lemma \ref{lem:hm-fix} gives a
corrected version of this analysis, with a slightly different bound.

\begin{restatable}{lem}{lemHMcorrection}
     \label{lem:hm-fix}
$$
|t^*|_\infty \le R_0 := \frac{1}{\gamma} \log \frac{\sum_S \Delta_S}{\Delta_S^{{\rm min}}} .
$$
\end{restatable}

\medskip
\noindent
{\bf Proof.}
Note that the argument of the logarithm is always at least $1$.
The analysis in \cite{HM} considers the function
$f(t) = \Phi(t)- \iprod{c}{t}$, which is the one we want to minimize, and establishes
an upper bound and a lower bound on $f(t^*) = -\Phi^*(c)$. The lower bound, which holds
for any $t$ (normalized as in the definition of $\gamma$),
as stated in \cite[Claim 31]{HM}, asserts that
$$
f(t) \ge -\log \frac{1}{\Delta_S^{{\rm min}}} + \gamma |t|_\infty .
$$
The upper bound in \cite{HM} is faulty, because it replaces $f(0)$ by $-f(0)$.
We offer the following alternative bound, which is
$$
f(t^*) \le f(0) =  \log \det Q(0) =  \log \left( \sum_S \Delta_S \right)
$$
(where, as usual, we use the Cauchy-Binet formula).
Combining the upper and lower bounds, we get
$$
\log \left( \sum_S \Delta_S \right) \ge
-\log \frac{1}{\Delta_S^{{\rm min}}} + \gamma |t^*|_\infty , \quad\text{or}
$$
$$
|t^*|_\infty \le \frac{1}{\gamma} \log \frac{\sum_S \Delta_S}{\Delta_S^{{\rm min}}} ,
$$
as asserted.
$\Box$


Unfortunately, the estimate of $|t^*|_\infty$ in Lemma \ref{lem:hm-fix} is weak, as it is sensitive
to the existence of even a single $d$-tuple $S$ of indices whose corresponding $d$-tuple
of vectors of $X$ is `almost linearly dependent', thus having a very small (albeit positive)
determinant $\Delta_S$. We give here a derivation of an alternative, considerably
improved bound on $|t^*|_\infty$.

See the comments at the end of Section~\ref{sec:HMimprove} for additional discussion of this approach.

\subsubsection{An improved bound for $|t^*|_\infty$}

\begin{lem} \label{lambda-ratio}
Assume that the configuration $X=\{x_i\}_{i=1}^n$ and the weight vector $c$
satisfy the condition \eqref{eq:definition-of-asterix-condition} (i.e., $c$ is $(\eta,\delta)$-deep
inside $K_X$, for prescribed parameters $\eta,\;\delta>0$).
Let $T$ be the positive definite map that sends $X$ to radial isotropic position,
and let the eigenvalues of $T$ be given by $0<\lambda_1 \le \lambda_2 \le \cdots \le \lambda_d$.
Then, for any $k=1,\ldots,d-1$,
\begin{equation} \label{lamd1}
\frac{\lambda_{k+1}}{\lambda_{k}} \le
\left(\frac{d-k+\eta k}{\eta k}\cdot\frac{1-\delta^2}{\delta^2} \right)^{1/2} , 
\qquad\text{and hence}\qquad
\frac{\lambda_{d}}{\lambda_{1}} \le
\left(\frac{8}{\eta\delta^2}\right)^{(d-1)/2}.
\end{equation}
\end{lem}

\noindent{\bf Proof.}
Fix $k$ and denote by $E$ a subspace spanned by all eigenvectors of $T$
with eigenvalues less than or equal to $\lambda_k$, with $\dim(E) = k$.
Note that if the eigenvalue $\lambda_k$ has multiplicity larger than one, 
the choice of $E$ is not unique.
Since $T$ sends $\{x_i\}_{i=1}^n$ to radial isotropic position, we have
\[
\sum_{i=1}^n \frac{c_i}{\left|T x_i \right|^2} (T x_i )\otimes (T x_i ) =  I_d ,
\]
and in particular, if we project both sides onto $E^\perp$, we get
\[
\sum_{i=1}^n \frac{c_i }{\left|T x_i \right|^2} (P_{E^{\perp}}T x_i )\otimes (P_{E^{\perp}} T x_i ) =  P_{E^\perp} .
\]
Taking the trace on both sides, we get
\begin{equation} \label{projmass}
\sum_{i=1}^n c_i\frac{\left|P_{E^{\perp}}Tx_i\right|^2}{\left|T x_i\right|^2} = d-k.
\end{equation}

On the other hand, every $x_i\in \sph^{d-1}$ can be written as $x_i = y_i + z_i$,
with $y_i \in E$, $z_i \in E^\perp$. When $x_i\notin E_\delta$, that is, $d(x_i, E)>\delta$,
we know also that $|z_i|>\delta$ and thus $|y_i|<\sqrt{1-\delta^2}$.
Note that in this case we have $Tx_i = T y_i + T z_i$ and $|Tx_i|^2 = |Ty_i|^2 + |Tz_i|^2$.
(The orthogonality of $Ty_i$ and $Tz_i$ follows from the fact that each of $E$ and $E^\perp$
is spanned by eigenvectors of $T$.)
Since $z_i \in E^\perp$, and since $E^\perp$ is spanned by eigenvectors with eigenvalues
at least $\lambda_{k+1}$, we know that
$$
|Tx_i|\ge |Tz_i| \ge \lambda_{k+1} |z_i|\ge \lambda_{k+1}\delta .
$$
Similarly, $|Ty_i|\le \lambda_k |y_i|\le \lambda_k (1-\delta^2)^{1/2}$. We thus have
\[
\left|P_{E^\perp} \frac{Tx_i}{|Tx_i|}\right|^2 =
1 - \left|P_{E} \frac{Tx_i}{|Tx_i|}\right|^2 =
1 - \frac{\left|T y_i\right|^2}{|Tx_i|^2}\ge 1 - \frac{\lambda_k^2 (1-\delta^2)}{\lambda_{k+1}^2\delta^2}.
\]
We sum these inequalities only over those $x_i$ that do not belong to $E_\delta$.
By Condition \eqref{eq:definition-of-asterix-condition}, the total ``mass'' (namely sum of the corresponding $c_i$'s)
of those $x_i$'s is at least $d-k(1-\eta)$. Hence we get that
\[
\sum_{i=1}^n c_i \frac{|P_{E^\perp} Tx_i|^2}{|Tx_i|^2} \ge
\sum_{\{i\mid x_i \in E_\delta\}} c_i \frac{|P_{E^\perp} Tx_i|^2}{|Tx_i|^2} \ge
(d-k(1-\eta))\left(1 - \frac{\lambda_k^2 (1-\delta^2)}{\lambda_{k+1}^2\delta^2} \right).
\]
Combining this with (\ref{projmass}), we obtain the inequality
\[
d-k \ge (d-k+\eta k)\left(1 - \frac{\lambda_k^2 (1-\delta^2)}{\lambda_{k+1}^2\delta^2} \right) ,
\]
which can be rewritten as
\[
\frac{\lambda_{k+1}^2}{\lambda_k^2} \le \frac{d-k+\eta k}{\eta k}\cdot\frac{1-\delta^2}{\delta^2} ,
\]
which establishes the first inequality in the lemma.

The second inequality now follows by taking the product of these estimates over
$k= 1, \ldots, d-1$, obtaining
\[
\prod_{k=1}^{d-1}\frac{\lambda_{k+1}}{\lambda_{k}} \le
\left(\frac{1-\delta^2}{\delta^2}\right)^{\frac{d-1}{2}} 
\prod_{k=1}^{d-1} \left(\frac{d-k+\eta k}{\eta k} \right)^{1/2},
\]
and the latter can be bounded as follows: take the product on the right only for 
those value of $k$ for which the terms are at least $2$, namely for $k\le d/(1+\eta)$. 
Compensate by $2$ for 
each terms we did not consider. The terms that we do consider can be upper bounded, each in turn, 
by $2\frac{d-k}{\eta k}$. We are thus left with the product
\[
\prod_{k=1}^{d-1}\frac{\lambda_{k+1}}{\lambda_{k}} \le
\frac{1}{\delta^{d-1}} \frac{1}{\eta^{\frac{d-1}{2}}} 2^{d-1} 
\prod_{k=1}^{\lfloor d/(1+\eta) \rfloor} \left(\frac{d-k}{k} \right)^{1/2} =    
\left(\frac{4}{ \eta\delta^2 }\right)^{\frac{d-1}{2}} \binom{d-1}{\lfloor d/(1+\eta) 
\rfloor}^{1/2}\le  \left(\frac{8}{ \eta\delta^2 }\right)^{\frac{d-1}{2}} . 
\]
\qed

\smallskip

Applying this analysis to the extremizing vector $t^*$, 
we seek an upper bound on $\max_i t^*_i$, where
(a) $Q^{-1/2}(t^*)={\Sigma}^{-1}V^T$ maps the
configuration to radial isotropic position, and (b) $\min_i t^*_i = 0$. Recall that
$Q(t^*) = X(t^*)X(t^*)^T=\sum_{i=1}^n e^{t^*_i} x_i \otimes x_i$, where
${X}(t^*) = \left\{ e^{t^*_i/2} x_i \right\}_{i=1}^n$, and that
${\Sigma}$ and $V$ come from the SVD of ${X}(t^*)^T$.
Hence the eigenvalues $0<\lambda_1 \le \lambda_2 \le \cdots \le \lambda_d$
of the matrix $T$ in Lemma~\ref{lambda-ratio}, which is $Q^{-1/2}(t^*)$,
are the diagonal entries of ${\Sigma}^{-1}$, namely $\lambda_i = \frac{1}{\sigma_i}$
for $i=1,\ldots,d$, where $\sigma_1 \ge \sigma_2 \ge \cdots \ge \sigma_d$
are the singular values of ${X}(t^*)$.

Assume without loss of generality that the entries of $t^*$ are sorted in increasing order,
so $0 = t^*_{\rm min} := \min_i t^*_i = t^*_1$ and $t^*_{\rm max} := \max_i t^*_i = t^*_n$.
Since $Q(t^*)$ is symmetric and positive definite, its eigenvalues are easily
seen to be $\frac{1}{\lambda_1^2}\ge\cdots\ge\frac{1}{\lambda_d^2}$, i.e.,
$\sigma_1^2\ge\cdots\ge\sigma_d^2$. We have
\[
\frac{1}{\lambda_1^2} = \max_{|y|=1} \iprod{Q(t^*)y}{y}
= \max_{|y|=1} \sum_{i=1}^n e^{t^*_i} \iprod{x_i x_i^Ty}{y} 
= \max_{|y|=1} \sum_{i=1}^n e^{t^*_i} \iprod{x_i}{y}^2
\ge \max_{|y|=1} e^{t^*_n} \iprod{x_n}{y}^2 = e^{t^*_n} .
\]
On the other side, to obtain a lower bound on $\lambda_d$, we use the property
that $t^*$ is the extremizing vector. In the SVD interpretation, since $t_1^* = 0$, we have
$$
u_1 = {\Sigma}^{-1} V^T \left(e^{t^*_1/2} x_1 \right) = {\Sigma}^{-1} V^T x_1 .
$$
Since $t^*$ is extremizing, we have $|u_1|^2 = c_1$. That is,
writing $y_1 = V^Tx_1$, we have
${\displaystyle c_1 = \sum_{k=1}^d \frac{y_{1k}^2}{\sigma_k^2}}$.
Since $|y|=1$, at least one of the $\sigma_k$'s must be at most $1/\sqrt{c_1}$.
In particular, we have $\sigma_d \le 1/\sqrt{c_1}$. We thus have
${\displaystyle \frac{\lambda_d^2}{\lambda_1^2} \ge c_{\rm min} e^{t^*_n}}$, 
where $c_{\rm min} = \min_i c_i$. Combining this with (\ref{lamd1}), we get
${\displaystyle e^{t^*_{\rm max}} \le \frac{1}{c_{\rm min}}  \left(\frac{8}{ \eta \delta^2 }\right)^{d-1}}$.
That is, we have shown
\begin{lem} \label{R-cll}
Let $\delta$, $\eta>0$ be parameters for which $c$ satisfies Condition \eqref{eq:definition-of-asterix-condition}
(for being $(\eta,\delta)$-deep inside $K_X$). Then the extremizing vector $t^*$, normalized so that
$t^*_{\rm min} = 0$, satisfies
\begin{equation} \label{our-tinf}
|t^*|_\infty \le \log \frac{1}{c_{\rm min}} + (d-1) \log \left(\frac{8}{\eta\delta^2}\right) .
\end{equation}
In the uniform case, namely when $c=\frac{d}{n}{\bf 1}$, this becomes
\begin{equation} \label{our-tinf-uniform}
|t^*|_\infty \le \log \frac{n}{d} + (d-1) \log \left(\frac{8}{\eta\delta^2}\right) .
\end{equation}
\end{lem}

It is interesting to note that the bound on $|t^*|_\infty$
that we obtain in Lemma~\ref{R-cll}, and Hardt and Moitra`s (fixed) bound
in Lemma~\ref{lem:hm-fix}, depend on $\frac{1}{c_{\rm min}}$ 
(see the comments at the end of Section \ref{sec:HMimprove}),
and deteriorate when the minimum entry in $c$ is too close to $0$.
However, the dependency of the bound of Lemma~\ref{R-cll}
on $c_{\rm min}$ is exponentially better than the dependency of
the bound of Hardt and Moitra.

In Section \ref{sec:large-t}, we give a simple example
of three vectors in the plane, and $c=\left(\frac{2}{3},\frac{2}{3},\frac{2}{3} \right)$, 
in which $|t^*|_\infty$ indeed grows proportionally to $\log \left(\frac{1}{\delta}\right)$. 
We leave the question of whether the other terms and factors in our bounds in Equations (\ref{our-tinf}) 
and (\ref{our-tinf-uniform}) are also worst-case essential, for future research.

\subsubsection{Improving the bound of Hardt and Moitra}
\label{sec:HMimprove}

In this subsection we show that Hardt and Moitra's condition (\ref{hm:inside})
implies our new condition \eqref{eq:definition-of-asterix-condition}, for suitable choices of $\delta$ and $\eta$,
which depend on the parameters $\gamma$ and $\Delta_S^{{\rm min}}$.
We then use Lemma~\ref{R-cll}, with these choices, to get a better bound
on $|t^*|_\infty$ than the bound of Hardt and Moitra in Lemma~\ref{lem:hm-fix},
in terms of the parameters of Hardt and Moitra.

We make use of the following lemma.
\begin{lemma} \label{smallvol}
    Let $X$ be a set of $n$ unit vectors spanning $\RR^d$.
    Let $E$ be a subspace of $\RR^d$ of some dimension $k<d$. Let
    $A\subset [n]$, $|A| = k+1$, be a set of indices such that
    $(x_i)_{i\in A}$ are linearly independent, and
    ${\rm dist}(x_i,E) \le \delta$ for each $i\in A$.
    Then one can complete $(x_i)_{i\in A}$, with $d-k-1$ additional vectors
    $(x_i)_{i\in B}$ from $X$, where $B\subset [n]$, $|B|=d-k-1$,
    to a basis $(x_i)_{i\in A\cup B}$ of $\RR^d$,
    so that the determinant of the basis vectors (the volume of the
    parallelepiped determined by these vectors) is at most $(1+\delta)^{k+1}-1$.
\end{lemma}
\noindent{\bf Proof.}
For each $i\in A$, denote by $y_i$ the orthogonal projection
of $x_i$ onto $E$ and let $z_i = x_i-y_i$. We have that $|z_i| \le \delta$.

Complete $(x_i)_{i\in A}$ into a basis by picking $d-k-1$ arbitrary
linearly independent vectors from $X$ that are independent of the vectors
in $(x_i)_{i\in A}$. Let $B$ denote the set of indices of these vectors, and
let $S=A\cup B$. We have that
$$
{\rm det} ((x_i)_{i\in S}) =
{\rm det} ((y_i+z_i)_{i\in A},(x_i)_{i\in B}) =
\sum_{I\subseteq A} {\rm det} (C_I(Y,Z)) ,
$$
where $C_I(Y,Z)$ is the matrix whose $j$-th column, for $j=1,\ldots,k+1$, is $y_j$ (resp., $z_j$)
for $j\in I$ (resp., for $j\notin I$), and where the columns $k+2,\ldots,d$
are the vectors $(x_i)_{i\in B}$.
Note that the determinant of $C_I(Y,Z)$ is $0$ for $I = A$,
as the vectors $(y_i)_{i\in A}$ are $k+1$ linearly dependent vectors in a $k$-dimensional subspace.
Note also that if $|I|=k+1-j$, for $j\ge 1$, then ${\rm det}(C_I(Y,Z))$
consists of $j$ columns of norm at most $\delta$ and
the rest of its columns are of norm at most $1$, so
${\rm det}(C_I(Y,Z))\le \delta^j$.

Summing up, we obtain that
$$
\sum_{I\subseteq A} {\rm det} (C_I(Y,Z)) \le (k+1)\delta + {k+1\choose 2}\delta^2 + \cdots +
{k+1\choose k+1}\delta^{k+1} = (1+\delta)^{k+1} - 1 \ ,
$$
as asserted.
$\Box$


\begin{restatable}{lem}{lemHMbetter}
     \label{HM-improved}
    Assume that Condition \eqref{hm:inside} holds for any vector $t$ whose minimum coordinate is $0$.
    Then Condition \eqref{eq:definition-of-asterix-condition} holds with 
    $\delta =\sqrt{\Delta_S^{{\rm min}}}/2d$ and $\eta= \gamma$.
    Furthermore, we have that
    \begin{equation} \label{our-tinf-hm}
    |t^*|_\infty \le \log \frac{1}{\gamma d} + (d-1)
    \log \left(\frac{32d^2}{\gamma \Delta_S^{{\rm min}}} \right) .
    \end{equation}
\end{restatable}

Note that the dependency of this bound on $\gamma$ is exponentially better 
than in the original result of Hardt and Moitra in  Lemma~\ref{lem:hm-fix}.


\medskip
\noindent{\bf Proof.}
Consider a subspace $E$ with ${\rm dim}(E) = k \le d-1$.

Set $\delta = \sqrt{\Delta_S^{{\rm min}}}/2d$. Then, by
Lemma \ref{smallvol}, $E_\delta$ contains at most $k$
linearly independent vectors. Indeed, Lemma \ref{smallvol} says that
if $E_\delta$ contains $k+1$ linearly independent vectors then there
is a set $S\subset [n]$, $|S|=d$ such that
$$
{\rm det} ((x_i)_{i\in S}) \le (1+\delta)^{k+1} -1 \le (1+\delta)^{d} -1
< e^{\delta d} -1 \le 2\delta d = \sqrt{\Delta_S^{{\rm min}}} ,
$$
which contradicts the definition of $\Delta_S^{{\rm min}}$.
(The last inequality follows since $e^x \le 1+ 2x$ for $x\le 1$;
our $x=d\delta$ is actually at most $1/2$, since $\Delta_S^{{\rm min}} \le 1$,
because all the vectors $x_i$ are unit vectors.)

Note that the argument just given implies in particular that $E_\delta$
cannot contain all of $X$, for then it would contain $d\ge k+1$
linearly independent vectors.

Let $t=(t_1,\ldots,t_n)$ be such that $t_i = 1$ if $x_i \in E_\delta$,
and $t_i = 0$ otherwise; by the comment just made, we have indeed
$t_{\rm min} = 0$. By applying Condition  \eqref{hm:inside}  to $t$ we get that
\[
\sum_{x_i\in E_\delta} c_i \le (1-\gamma) \max_{S\in \cal S} \iprod{{\bf 1}_S}{t} \le (1-\gamma)k ,
\]
where the last inequality follows since, as has just been
argued, $E_\delta$ contains at most $k$ linearly independent vectors.
It follows that Condition \eqref{eq:definition-of-asterix-condition} indeed holds for
$\eta=\gamma$ and $\delta = \sqrt{\Delta_S^{{\rm min}}}/2d$.
As stated in Comment (2) above, we have $c_{\rm min} \ge \gamma d$.

Substituting $\eta=\gamma$ and $\delta = \sqrt{\Delta_S^{{\rm min}}}/2d$,
and this lower bound on $c_{\rm min}$ into the bound of Equation \eqref{our-tinf}
of Lemma \ref{R-cll}, we get that
\begin{equation} \label{our-tinf-hmx}
|t^*|_\infty \le \log \frac{1}{\gamma d} + (d-1)
\log \left(\frac{32d^2}{\gamma \Delta_S^{{\rm min}}}
\right).
\end{equation}
\qed

\paragraph{Comments on Hardt and Moitra's bound.}
\label{rem:bound1overn-forgamma-isbestpossible}

{\bf (1)} We point out that in the reducible case, condition \eqref{hm:inside} does not hold. 
Indeed, let $X = Y\cup Z$, such that ${\rm span}(Y)= E$, ${\rm span}(Z) = F$ and $E\oplus F = \RR^n$. 
For the vector $t$ that is $0$ (resp., $1$) at coordinates $i$ such that $x_i \in Y$ (resp., $x_i \in Z$),
we get that $\iprod{c}{t} \le \dim F$, and also every $S\in {\cal S}$ satisfies the property
that precisely $\dim F$ of its elements lie in $Z$. Therefore, condition \eqref{hm:inside}
does not hold if $X$ is reducible. 
On the other hand, when $X$ is irreducible, and $c\in {\rm relint}(K_X)$, there will always 
be some $\gamma$ for which condition \eqref{hm:inside} holds.
Indeed, irreducibility means that the dimension of $K_X$ is $d-1$, and thus, for every 
fixed $t$ with non-negative coordinates such that not all of its coordinates are equal, 
the assumption of $c$ being in the relative interior of $K_X$ implies that
$\iprod{c}{t} < \max_{S\in \cal S} \iprod{{\bf 1}_S}{t}$.
Since the expressions on both sides of this inequality are homogeneous in $t$,
it suffices to require that $\gamma$ exists uniformly for only the vectors $t$ in the simplex $\{ t: t_i \ge 0 ,\sum t_i = 1\}$.
By compactness, there exists a positive value of $\gamma$ such that 
$1-\frac{\iprod{c}{t}}{\max_{S\in \cal S} \iprod{{\bf 1}_S}{t}} \ge \gamma$,
for all $t$ in the simplex. Hence (\ref{hm:inside}) holds with this choice of $\gamma$.

\medskip
\noindent{\bf (2)}
It is useful to note that any $\gamma$ that satisfies \eqref{hm:inside} must satisfy 
$\gamma \le \frac1n$. In fact, by picking $t = {\bf 1}- e_i$ (that is, $t_i=0$ and 
all other $t_j = 1$) and some $S$ which does not contain $i$ (there must be such an 
$S\in {\cal S}$ since otherwise the set $X$ is reducible, because 
$\RR^n={\rm span} \{x_i\} \oplus {\rm span} \{x_j\}_{j\neq i}$, in which case 
there is no $\gamma$ at all, as follows from the preceding remark.
The inequality then gives $\sum_{j\neq i} c_i \le (1-\gamma)d$, which can be 
rearranged, using that $\sum c_j = d$, to yield $c_i \ge \gamma d$. 
This was valid for any $i$, so $\gamma \le \frac 1d \min c_i$. 
This is stronger than our original claim, as clearly $\min c_i \le \frac{d}{n}$,
and equality holds if and only if $c$ is the uniform vector $\frac{d}{n}{\bf 1}$.

\subsubsection{Lower bound for $\gamma$ in general position}
\label{sec:gammageneralposlowerbounds}

The analysis in \cite{HM} caters for an arbitrary vector $c$.
Let us first assume that $c=\frac{d}{n}{\bf 1}$. In view of the preceding Comment (2),
the following lemma gives the best possible value of $\gamma$.
\begin{lem} \label{gamma:unic}
    For $c=\frac{d}{n}{\bf 1}$ and $X$  in general position we have $\gamma=1/n$.
\end{lem}

\noindent{\bf Proof.} 
Recall that by Comment (2) above, we must have $\gamma \le 1/n$.
Assume, without loss of generality, that the
coordinates of $t$ are sorted in decreasing order.
Since $X$ is in general position, $(1,2,\ldots,d)$ belongs to $\cal S$,
and it maximizes $\sum_{i\in S} t_i$ over all $S\in{\cal S}$.
It thus suffices to show that (recall that we only consider vectors with $t_n=0$)
$$
\frac{d}{n}\sum_{i=1}^{n-1} t_i \le \left(1-\frac{1}{n}\right) \sum_{i=1}^d t_i ,\quad\text{or}
$$
$$
\frac{1}{n-1}\sum_{i=1}^{n-1} t_i \le \frac{1}{d} \sum_{i=1}^d t_i ,
$$
which certainly holds since the $t_i$'s are in decreasing order.
$\Box$
\smallskip

Under the assumption of the points $\{x_i\}$ being in general position, this result 
can be extended for general coefficient vectors $c$, as asserted in the following lemma.
\begin{lem} \label{gamma:anyc}
    If $X$ is in general position, $n>d$, and $c$ is any vector (of coefficients in $(0,1)$
    that sum up to $d$) that satisfies $\max c_i \le 1-\frac{c_{\rm min}}{d}$,
    then we have $\gamma = \frac{c_{\rm min}}{d}$.
\end{lem}

\medskip
\noindent{\bf Proof.} 
Put $a = c_{\rm min}$, and note that $a\le d/n$.
(The assumption that $\max c_i \le 1-\frac{a}{d}$, which holds for the uniform $c$ when $d\le n-1$,
is not very restrictive.)
Using again Comment (2), we know that $\gamma \le \frac{c_{\rm min}}{d}$.
As before, sort the $t_i$'s in decreasing order, and assume $t_n=0$. To establish the desired
inequality, under the general position assumption, we need to show that
$$
\iprod{c}{t} = \sum_{i=1}^{n-1} c_it_i \le \left(1-\frac{a}{d}\right) \sum_{i=1}^d t_i ,\quad\text{or}
$$
$$
\frac{1}{d-a} \sum_{i=1}^{n-1} c_it_i \le \frac{1}{d} \sum_{i=1}^d t_i .
$$
Write the left-hand sum as $C_1+C_2$, where
\[
C_1 = \frac{1}{d-a} \sum_{i=1}^{d} c_it_i \qquad\text{and}\qquad 
C_2 = \frac{1}{d-a} \sum_{i=d+1}^{n-1} c_it_i .
\]
The assumption that $\max_i c_i \le 1 -\frac{a}{d}$ implies that, for any $i=1,\ldots,d$, we have
${\displaystyle \frac{c_i}{d-a} \le \frac{1}{d}}$,
so ${\displaystyle C_1 \le \frac{1}{d}\sum_{i=1}^d t_i}$, and we have
$$
\frac{1}{d}\sum_{i=1}^d t_i - C_1 = \sum_{i=1}^d \left( \frac{1}{d} - \frac{c_i}{d-a} \right) t_i
\ge \sum_{i=1}^d \left( \frac{1}{d} - \frac{c_i}{d-a} \right) t_d
= \left( 1 - \frac{\sum_{i=1}^d c_i}{d-a} \right) t_d .
$$
Since ${\displaystyle \sum_{i=1}^{n-1} c_i = d - c_n \le d - c_{\rm min} = d-a}$, this is
$$
\ge \left( 1 - \frac{\sum_{i=1}^d c_i}{\sum_{i=1}^{n-1} c_i} \right) t_d
= \frac{\sum_{i=d+1}^{n-1} c_i}{\sum_{i=1}^{n-1} c_i} t_d
\ge \frac{\sum_{i=d+1}^{n-1} c_it_i}{\sum_{i=1}^{n-1} c_i}
\ge \frac{\sum_{i=d+1}^{n-1} c_it_i}{d-a} =C_2 .
$$
We have thus shown that
$\frac{1}{d}\sum_{i=1}^d t_i \ge C_1+C_2$, which is the inequality asserted in the lemma.
$\Box$

\medskip
\noindent{\bf Remark.}
In general, when $X$ is not in general position, the lemma fails. 
In fact, it might even be the case that $c$ does not lie in $K_X$ at all.

To recap, the condition in (\ref{hm:inside}) gives an alternative definition for being
deep inside $K_X$, expressed in terms of the parameter $\gamma$ introduced in Hardt and Moitra~\cite{HM}.
The preceding analysis (a) connects between $\gamma$ and our parameters $\eta$, $\delta$,
(b) yields an improved bound on $t^*_\infty$, and (c) yields a reasonably good lower bound
for $\gamma$ for sets $X$ in general position, which in turn leads to an improved bound on $t^*_\infty$. 

\subsection{The Hessian (estimating $\beta$)} \label{sec:beta}

Since $\Phi$ and $f$ differ by a linear function, they share the same
Hessian $H(t)=H^\Phi(t) = H^f(t)$ (or in an alternative standard notation $\nabla^2 \Phi(t) = \nabla^2 f(t)$).

Using \eqref{eq:gradient-f} we write
$$
\frac{\partial \Phi}{\partial t_j} (t)= e^{t_j} x_j^T Q^{-1}(t) x_j\ ,
\quad\text{for $j=1,\ldots,n$} .
$$
Its derivative with respect to $t_k$ satisfies
\begin{equation} \label{eq:gradient-f2}
    \frac{\partial^2 \Phi }{\partial t_j \partial t_k}(t)= \begin{cases}
 e^{t_j} x_j^T \frac{\partial }{\partial t_k}Q^{-1}(t) x_j & \text{for $k\ne j$} \\
 e^{t_j} x_j^T Q^{-1}(t) x_j +
 e^{t_j} x_j^T \frac{\partial }{\partial t_j}Q^{-1}(t) x_j & \text{for $k= j$} .
\end{cases}
\end{equation}
To obtain $\frac{\partial }{\partial t_k}Q^{-1}(t)$, we start with the identity
\[
Q(t)Q^{-1}(t)= I_d\ ,
\]
take the derivative of both sides with respect to $t_k$, and get that
\[
\frac{\partial Q}{\partial t_k} (t)Q^{-1}(t) + Q(t) \frac{\partial Q^{-1}}{\partial t_k} (t) = 0 \ .
\]
Rearranging, we get that
\begin{equation} \label{eq:partialX-1}
\frac{\partial Q^{-1} }{\partial t_k}(t) = -Q^{-1}(t)\frac{\partial Q }{\partial t_k}(t) Q^{-1}(t) .
\end{equation}
Substituting Equation  \eqref{eq:partialX-1}  into Equation  \eqref{eq:gradient-f2}, we get that
\begin{equation} \label{eq:gradient-f1}
    \frac{\partial^2 f}{\partial t_j \partial t_k}(t) =
\begin{cases}
-e^{t_j} x_j^T Q^{-1}(t)\frac{\partial Q }{\partial t_k}(t) Q^{-1}(t)  x_j\ & \text{for $k\ne j$} \\
 e^{t_j} x_j^T Q^{-1}(t) x_j -
e^{t_j} x_j^T Q^{-1}(t)\frac{\partial Q }{\partial t_j}(t) Q^{-1}(t)  x_j\ & \text{for $k = j$} .
\end{cases}
\end{equation}
But we have, by construction,
\[
\frac{\partial Q}{\partial t_k} = e^{t_k} x_k  x_k^T ,
\]
which leads to, for $k\ne j$,
\begin{align*} \label{eq:gradient-f11}
    \frac{\partial^2 f }{\partial t_j \partial t_k}(t) & =
    -e^{t_j} x_j^T Q^{-1}(t)e^{t_k} x_k   x_k^T Q^{-1}(t)  x_j \\
    & = -\left( e^{t_j/2} x_j^T Q^{-1}(t)e^{t_k/2} x_k \right) \cdot
        \left( e^{t_k/2}x_k^T Q^{-1}(t) e^{t_j/2} x_j \right)^T \\
    & = -\left( e^{t_j/2} x_j^T Q^{-1}(t)e^{t_k/2} x_k \right)^2 ,
\end{align*}
and, for $k=j$ we similarly have
\begin{equation} \label{eq:gradient-f19}
    \frac{\partial^2 f }{\partial t_j^2}(t) =
    e^{t_j} x_j^T Q^{-1}(t) x_j -
    \left( e^{t_j/2} x_j^T Q^{-1}(t) x_j \right)^2 .
\end{equation}

Recall that the vectors $u_j = Q^{-1/2}(t)e^{t_j/2} x_j$ are
the column vectors of $U^T$ in the SVD decomposition $U\Sigma V^T$
of $\left\{e^{t_j/2}x_j\right\}^T$, provided that we use the definition
$Q^{-1/2}(t) = \Sigma^{-1}V^T$.
This allows us to rewrite the last pair of equations as
\begin{equation} \label{eq:gradient-f111}
    \frac{\partial^2 f }{\partial t_j \partial t_k}(t) =
\begin{cases}
-\iprod{u_k}{u_j}^2 & \text{for $k\ne j$} \\
|u_j|^2 - |u_j|^4 & \text{for $k=j$} ,
\end{cases}
\end{equation}
where $u_k$ is the $k$-th row of $U$, for $k=1,\ldots,n$.
(Any other choice of $Q^{-1/2}$ simply rotates all the $u_j$'s by the same
orthonormal matrix, which does not affect the equations  \eqref{eq:gradient-f111}).

In other words, Equation \eqref{eq:gradient-f111} shows that the Hessian $H$ can be expressed as
\begin{align*}
H_{jj} & = |u_j|^2 - |u_j|^4, \qquad\text{for $j=1,\ldots,n$, and} \\
H_{jk} & = H_{kj} = -\iprod{u_k}{u_j}^2, \qquad\text{for $j\not= k = 1,\ldots,n$} .
\end{align*}

Since $f(t+a{\bf 1}) = f(t)$ for any $t$ and any scalar $a$,
we know that $H{\bf 1} = {\bf 0}$ for any $t$. That is, for each $j$ we have
\begin{equation} \label{eq:hessian:sum}
\left(H{\bf 1} \right)_j = \sum_{k=1}^n H_{jk} = -|u_j|^2 + \sum_{k=1}^n\iprod{u_k}{u_j}^2 = 0,
\qquad\text{or}\qquad
\sum_{k=1}^n \iprod{u_k}{u_j}^2 = |u_j|^2 .
\end{equation}

Since $\Phi$ is convex, all diagonal elements of $H$ must be positive, implying that
$|u_j|^2 - |u_j|^4\ge 0$ which means $|u_j| \le 1$ for every $j$.
(This is again obvious once it is noted that $\sum u_j \otimes u_j = I_d$.)
\footnote{%
  Another way of seeing that $|u_i|\le 1$ for every $i$ is to note that
  the columns of $U$ form an orthonormal system in $\RR^n$, which we can
  complete to an orthonormal basis, by adding $n-d$ columns to $U$.
  Every extended row of $U$ has thus norm $1$, so the norm of the
  original row is also at most $1$.}

\begin{lem} \label{lem:hnorm}
The spectral norm $\|H\|_2$ of $H$ satisfies $\|H\|_2\le 1/2$.
\end{lem}
\noindent{\bf Proof.}
The spectral norm $\|H\|_2$ of $H$ is, by definition $\|H\|_2 = \max \{ y^T H y \mid |y|=1 \}$.
For any unit vector $y$ we have
$$
y^T H y = \sum_{j,k} H_{jk} y_jy_k \le \sum_{j,k} |H_{jk}| |y_j| |y_k| 
= \sum_{j\ne k} (-H_{jk}) |y_j| |y_k| + \sum_{j=1}^n H_{jj} y_j^2 .
$$
We have $|y_j| |y_k| \le \frac12 \left(y_j^2 + y_k^2\right)$, and the symmetry of $H$ then implies

\begin{align*}
y^T H y & \le -\sum_{j\ne k} y_j^2 H_{jk} + \sum_{j=1}^n H_{jj} y_j^2
= \sum_{j=1}^n y_j^2 \left(  H_{jj}-\sum_{k\ne j} H_{jk}  \right) \\
& = 2\sum_{j=1}^n H_{jj}y_j^2 =
2\sum_{j=1}^n \left(|u_j|^2-|u_j|^4\right) y_j^2 \le \frac12 \sum_{j=1}^n y_j^2 = \frac12 .
\end{align*}
We have thus established that, for the Hassian matrix of $\Phi$ (and of $f$) at any point $t$, we have
\[
\|H\|_2\le 1/2,
\]
as asserted.
$\Box$

In particular, the largest eigenvalue $\beta$ of $H$ satisfies $\beta \le \frac{1}{2}$.

\subsection{How strongly convex is $\Phi$? (estimating $\alpha$)}
\label{sec:alpha}

We now turn to quantify and exploit the strong convexity of $f$
(that is, of $\Phi$) in order to 
establish a worst case bound on the performance of gradient
descent that depends on $\log (1/\eps)$ rather than
on $1/\eps$, where $\eps$ is our approximation parameter.

We carry out the analysis only for the case where $\Phi$ is irreducible,
as justified in Section \ref{sec:shiri}.

As discussed at the beginning of this section, we need to quantify the strong convexity of
(the irreducible) $\Phi$ in directions orthogonal to the direction ${\bf 1}$
of linearity of $\Phi$. Denoting, as above, the orthogonal complement of ${\bf 1}$
as $E_0$, we seek a parameter $\alpha>0$ that satisfies
$$
\Phi(y) - \Phi(x) \ge \nabla\Phi(x)^T (y-x) + \frac{\alpha}{2} |y-x|^2 ,
$$
for any pair of points $x$, $y$ in the projection $K_0$ of $K$ onto $E_0$. 
In the differentiable case, which holds for our function, we
need that the eigenvalues of $H$ restricted to $E_0$
be lower bounded by $\alpha > 0$ uniformly over $K_0$.
So let $y$ be a unit vector in $E_0$. We need to derive a lower bound
for $y^TH y$. That is, using our representation of the Hessian from Section \ref{sec:beta} (Equation (\ref{eq:gradient-f111}))
 we seek a lower bound for
\begin{align*}
y^TH y & = \sum_{i,j=1}^n H_{ij}y_iy_j, \qquad\text{which we can write as} \\
& = \sum_{i=1}^n |u_i|^2y_i^2 - \sum_{i,j=1}^n \iprod{u_i}{u_j}^2y_iy_j ,
\end{align*}
where $u_i = Q^{-1/2}(t)e^{t_i/2} x_i$.
We recall (see  \eqref{eq:hessian:sum}) that
$|u_i|^2 = \sum_{j=1}^n \iprod{u_i}{u_j}^2$,
for each $i$. Multiplying by $y_i^2$ and summing over $i$, we obtain
\begin{align*}
\sum_{i=1}^n |u_i|^2y_i^2 & = \sum_{i,j=1}^n \iprod{u_i}{u_j}^2y_i^2 ,
\qquad\text{and, by symmetry, we also have} \\
\sum_{i=1}^n |u_i|^2y_i^2 & = \sum_{i,j=1}^n \iprod{u_i}{u_j}^2y_j^2 ,
\end{align*}
so we get
\begin{equation} \label{eq:ythy}
y^TH y = \frac12 \sum_{i,j=1}^n (y_i^2+y_j^2-2y_iy_j) \iprod{u_i}{u_j}^2
= \frac12 \sum_{i,j=1}^n (y_i-y_j)^2 \iprod{u_i}{u_j}^2 .
\end{equation}
All the terms in this sum are nonnegative. In the extreme case we could make the sum equal to $0$ by
choosing $y_i=y_j$ whenever $u_i$ and $u_j$ are not orthogonal. This however is possible only when
the problem is reducible, which we have assumed not to be the case. To see what the issues are,
consider a concrete example, where we take two complementary subspaces $E_1$, $E_2$, put some
of the vectors $x_i$ in $E_1$ and the rest in $E_2$. As we defined, $u_i = Q^{-1/2}(t)e^{t_i/2} x_i$
for each $i$. Hence some of the $u_i$'s lie in the subspace $Q^{-1/2}(t)(E_1)$
and the rest in the subspace $Q^{-1/2}(t)(E_2)$. As follows from
Lemma~\ref{orth:iniso}, these subspaces must be orthogonal.
Hence, assigning the same value of $y_i$ to all the $u_i$'s in one subspace, and a
different value for the $u_i$'s in the complementary subspace (so that $\iprod{y}{{\bf 1}}=0$
and $|y|=1$), we get $y^T H y = 0$. This is precisely an example of a reducible set $X$,
which we have ruled out in the present analysis. The converse direction, that when the
sum is $0$ we have reducibility, follows using similar reasoning.\footnote{%
  Observe that the characterization of $K_X$ in Proposition~\ref{cll-version} dictates
  how many $x_i$'s can be placed in each subspace. Concretely, in the above decomposition, we must have
  $\sum_{x_i\in E_1} c_i = {\rm dim}(E_1)$ and $\sum_{x_i\in E_2} c_i = {\rm dim}(E_2)$.}

Since we assume that $X$ is irreducible (we even assume the stronger property that
$c$ is $(\eta,\delta)$-deep inside $K_X$), the sum in  \eqref{eq:ythy}  cannot be $0$, but it can
get close to it.  To obtain a lower bound, we first ``get rid'' of the factors $(y_i-y_j)^2$.
For this, we note that the maximum absolute value of the $y_i$'s is at least $1/\sqrt{n}$
(and it is larger when some $y_i$'s are very close to $0$). Moreover, there are some positive $y_i$'s
and some negative ones (or else $y$ would not be orthogonal to ${\bf 1}$), so we conclude that
$$
\max_i y_i - \min_i y_i \ge \frac{1}{\sqrt{n}} .
$$
In particular, assuming that the $y_i$'s are sorted in increasing order, there exists an index
$k$ such that $y_{k+1}-y_k \ge \frac{1}{n^{3/2}}$.
Denote by $\sigma^-$ and $\sigma^+$ the subsets $[1,k]$ and $[k+1,n]$ of $[n]$.
It follows that $y_i-y_j \ge \frac{1}{n^{3/2}}$ for every $i\in \sigma^+$ and $j\in\sigma^-$.
In particular, we have
$$
y^TH y = \frac12 \sum_{i,j=1}^n (y_i-y_j)^2 \iprod{u_i}{u_j}^2
\ge \sum_{i\in\sigma^+} \sum_{j\in\sigma^-} (y_i-y_j)^2 \iprod{u_i}{u_j}^2
\ge \frac{1}{n^3} \sum_{i\in\sigma^+} \sum_{j\in\sigma^-} \iprod{u_i}{u_j}^2 .
$$
We define 
\begin{equation} \label{xis}
\Xi(\sigma^+,\sigma^-) = \sum_{i\in \sigma^-} \sum_{j\in \sigma^+} \iprod{u_i}{u_j}^2 ,
\end{equation}
and then define $\Xi(X)$ as the minimum value of $\Xi(\sigma^+,\sigma^-)$, over all points $t\in K_0$
and over all possible partitions of $[n]$ into two nonempty subsets $\sigma^+$, $\sigma^-$.

The preceding discussion implies that
\begin{equation} \label{alpha:xi}
\alpha = \min \{ y^TH(t) y \mid t\in K_0,\;y \bot {\bf 1},\; |y|=1 \}
\ge \frac{1}{n^3} \Xi(X) .
\end{equation}

\subsubsection{Estimating $\Xi(X)$}

We assume that $c$ is $(\eta,\delta)$-deep inside $K_X$. Our
bound depends on $\eta$ and $\delta$. We use the following lemma.

\begin{lem} \label{yzperp}
Let $0<\delta<1$, and let
 $Y$ and $Z$ be a pair of nonempty sets
of vectors in $\RR^d$ of norm at most $1$
 such that 
\begin{equation} \label{eq:gamma}
\sum_{y\in Y} \sum_{z\in Z} \iprod{y}{z}^2 \le \gamma ,\qquad\text{for}\qquad
\gamma  = \frac{\delta^2} {d\left( 1 + \frac{1}{\delta} \right)^{2d}} \ .
\end{equation}
Then there exist two complementary subspaces
$E$, $E^\perp$ of $\RR^d$, such that
$Z\subset E_{\delta}$ and $Y\subset E^\perp_{\delta}$.
\end{lem}
\noindent{\bf Proof.}
We may assume that the vectors in $Z$ and $Y$ are of norm 
at least $\delta$. Vectors of norm smaller than $\delta$ 
can be discarded as they belong to $E_\delta$ for
any subspace $E$.

We construct
$E$
by 
 applying the  the Gram-Schmidt procedure as follows.
We use $\delta$ as a threshold parameter, and iterate over $z\in Z$,
in an arbitrary order,  starting at some vector $z_1$.
At each step of the iteration, we maintain a subset $Z_0$ of vectors of $Z$ (we will show
that its size is at most $d-1$). We initialize the process by taking $Z_0$ to be the
singleton set consisting of the first vector $z_1$. At each step of the procedure,
we take the next $z\in Z$, and apply to it the Gram-Schmidt operator with respect to the
current $Z_0$, which turns $Z_0\cup\{z\}$ into an orthogonal set. Concretely,
assuming that we have $Z_0 = \{z_1,\ldots,z_j\}$, and putting $z_{j+1}:=z$, we compute,
for $i=1,2,\ldots,j+1$,
\begin{align*}
u_1 & = z_1 \\
u_2 & = z_2 - \frac{\iprod{u_1}{z_2}} {\iprod{u_1}{u_1}} u_1 \\
& \cdots \\
u_i & = z_i - \frac{\iprod{u_1}{z_i}} {\iprod{u_1}{u_1}} u_1 - \frac{\iprod{u_2}{z_i}} {\iprod{u_2}{u_2}} u_2
- \cdots - \frac{\iprod{u_{i-1}}{z_i}} {\iprod{u_{i-1}}{u_{i-1}}} u_{i-1} \\
& \cdots .
\end{align*}
We then consider $|u_{j+1}|$. If it is at most $\delta$, then ${\rm dist}(z,\,{\rm span}(Z_0)) \le \delta$,
and we skip $z$ (do not add it to $Z_0$). Otherwise, we add $z$ to $Z_0$, and repeat the step in either case.

Note that, regardless of whether we add $z$ to $Z_0$ or not, the vectors $u_1,\ldots,u_j$
do not change in the next iteration. If we do add $z$, the vector $u_{j+1}$ is added to
this pool of vectors and also does not change later. We therefore maintain a set $B_0$
of the current vectors $u_1,\ldots,u_j$, and we only need to compute $u_{j+1}$ (where $j$
is the size of the current $Z_0$) when we inspect a new vector $z$.

The vectors of $B_0$ are mutually orthogonal. We thus have, upon
termination of the procedure, $k:=|Z_0|=|B_0| \le d$, but we
will shortly argue that $k$ can be at most $d-1$. By construction,
we have $|u_1|=|z_1|\ge\delta$, and each subsequent vector also satisfies $|u_i| \ge\delta$.
Since the $z_i$'s are of norm at most $1$, and the $u_i$'s are their projections to corresponding subspaces, we also have $|u_i|\le 1$.
Let $E$ denote ${\rm span}(Z_0) = {\rm span}(B_0)$.
We have that $Z\subset E_\delta$ by construction.

Let $e_1,\ldots,e_k$ denote the
orthonormal basis of $E$ obtained by setting $e_i = u_i/|u_i|$, for $i=1,\ldots,k$.
To show that $Y\subset E^\perp_{\delta}$ we need the following lemma.

\begin{lem} \label{small:lincomb}
Let $w=\sum_{i=1}^k w_i e_i$ be a vector in $E$ satisfying $|w|_\infty = 1$.
Then we can write $w$ as a (unique) linear combination
$w = \sum_{i=1}^k \beta_i z_i$ of the vectors $z_1,\ldots,z_k$ of $Z_0$, so that
\begin{equation} \label{beta-bd}
|\beta_j| \le \zeta (1+\zeta)^{k-j},\qquad\text{for $j=1,\ldots,k$} ,
\end{equation}
for $\zeta = 1/\delta$.
\end{lem}
\noindent{\bf Proof.}
We need to solve the system of equations (in the variables $\beta_1,\ldots,\beta_k$), given by
$$
w_1e_1 + w_2e_2 + \cdots + w_ke_k =
\beta_1 z_1 + \beta_2 z_2 + \cdots + \beta_k z_k .
$$
Taking the inner product with $e_i$, for each $i=1,\ldots,k$, we get
\begin{align*}
w_i & = \beta_1 \iprod{z_1}{e_i} + \beta_2 \iprod{z_2}{e_i} + \cdots + \beta_k \iprod{z_k}{e_i} \\
& = \frac{1}{|u_i|} \Bigl( \beta_1 \iprod{z_1}{u_i} + \beta_2 \iprod{z_2}{u_i} + \cdots + \beta_k \iprod{z_k}{u_i} \Bigr) .
\end{align*}
By the properties of the Gram-Schmidt procedure, $u_i$ is orthogonal to $z_1,\ldots,z_{i-1}$,
so we get the following triangular system of $k$ linear equations in $\beta_1,\ldots,\beta_k$.
\begin{align*}
\beta_1 \iprod{z_1}{u_1} + \beta_2 \iprod{z_2}{u_1} + \cdots + \beta_k \iprod{z_k}{u_1} & = |u_1|w_1 \\
\beta_2 \iprod{z_2}{u_2} + \cdots + \beta_k \iprod{z_k}{u_2} & = |u_2|w_2 \\
\cdots & \\
\beta_k \iprod{z_k}{u_k} & = |u_k|w_k .
\end{align*}
Another property of the Gram-Schmidt procedure, evident from the form given above, is that $\iprod{z_i}{u_i} = |u_i|^2$,
for each $i=1,\ldots,k$.
Hence the diagonal entries of the matrix of this system are $|u_i|^2$, for $i=1,\ldots,k$.
As observed earlier, all these values are between $\delta^2$ and $1$.

Working out the solution backwards, we get
\begin{align*}
\beta_k & = \frac{w_k}{|u_k|} \\
\beta_{k-1} & = \frac{w_{k-1}}{|u_{k-1}|} - \frac{\beta_k}{|u_{k-1}|^2} \iprod{z_k}{u_{k-1}} \\
& \cdots \\
\beta_{j} & = \frac{w_j}{|u_j|} - \sum_{\ell=j+1}^{k}\frac{\beta_\ell}{|u_j|^2} \iprod{z_\ell}{u_j} \\
& \cdots .
\end{align*}
Taking absolute values, and noting that
$$
|\iprod{z_\ell}{u_j}| \le |z_\ell|\cdot|u_j| \le |u_j| ,
$$
for any $\ell$ and $j$, we obtain
\begin{align*}
|\beta_k| & = \frac{|w_k|}{|u_k|} \\
|\beta_{k-1}| & \le \frac{|w_{k-1}|}{|u_{k-1}|} + \frac{|\beta_k|}{|u_{k-1}|} \\
& \cdots \\
|\beta_{j}| & \le \frac{|w_j|}{|u_j|} + \sum_{\ell=j+1}^{k}\frac{|\beta_\ell|}{|u_j|} \\
& \cdots .
\end{align*}
Furthermore, we have, for every $j$, $|u_j|\ge\delta$ and, by our assumption, $|w_j|\le 1$.
As in the lemma statement, put $\zeta = \frac{1}{\delta}$, to obtain the recurrence
\begin{align*}
|\beta_k| & \le \zeta \\
|\beta_{k-1}| & \le \zeta + \zeta|\beta_k| \\
& \cdots \\
|\beta_{j}| & \le \zeta + \zeta \sum_{\ell=j+1}^{k}|\beta_\ell| \\
& \cdots .
\end{align*}
A simple induction shows that the solution of this recurrence is
$$
|\beta_j| \le \zeta (1+\zeta)^{k-j},\qquad\text{for $j=1,\ldots,k$} ,
$$
as asserted.
$\Box$
\medskip 

We now continue with the proof of Lemma~\ref{yzperp}
and show that $Y\subset E^\perp_\delta$.
 Recall that we have, for each $y\in Y$,
\begin{align*}
\sum_{z\in Z} \iprod{y}{z}^2 & \le \gamma ,\qquad\text{so, in particular,} \\
\sum_{i=1}^k \iprod{y}{z_i}^2 & \le \gamma .
\end{align*}
Using lemma~\ref{small:lincomb}, for each of the basis vectors $e_j$ of $E$, $j=1,\ldots,k$, write
$$
e_j = \sum_{i=1}^k \beta_{ji} z_i ,
$$
with the coefficients $\beta_{ji}$ satisfying  \eqref{beta-bd}. Thus,
\begin{align*}
\iprod{y}{e_j} & = \sum_{i=1}^k \beta_{ji} \iprod{y}{z_i} ,\qquad\text{so} \\
|\iprod{y}{e_j}| & \le \left( \sum_{i=1}^k \beta^2_{ji} \right)^{1/2} \cdot
\left( \sum_{i=1}^k \iprod{y}{z_i}^2 \right)^{1/2} ,
\end{align*}
for each $j$. Using the bounds in  \eqref{beta-bd},
and summing up the resulting geometric series, we get
\begin{align*}
|\iprod{y}{e_j}| & \le \left( \zeta^2 \sum_{i=1}^k (1+\zeta)^{2i-2} \right)^{1/2} \cdot
\left( \sum_{i=1}^k \iprod{y}{z_i}^2 \right)^{1/2} \\
& \le \frac{ \zeta (1+\zeta)^{k} }{\sqrt{2\zeta + \zeta^2}} \cdot \sqrt{\gamma} \\
& < (1+\zeta)^{k} \sqrt{\gamma} .
\end{align*}
We therefore have
\begin{equation} \label{kltd}
\sum_{j=1}^k |\iprod{y}{e_j}|^2 < k(1+\zeta)^{2k} \gamma \le \delta^2\ ,
\end{equation}
which follows from the definition of $\gamma$ (Equation (\ref{eq:gamma})).
This implies
that $Y\subset (E^\perp)_{\delta}$.
Note that, since we assumed that $|y|\ge \delta$, it follows that $k$ must be smaller than $d$,
for otherwise the left-hand side of (\ref{kltd}) would be $|y|^2 \ge \delta^2$, and then
the inequality in (\ref{kltd}) would be impossible.
This completes the proof of Lemma~\ref{yzperp}.
$\Box$
\medskip

We now apply this machinery to the preceding analysis of the strong convexity parameter $\alpha$. 
Recall that this analysis has lead to a partition $X = X^-\cup X^+$, where
$X^- = \{x_i \mid i\in \sigma^-\}$ and $X^+ = \{x_i \mid i\in \sigma^+\}$,
for some pair of complementary subsets $\sigma^-$, $\sigma^+$ of $[n]$, and
our goal is to establish a lower bound on
$$
\Xi(\sigma^+,\sigma^-) = \sum_{i\in \sigma^-} \sum_{j\in \sigma^+} \iprod{u_i}{u_j}^2 ,
$$
where $u_i = \Sigma^{-1}V^T \left( e^{t_i/2}x_i \right)$, for $i=1,\ldots,n$,
and where $t$ is an arbitrary vector in the domain $K_0$.
(Recall that the $u_i$'s are the rows of the $n\times d$ matrix $U$ 
in the SVD $U\Sigma V^T$ of $\left( \{e^{t_i/2}x_i\}_{i=1}^n \right)^T$.

We apply Lemma~\ref{yzperp} to the sets
\begin{align*}
U^- & = \left\{u_i = \Sigma^{-1}V^T \left( e^{t_i/2}x_i \right) \mid i\in \sigma^- \right\} \\
U^+ & = \left\{u_i = \Sigma^{-1}V^T \left( e^{t_i/2}x_i \right) \mid i\in \sigma^+ \right\}
\end{align*}
of rows of $U$.
Recall that $|u_i| \le 1$ for all $i$ (see
Section~\ref{sec:beta}).

Before applying the lemma, we need the following auxiliary step.
\begin{lem} \label{ifu-thenx}
Let $i\in [n]$ and let $x_i$ and $u_i$ be as defined above, for some $t\in K_0$. 
Let $E$ be some subspace of $\RR^d$. Then, putting $t_{\rm min} := \min_i t_i$,
$$
{\rm dist}(x_i, V\Sigma(E)) \le \frac{\sigma_{\rm max}}{e^{t_i/2}}
{\rm dist}(u_i, E ) \le \frac{\sigma_{\rm max}}{ e^{t_{\rm min}/2}} {\rm dist}(u_i, E ) .
$$
\end{lem}
\noindent{\bf Proof.}
By definition, there exists a vector $q_i\in E$ such that
$$
{\rm dist}(u_i, E ) = |u_i - q_i| .
$$
We can write $q_i = \Sigma^{-1}V^T p_i$, for some $p_i\in V\Sigma(E)$. Then
$$
u_i - q_i = \Sigma^{-1}V^T \left( e^{t_i/2}x_i - p_i \right)
= \Sigma^{-1}V^T \left( e^{t_i/2}(x_i - p'_i) \right) ,
$$
where $p'_i = e^{-t_i/2}p_i$ also belongs to $V\Sigma(E)$. Equivalently, we have
$$
e^{t_i/2}(x_i - p'_i) = V\Sigma (u_i - q_i) .
$$
Writing $z_i = u_i-q_i$, we get
$$
e^{t_i/2} {\rm dist}(x_i, V\Sigma(E)) \le e^{t_i/2}|x_i - p'_i| = \left| V\Sigma (u_i - q_i) \right| =
\left| \Sigma z_i \right| ,
$$
since $V$ is orthonormal. We have
\[
\left| \Sigma z_i \right| = \left| \left( \sigma_1 z_{i1} ,\ldots, \sigma_d z_{id} \right) \right| 
= \left( \sum_{k=1}^d \sigma_k^2 z_{ik}^2 \right)^{1/2} 
\le \sigma_{\rm max} \left( \sum_{k=1}^d z_{ik}^2 \right)^{1/2} \le \sigma_{\rm max} |z_i| .
\]
That is, we have shown that
$$
e^{t_i/2} {\rm dist}(x_i, V\Sigma(E)) \le \sigma_{\rm max} |z_i| =
\sigma_{\rm max} {\rm dist}(u_i, E ) .
$$
This completes the proof of the lemma.
$\Box$
\medskip

We now take the $\delta$ in Condition \eqref{eq:definition-of-asterix-condition}, and put
\begin{equation} \label{delta1}
\delta_1 = \frac{e^{t_{\rm min}/2 }}{ \sigma_{\rm max} } \delta .
\end{equation}
We apply Lemma~\ref{yzperp} with $\delta_1$ to $Y = U^-$ and $Z = U^+$.
It asserts that if
$$
\Xi(\sigma^+,\sigma^-)=\sum_{i\in \sigma^-} \sum_{j\in \sigma^+} \iprod{u_i}{u_j}^2 \le 
\gamma = \frac{\delta_1^2} {d\left( 1 + \frac{1}{\delta_1} \right)^{2d}}
$$
then there would exist two complementary subspaces
$E$, $E^\perp$ of $\RR^d$, such that
$U^+\subset E_{\delta_1}$ and $U^-\subset E^\perp_{\delta_1}$.
By Lemma~\ref{ifu-thenx} and Equation (\ref{delta1}), this would imply that
$X^+ \subset (F^+)_{\delta}$ and $X^- \subset (F^-)_{\delta}$, where
$F^+ = V\Sigma E^+$ and $F^- = V\Sigma E^-$.
Condition \eqref{eq:definition-of-asterix-condition} would then imply that
$$
\sum_{x_i\in X^-} c_i \le {\rm dim}(F^+)(1-\eta) \qquad\text{and}\qquad
\sum_{x_i\in X^+} c_i \le {\rm dim}(F^-)(1-\eta) ,
$$
which is a contradiction, since the left-hand sides sum up to $d$ and the
right-hand sides sum up to $d(1-\eta)$.

It follows (using the expression for $\delta_1$ given in \eqref{delta1}) that
\begin{equation} \label{eq:fixedt}
\Xi(\sigma^+,\sigma^-)  > \gamma =
\frac{\delta_1^2} {d\left( 1 + \frac{1}{\delta_1} \right)^{2d}} =
\frac{\delta^2 e^{t_{\rm min}}} {d\sigma_{\rm max}^2 \left( 1 + \frac{\sigma_{\rm max} }{\delta e^{t_{\rm min}/2}} \right)^{2d}} ,
\end{equation}
for every partition into non-empty subsets $\sigma^+,\sigma^-$ of $[n]$.

This discussion so far was for a fixed $t\in K_0$ which defined the set of the
$u_i$'s, and indeed $\sigma_{\rm max}$ in \eqref{eq:fixedt} depends on $t$.

 
To get a lower bound on $\Xi(X)$ we need
an upper bound on $\sigma_{\rm max}$, over $t\in K_0$. A simple upper bound of this kind is
$\sigma_{\rm max} \le \sqrt{n}e^{t_{\rm max}/2}$, because each $\sigma_k^2$
is the sum of the squared projections of the vectors $e^{t_i/2}x_i$ in some direction $v$;
that is, we have, for some $v$,
$$
\sigma_{\rm max}^2 = \sum_{i=1}^n e^{t_i} \iprod{x_i}{v}^2 \le
\sum_{i=1}^n e^{t_i} |x_i|^2 =
\sum_{i=1}^n e^{t_i} \le ne^{t_{\rm max}} ,
$$
from which the claim follows. Substituting this bound in (\ref{eq:fixedt}),
and assuming there that $t_{\rm min} = 0$, we obtain
\[
\alpha \ge \frac{1}{n^3}\Xi(X) >
\frac{\delta^2} {dn^4e^{2t_{\rm max}} \left( 1 + \frac{\sqrt{n}e^{t_{\rm max}} }{\delta} \right)^{2d}} ,
\]
which is Inequality (\ref{uiuj-lb}) reviewed earlier. This readily leads to Theorem~\ref{thm:strong},
which we restate here for the convenience of the reader.

\strongtheorem*

\subsection{A different path to strong convexity}\label{App-more-on-strong-conv}

In this section we present a different estimate for the smallest positive
eigenvalue of the Hessian $\nabla^2 \Phi(t) = \nabla^2 f(t)$. 
We have the following lower bound for the Hessian of $f$ restricted to $y\perp {\bf 1}$. 
\begin{thm} \label{alt:alpha}
Let $X = \{ x_i\}_{i=1}^n\subset \sph^{d-1}\subset \RR^d$ and assume each $d$-tuple is linearly independent. 
Let $\Delta_S = \det^2 \left(\{x_i\}_{i\in S}\right)$, and $\Delta_S^{\rm min} = \min_S \Delta_S>0$. 
Let $t\in K_0$  and let $|y|=1$ with $\sum y_i =0 $. Then 	
\[ 
y^T \nabla^2 \Phi (t)y \ge 
\frac{(\Delta_S^{\rm min})^2}{e^{4d|t|_\infty}}\frac{d(n-d)}{n(n-1) } .  
\] 
\end{thm}
\noindent{\bf Proof.}
Denote $g(t) = \exp (\Phi(t))$, so that 
\[ 
\nabla \Phi(t)  = \frac{\nabla g(t)}{g(t)}, \qquad \nabla^2 \Phi(t) = \frac{g(t)\nabla^2 g(t) - \nabla g(t) \otimes \nabla g(t)}{g(t)^2}, 
\] 
and recall that 
\[ 
g (t)  =  \sum_{|S|=d} e^{\sum_{i\in S} t_i} \Delta_S. 
\]
As before, let ${\bf 1}_S$ stand for the $n$-dimensional vector with $1$ 
at coordinates in $S$ and $0$ outside $S$. Thus we may write 
\[ 
g (t) = \sum_{|S|=d} \Delta_S e^{\iprod{t}{{\bf 1}_S}}.
\]
In particular, 
\[ 
\nabla g (t)= \sum_{|S|=d} \Delta_S  e^{\iprod{t}{{\bf 1}_S}} {\bf 1}_S, 
\qquad \text{and}\qquad 
\nabla^2 g(t) = \sum_{|S|=d} \Delta_S  e^{\iprod{t}{{\bf 1}_S}} {\bf 1}_S\otimes {\bf 1}_S. 
\] 
We may thus write that 
\[
\nabla^2 \Phi (t)= \frac{\left( \sum_{|S_1|=d} \Delta_{S_1} e^{\iprod{t}{{\bf 1}_{S_1}}} \right)\left(\sum_{|S_2|=d} \Delta_{S_2}
	e^{\iprod{t}{{\bf 1}_{S_2}}} {\bf 1}_{S_2}\otimes {\bf 1}_{S_2}\right)- \nabla g(t) \otimes \nabla g(t)}{g^2(t)}, 
\]
that is, 
\[
\nabla^2 \Phi (t) = \frac{ \sum_{|S_1|=d, |S_2|=d} \Delta_{S_1}\Delta_{S_2}  
	e^{\iprod{t}{{\bf 1}_{S_1}+{\bf 1}_{S_2}}}\left({\bf 1}_{S_2}\otimes {\bf 1}_{S_2} 
	- {\bf 1}_{S_1} \otimes  {\bf 1}_{S_2} \right) }{g^2(t)}, 
\]
which can be rewritten as 
\[  
\nabla^2 \Phi (t)= \frac{ \sum_{|S_1|=d, |S_2|=d} \Delta_{S_1}\Delta_{S_2}  
	e^{\iprod{t}{{\bf 1}_{S_1} + {\bf 1}_{S_2}}}\left({\bf 1}_{S_1} - {\bf 1}_{S_2}\right)\otimes 
	\left({\bf 1}_{S_1} - {\bf 1}_{S_2}\right) }{2 g^2(t)}. 
\]

This observation allows us to lower bound the Hessian (in the sense of positive definite matrices) 
in the case, under consideration, where $X$ is in general position. 
That is, recalling that $t\in E_0$ and reasoning as in the preceding subsection, we have
\[  
\nabla^2 \Phi (t) \ge \frac{(\Delta_S^{\rm min})^2}{2 g^2(t)e^{2d|t|_\infty}} \sum_{|S_1|=d, |S_2|=d}    
\left({\bf 1}_{S_1}-  {\bf 1}_{S_2}\right)\otimes \left({\bf 1}_{S_1} - {\bf 1}_{S_2}\right) . 
\]
The expression $g(t)$ in the denominator can be upper bounded by $g(0)e^{d|t|_\infty}$, 
which in turn is at most $\binom{n}{d}e^{d|t|_\infty}$, 
as $g(0) = \sum \Delta_S$ and each $\Delta_S$ is at most $1$.

The matrix in the numerator can easily be computed. Since, by construction, it is fully symmetric in all
coordinates, it must be of the form $a I_n + b {\bf 1}_n$ (where $I_n$ is the $n\times n$ identity matrix and 
${\bf 1}_n = {\bf 1}_{[n]}\otimes {\bf 1}_{[n]}$ is the all $1$ matrix). Note also that
the sum of its entries must be $0$, as this is the case for each matrix in the sum.
Thus, to find $a$ and $b$, it suffices to compute the trace of this matrix, which is 
$n(a+b)$; then we can figure out $a$ and $b$ using the fact that $n(a+b)+n(n-1)b=0$.

Since
\[ 
{\rm trace} \left(\left({\bf 1}_{S_1}-  {\bf 1}_{S_2}\right)\otimes \left({\bf 1}_{S_1} - {\bf 1}_{S_2}\right)\right) 
= |S_1\setminus S_2| + |S_2\setminus S_1| = 2|S_1\setminus S_2|, 
\]
we have, for any fixed $S_0$,
\[ 
\sum_{|S_1|=d, |S_2|=d} {\rm trace} \left({\bf 1}_{S_1} - {\bf 1}_{S_2}\right)\otimes 
\left({\bf 1}_{S_1} - {\bf 1}_{S_2}\right) = 2\binom{n}{d} \sum_{|S|=d} |S_0\setminus S| 
= 2\binom{n}{d}\sum_{j=0}^d j \binom{d}{d-j}\binom{n-d}{j} ,
\] 
which, by a classical combinatorial formula, gives
\[ 
{\rm trace} \left(\sum_{|S_1|=d, |S_2|=d} \left({\bf 1}_{S_2} - {\bf 1}_{S_1}\right)\otimes 
\left({\bf 1}_{S_1} - {\bf 1}_{S_2}\right)\right) = 2\binom{n}{d} (n-d)\binom{n-1}{d-1} .
\]
This means that the diagonal entries, $a+b$, are just $2\binom{n}{d} \frac{n-d}{n}\binom{n-1}{d-1}$. 
The off diagonal entries must then satisfy $n(n-1)b + 2\binom{n}{d}(n-d)\binom{n-1}{d-1}=0$,
so we get 
\[ 
b = - 2\binom{n}{d} \binom{n-1}{d-1}\frac{n-d}{n(n-1)}, 
\]
which in turn gives 
\[ 
a = (a+b)-b = 2\binom{n}{d} \frac{n-d}{n}\binom{n-1}{d-1} +2\binom{n}{d} \frac{n-d}{n(n-1)}\binom{n-1}{d-1}  
= 2\binom{n}{d} \binom{n-1}{d-1}  \frac{n-d}{n-1 } .
\]
In conclusion, we have
\[ 
\sum_{|S_1|=d, |S_2|=d} \left({\bf 1}_{S_1} - {\bf 1}_{S_2}\right)\otimes \left({\bf 1}_{S_1} - {\bf 1}_{S_2}\right) = 
2\binom{n}{d} \binom{n-1}{d-1} \frac{n-d}{n-1 } \left( I_n - \frac{1}{n} {\bf 1}_n\right) .
\]
Note that $I_n - \frac{1}{n} {\bf 1}_n$ is simply the 
orthogonal projection onto the subspace $\sum t_i = 0$. This corresponds to the fact
that we do not expect strict convexity of $\Phi$ in the direction orthogonal to this subspace.

Summing up, using our assumptions that $y \perp {\bf 1}_{[n]}$, we obtain 
\[ 
y^T \nabla^2 \Phi (t)y \ge \frac{(\Delta_S^{\rm min})^2}{2 g^2(t)e^{2d|t|_\infty}} 2\binom{n}{d} \binom{n-1}{d-1} \frac{n-d}{n-1 } |y|^2 , 
\] 
where $g = \exp(\Phi)$ and $\Delta_S^{\rm min} = \min \Delta_S$. 

Using the inequality $g(t) \le e^{d|t|_\infty}g(0) \le \binom{n}{d} e^{d|t|_\infty}$, 
we get that, for $y \perp {\bf 1}_{[n]}$, and $t\in K_0$, we have 
\[ 
y^T \nabla^2 \Phi (t)y \ge \frac{(\Delta_S^{\rm min})^2}{e^{4d|t|_\infty}} 
\frac{\binom{n-1}{d-1}}{\binom{n}{d}} \frac{n-d}{n-1 } |y|^2 = 
\frac{(\Delta_S^{\rm min})^2}{e^{4d|t|_\infty}}\frac{d(n-d)}{n(n-1) } |y|^2 , 
\] 
as claimed.
$\Box$

Theorem \ref{alt:alpha} implies that, when $X$ is in general position, the smallest eigenvalue
$\alpha$ of $H$ within $E_0$ satisfies
\[
\alpha \ge \frac{(\Delta_S^{\rm min})^2}{e^{4d|t|_\infty}}\frac{d(n-d)}{n(n-1) } , 
\] 
which yields the alternative bound in Theorem~\ref{thm:strong} for sets $X$ in general position.


\paragraph{Acknowledgements.}
The authors wish to express their gratitude to Saugata Basu, Shachar Lovett, Shay Moran and Sivan Toledo
for useful discussions concerning certain aspects of the problem.


\appendix

\section{Putting a set in radial isotropic position:\\
    Characterization and properties} \label{sec:shiri-full}

We recall the definition of the basic
polytope associated with a set $X$ of vectors.\footnote{%
  Please note that the statements (definitions, lemmas, propositions) given in
  the appendices come with two distinct numberings: Those already appearing in the main
  part of the paper are given with their original numbering, whereas statements introduced
  only in the appendix are numbered by the appendix section.}

\defbasicpoly*

We denote by $\cal S$ the collection of all linearly independent $d$-tuples $S$ (those
are the extremal vectors of $K_X$). Define, for a set $S\subseteq [n]$, $|S|=d$,
\begin{equation} \label{def:deltaS}
\Delta_S = \det( (x_i)_{i \in S})^2 = \det\left(\sum_{i\in S} x_i\otimes x_i\right) .
\end{equation}
Clearly $\Delta_S > 0$ if and only if $S\in \cal S$.

\medskip
\noindent
Some authors (e.g., Forster~\cite{For}) consider only the special case
of \emph{general position}, where every $d$ distinct vectors in $X$
are linearly independent. In this case
the basic polytope is simply
\[
K_X =\conv \Bigl\{ {\bf 1}_S \mid S\subseteq[n], \; |S| = d \Bigr\}.
\]
As is easy to check, this polytope (in this special case)
is simply the cross-section $H$ of the unit cube $[0,1]^n$ by the hyperplane $\sum_{i=1}^n z_i = d$.
This implies (see Theorem \ref{th:barth} below) that
$X$ can  be brought to radial $c$-isotropic position
for any $c$ such that $c_i\in (0,1)$ and $\sum_i c_i = d$.

Barthe's celebrated theorem, whose proof is given in this section, along with equivalent 
formulations and some other relevant material, is as follows.

\begin{thm}[Barthe~\cite{Bar}] \label{th:barth}
    Let $c\in(\R^+)^n$. A set $X = \{x_1,\dots,x_n\}$ of unit vectors in $\RR^d$ can be put in
    radial $c$-isotropic position if and only if $c \in {\rm relint} (K_X)$.
\end{thm}
Here ${\rm relint} (A)$ denotes the relative interior of a set $A$, namely
its interior with respect to its affine hull. We remind the reader that any vector $c$
for which the theorem can be applied must also satisfy $\sum_{i=1}^n c_i = d$; we will
generally not mention this condition explicitly in what follows.
%
We also mention the obvious fact that
if the vectors in $X$ do not span $\RR^d$ then $K_X$ is empty and the vectors cannot be
put in radial $c$-isotropic position, for any vector $c$.
\medskip

Here is a brief overview of the contents of this section.
In Section \ref{sec:cll} we establish an equivalent representation of the basis polytope, 
due to Carlen, Lieb and Loss~\cite{CLL}. 
In Section \ref{sec:cb} we discuss the basic properties of the function $\Phi$, which
is the key to mapping $X$ into radial $c$-isotropic position, for suitable values of $c$.
Then, in Section \ref{sec:Legduality}, we discuss the Legendre dual $\Phi^*$ of $\Phi$,
and characterize its domain, consisting of those vectors $c$ for which $\Phi^*(c)$ is finite.  
Section \ref{sec:attained} characterizes the vectors $c$ for which there exists some $t$ 
satisfying $\Phi^*(c)+ \Phi(t) = \iprod{t}{c}$; that is, $\Phi^*$ is finite and is attained 
at the corresponding vector $t$. These turn out, as we show in Section \ref{sec:isoandphistar}, 
to be precisely the vectors $c$ for which $X$ can be mapped into $c$-radial isotropic position.
Moreover, the $t$ at which $\Phi^*(c)$ is attained determines the linear transformation that
sends $X$ to this position. We thus essentially complete the proof of Barthe's theorem in 
this subsection. In Sections \ref{sec:equivrel} and \ref{sec:transitivity} we discuss the 
dimension of $K_X$, and the notion of irreducibility of a system $\{x_i\}$, which ensures 
that $K_X$ is $(n-1)$-dimensional. In Section \ref{sec:unique:rip} we address uniqueness 
of the radial isotropic position. Finally, we connect radial isotropy with the notion of 
maximal entropy in Section \ref{sec:good-combination}.

\subsection{Equivalent representation of $K_X$}
\label{sec:cll}

As mentioned in the main text, the basic polytope has an equivalent representation, which also served as our motivation for 
the notion of $(\eta,\delta)$-deepness given in Definition \ref{def:etadelta}. We include, for completeness, the proof of this equivalence. 
%
\propcllversion*

\noindent{\bf {Proof.~}}
Let $\tilde{K}_X$ denote the right-hand side of the identity asserted in the proposition.
We make use of the following notion: given a vector $c\in\tilde{K}_X$, we
say that a subset $J\subset [n]$ is \emph{critical} for $c$ if
$\sum_{i\in J} c_i = \dim ({\rm span~} \{x_i\}_{i\in J})$.
For example, if $c_i = 1$ then $\{i\}$ is critical for $c$.
For the standard case $c =\frac{d}{n}{\bf 1}$, $J$ is critical for $c$ when
$$
|J| = \frac{n}{d} \dim ({\rm span~} \{x_i\}_{i\in J}) .
$$
Note that for this uniform vector $c$, we always have, by definition, 
$|J| \le \frac{n}{d} \dim ({\rm span~} \{x_i\}_{i\in J})$, for every $J\subset [n]$.

We claim that the following property holds. Let $c\in \tilde{K}_X$ be a vector
for which there exists some critical set $J$, and let $J_0$ be a subset of $J$.
Then there exists a minimal $J^*$ that is critical for $c$ and contains $J_0$.
$J^*$ is minimal not only in that it has minimal cardinality, but in that every 
other set $J_1$ that is critical for $c$ and contains $J_0$ must also contain $J^*$.

To establish this property, take $J^*$ to be the set of minimal cardinality
that contains $J_0$ and is critical for $c$. Assume, towards a contradiction,
that there exists some $c$-critical $J_1$ that contains $J_0$ but not $J^*$.
Consider the two sets $J_1\cap J^*$ and $J_1 \cup J^*$. Denoting
$W_1 = {\rm span} (x_i)_{i\in J_1}$ and $W^* = {\rm span} (x_i)_{i\in J^*}$,
we clearly have that
\[
{\rm span} (x_i)_{i\in J_1\cap J^*} \subset W_1 \cap W^*,
{\rm ~and~}{\rm span} (x_i)_{i\in J_1\cup J^*} \subset W_1 + W^* ,
\]
and at the same time
\begin{eqnarray*}
    \dim( {\rm span} (x_i)_{i\in J_1\cap J^*}) &+ & \dim ({\rm span} (x_i)_{i\in J_1\cup J^*})
    \le \dim (W_1\cap W^*)+ \dim (W_1 + W^*)\\
    & = & \dim (W_1)+ \dim(W^*) = \sum_{i\in J_1} c_i +\sum_{i\in J^*} c_i\\
    & = & \sum_{i\in J_1\cap J^*} c_i + \sum_{i\in J_1\cup J^*} c_i .
\end{eqnarray*}
But an opposite inequality holds for each of the two sums in the final expression, since $c\in \tilde{K}_X$.
We therefore must have a term by term equality, which means that $J_1\cap J^*$
is a $c$-critical set that contains $J_0$, contradicting the minimality of $J^*$.

With these preparations, we now proceed to the proof itself.
The fact that $K_X\subseteq \tilde{K}_X$ follows from the facts that $\tilde{K}_X$
is convex (as an intersection of halfspaces), and that for every $S$ with $|S| = d$
and with $\{x_i\mid i\in S\}$ independent, the vector ${\bf 1}_S$ clearly satisfies
the conditions of $\tilde{K}_X$ (in fact, as equalities). Therefore, their entire
convex hull is contained in $\tilde{K}_X$, and we have the first inclusion.

For the opposite inclusion, we show that the extreme points of $\tilde{K}_X$
are precisely the extreme vectors ${\bf 1}_S$. Clearly, each such vector is extreme
(it is a vertex of the cube $[0,1]^n$ and $\tilde{K}_X\subseteq [0,1]^n$).
Let then $c$ be an extreme vector of $\tilde{K}_X$. We need to show that each
of its entries is either $0$ or $1$. The rough idea is that if it has an entry
$0<c_k<1$ then, since $\sum_\ell c_\ell = d$, there must exist another non-integer
entry $c_j$, and then one would like to change both of them simultaneously to
$c_k \pm \eps$ and $c_j\mp \eps$, for $\eps>0$ sufficiently small, in such a
way that the inequalities defining $\tilde{K}_X$ are met. The obstacle are of course
the critical sets $J$ for which the conditions of $\tilde{K}_X$ are met as equalities.
For any such $J$, we must make sure that the two indices $k$ and $j$ are either
both in $J$, or both outside $J$.

Formally, let $J$ be some $c$-critical set that contains some index $k$ for
which $0<c_k<1$. By the claim at the beginning of the proof, there exists a
minimal $c$-critical set $J^*$ that contains $k$ and is contained in every
$c$-critical set that contains $k$ (such as $J$). Since $J^*$ is $c$-critical
and $\sum_{i\in J^*} c_i$ is an integer, $J^*$ must contain another index $j$
with $0<c_j<1$, so $\{j,k\}$ is contained in every $c$-critical set
containing $k$. Thus the argument offered above shows that $c$ is not
extreme, a contradiction implying that all entries of $c$ are $0$ and $1$.
Since $\sum_{i=1}^n c_i = d$, by assumption, it follows that $c={\bf 1}_S$,
for some $d$-element subset $S$, which is easily seen to be independent.

This establishes the inclusion $\tilde{K}_X\subseteq {K}_X$, and thus completes the proof.
\qed

\subsection{The function $\Phi$ and its basic properties}
\label{sec:cb}

The proof of Theorem \ref{th:barth} proceeds via the analysis of the
\emph{Legendre transform} of the function $\Phi: \R^n\to \RR$, defined by
\[
\Phi(t_1, \ldots, t_n)= \log \det \left( \sum_{i=1}^n e^{t_i} x_i \otimes x_i \right).
\]
We denote $t=(t_1,\ldots,t_n)$ and put
\[
Q(t) = \sum_{i=1}^n e^{t_i} x_i \otimes x_i .
\]
With this notation we have that
\[
\Phi(t) =  \log \det (Q(t)) .
\]
As long as $X=\{x_i\}_{i=1}^n$ spans $\RR^d$, an assumption that we will make
throughout this paper, the determinant of $Q(t)$ is positive, the matrix $Q(t)$ is
positive definite and invertible, and $\Phi$ is everywhere defined and finite.
The function $\Phi$
is of crucial importance in our proof.
The following lemma gives its basic properties.

\lemcbrestated*
\noindent{\bf Proof.}
Consider first the representation of $\Phi$, which we can write as
$$
\sum_{i=1}^n a_i x_i\otimes x_i = AB ,
$$
with $a_i = e^{t_i}$ for $i=1,\ldots,n$,
where $A$ is the $d\times n$ matrix whose $i$th column is $a_ix_i$,
and $B$ is the $n\times d$ matrix whose $i$th row is $x_i^T$, for $i=1,\ldots,n$
(note that $B^T=X$, regarded as a sequence of column vectors). The 
\emph{Cauchy-Binet formula} for determinants asserts that, for an arbitrary 
$d\times n$ matrix $A$ and an arbitrary $n\times d$ matrix $B$, we have
$$
\det (AB) = \sum_{|S|=d} \det(A^S) \det(B_S) ,
$$
where the sum extends over all subsets $S\subseteq [1,\ldots, n]$ of size $d$,
and where $A^S$ (resp., $B_S$) is the square $d\times d$ matrix composed
of the $S$-columns of $A$ (resp., the $S$-rows of $B$). See  \cite{BW} for details.
Applying this formula in our context, one can easily verify,
by construction, that
$$
\det(A^S) = \left( \prod_{i\in S} a_i \right) \det ((B^T)^S) =
\left( \prod_{i\in S} a_i \right) \det (B_S) ,
$$
which implies that
$$
\det\left( \sum_{i=1}^n a_i x_i \otimes x_i \right)
= \sum_{|S|=d} a_S \Delta_S,
$$
where $a_S = \prod_{i\in S} a_i$ and $\Delta_S$ is as defined in (\ref{def:deltaS}).
Thus
\begin{equation} \label{detx:cb}
\det\left(Q(t)\right) = \det\left(\sum_{i=1}^ne^{t_i}x_i\otimes x_i\right) = \sum_{|S|=d} e^{\sum_{i\in S} t_i} \Delta_S ,
\end{equation}
as asserted.
We next establish the convexity of $\Phi$.
Using H\"{o}lder's inequality, we have, for any $\lambda\in (0,1)$,
\begin{eqnarray*}
    \Phi \left((1-\lambda)t+\lambda s \right)  &=& \log \left( \sum_{|S|=d}
    e^{\sum_{i \in S} (1-\lambda)t_i+\lambda s_i } \Delta_S \right) \\
    &=&\log \left( \sum_{|S|=d} \left(\Delta_S e^{\sum_{i \in S} t_i }\right)^{1-\lambda}
    \left(\Delta_S e^{\sum_{i \in S} s_i }\right)^{\lambda} \right) \\
    &\le& \log \left( \left( \sum_{|S|=d}\Delta_S e^{\sum_{i \in S} t_i}\right)^{1-\lambda}
    \left(\sum_{|S|=d}\Delta_S e^{\sum_{i \in S} s_i}\right)^{ \lambda }  \right) \\
    &=& (1-\lambda)\Phi(t) + \lambda\Phi(s).
\end{eqnarray*}
It follows from the representation of $\Phi$ that it is
monotonically increasing in each coordinate, is everywhere differentiable, 
its derivatives are as stated in the lemma, and  they are all positive and sum up to $d$.
$\Box$


Regarding strict convexity, one easily sees (see also Lemma \ref{fact:lineardir} below)
that, restricting $\Phi$ to any line in $\RR^n$ parallel to the direction
${\bf 1}  = (1, \ldots, 1)$ yields a linear function, so $\Phi$ is not strictly convex.
In fact, the directions on which $\Phi$ is not strictly convex (or, more concretely, linear)
are intimately connected with the issue of \emph{irreduciblity} of the set $X$
and of the dimension of $K_X$, which we address in Section~\ref{sec:equivrel}.


\medskip 
It will be useful to have yet another, different, explicit representation of $\nabla \Phi$, 
in terms of the square root of the matrix $Q(t)$ (which is positive definite, thus admits 
a unique positive definite square root).
\lemdiffoflogdetsum*
\noindent{\bf Proof.}
Indeed, we clearly have
\[
\frac{\partial Q(t)}{\partial t_j} = e^{t_j}x_j \otimes x_j ,
\]
for $j=1,\ldots,n$. We use the notion of
differentiating a determinant $\det A$ ``in the direction of a matrix $B$'', for a
nonsingular matrix $A$, which is given by
\begin{align*}
\frac{\partial}{\partial B}  \det (A) & = \lim_{\eps\to 0} \frac{\det(A+\eps B)-\det (A)}{\eps} \\
& = \lim_{\eps\to 0} \frac{\det(A(I_d +\eps A^{-1}B))-\det (A)}{\eps} \\
& = \lim_{\eps\to 0} \frac{\det(A) \Bigl(\det(I_d +\eps A^{-1}B) - 1 \Bigr)}{\eps} .
\end{align*}
The first-order terms (in $\eps$) of the expansion of $\det (I_d + \eps A^{-1}B)$ are
precisely those that arise from the product of all the diagonal
elements, where all but one of the factors are $1$. (The power of $\eps$ in any other product
of elements is at least $2$, except for the diagonal product where all factors are $1$, a product
that is cancelled by the term $-1$ in the numerator). This implies that
\[
\frac{\partial}{\partial B} \det (A) =
\det(A){\rm tr}(A^{-1}B) .
\]
Hence,
\[
\frac{\partial}{\partial t_j} \Phi(t) =
\frac{\partial}{\partial t_j} \log \det (Q(t)) =
\frac{\frac{\partial}{\partial t_j}\det (Q(t))}{\det (Q(t))}
= \lim_{\eps\to 0} \frac{\det( Q(t+\eps \delta_j)) - \det (Q(t))}{\eps \det (Q(t))} ,
\]
where $\delta_j$ is the $j$th standard unit vector in $\R^n$.
By definition of $Q(t)$, this is equal, up to first-order terms, to
\[
\lim_{\eps\to 0} \frac{\det\left( Q(t) + \eps e^{t_j} x_j\otimes x_j\right) - \det (Q(t))}{\eps \det (Q(t))} .
\]
Using the above formula for the derivative of a determinant in the direction of a matrix, we thus get
\begin{align*}
\frac{\partial}{\partial t_j} \log \det (Q(t)) & =
{\rm tr}\left(Q^{-1}(t)e^{t_j}x_j \otimes x_j \right)
= e^{t_j} \iprod{Q^{-1}(t)x_j}{x_j} \\
& = e^{t_j} x_j^T Q^{-1}(t) x_j
= e^{t_j} \left| Q^{-1/2}(t) x_j\right|^2 ,
\end{align*}
where we have used the fact that $Q(t)$ and its inverse are symmetric.
$\Box$

\subsection{The Legendre dual of $\Phi$}\label{sec:Legduality}

The \emph{Legendre dual} of $\Phi$, denoted by $\Phi^*$, is a map from $\RR^n$ to $\RR\cup\{+\infty\}$,
given at a point $\xi\in \RR^n$ by
\[
\Phi^*(\xi) = \sup_{t\in \RR^n} \Bigl\{ \iprod{t}{\xi} - \Phi(t) \Bigr\} .
\]
We review some useful highlights from the classical
theory of Legendre (or Legendre-Fenchel) duality; see \cite{Rock}.

In general, given a convex lower semi-continuous function
$\psi: \RR^n \to \RR \cup \{+\infty\}$, we define the \emph{domain} of $\psi$ by
$$
\dom(\psi) = \{ u \in \RR^n \mid \psi(u) <\infty \} .
$$
Clearly, this is a convex set. At any point $t\in \dom(\psi)$ we define the \emph{subdifferential}
(or \emph{subgradient}) of $\psi$ at $t$ by
\[
\partial \psi (t) = \{ \xi \in \RR^n \mid \psi(t) + \iprod{v-t}{\xi} \le \psi(v) {\rm~~for~every~} v\in\R^n \}.
\]
If $\psi$ is differentiable at $t$ then $\partial\psi(t) = \{\nabla \psi(t)\}$,
as easily follows from the definition. In general, the subgradient consists of
all $\xi$ for which $(-\xi,1)$ is a normal of a hyperplane that
supports the graph of $\psi$ at $(t,\psi(t))$.

As $\psi$ is convex, the set of supporting hyperplanes to its graph
is non-empty at every point in the interior of its domain, that is,
if $t\in \rm{int}( \dom (\psi))$, then $\partial \psi (t) \neq \emptyset$.
At a point of non-differentiability, $\partial \psi (t)$ is no longer a singleton,
although one can easily check that it is always convex.

The \emph{Legendre dual} $\psi^*$ of a general function $\psi$ is defined, as above, as
\[
\psi^*(\xi) = \sup_{t\in \RR^n} \Bigl\{ \iprod{\xi}{t} - \psi(t) \Bigr\} .
\]
It is always lower semi-continuous and convex. If the same holds also
for $\psi$ then $\psi^{**} = \psi$. For any $t$ and $\xi$, we have that
\[
\psi^*(\xi) + \psi(t) \ge  \iprod{\xi}{t} .
\]
Moreover,
\begin{equation}\label{eq:Legadresubdiffs}
\psi^*(\xi) + \psi(t) = \iprod{\xi}{t} \iff \xi \in \partial \psi (t) \iff t \in \partial \psi^*(\xi).
\end{equation}

Note that these ``duality relations'' imply that  $\psi$, say, is strictly convex (that is,
$t_1\neq t_2$ implies $\partial \psi (t_1)\cap \partial \psi (t_2)= \emptyset$)
if and only if $\psi^*$ is differentiable (that is, there is no $\xi$ for which
$\partial \psi^*(\xi)$ consists of more than one point).

For our function $\Phi$, whose domain is all of $\RR^n$ and which  is everywhere differentiable,
we may conclude that $\Phi^*$ is strictly convex. A key point will be to characterize
the domain of $\Phi^*$. Before doing that, we mention some simple properties.

\begin{lemma}\label{fact:lineardir}
    For any $\alpha\in \RR$, we have that $\Phi (t+ \alpha {\bf 1}) = \alpha d + \Phi(t)$.
    In particular, $\nabla \Phi (t) = \nabla \Phi (t+ \alpha {\bf 1})$ for any $\alpha$.
\end{lemma}
\noindent{\bf Proof.}
Indeed,
\begin{eqnarray*}
    \Phi(t+\alpha {\bf 1})&= &\log \det \left( \sum_{i=1}^n e^{t_i+\alpha} x_i \otimes x_i \right) =
    \log \det \left( e^\alpha \sum_{i=1}^n e^{t_i} x_i \otimes x_i \right) \\
    & =& \log \left( e^{\alpha d}\det \left( \sum_{i=1}^n e^{t_i} x_i \otimes x_i \right) \right) =
    \alpha d  + \Phi (t). \quad \Box
\end{eqnarray*}

\begin{lemma}
    $\dom (\Phi^*) \subseteq \{\xi \in \RR^n \mid \sum \xi_i = d,\;\xi_i\ge 0\;\text{for each $i$} \}$.
\end{lemma}
\noindent{\bf Proof.}
Indeed, using Lemma \ref{fact:lineardir},
\begin{eqnarray*}
    \Phi^*(\xi) &=& \sup_{t\in \RR^n} \left\{ \iprod{t}{\xi} - \Phi(t) \right\} =
    \sup_{t\in \RR^n, \alpha \in \RR} \left\{ \iprod{t+ \alpha {\bf 1}}{\xi} - \Phi(t+\alpha {\bf 1}) \right\}\\
    & = & \sup_{t\in \RR^n} \sup_{\alpha \in \RR}
    \left\{ \iprod{t}{\xi} + \alpha  \sum_{i=1}^n \xi_i - \Phi(t) -\alpha d\right\} =
    \Phi^*(\xi) + \sup_{\alpha \in \RR}  \left\{  \alpha  \left(\sum_{i=1}^n \xi_i - d \right)\right\}.
\end{eqnarray*}
Clearly the supremum is $\infty$ unless $\sum \xi_i = d$.

Similarly, if $\xi\in\RR^n$ has some negative coordinate $\xi_i$, then
letting $t_i\downarrow -\infty$ decreases $\Phi$ (by Lemma~\ref{lem:cb}),
and sends $\xi_it_i$ to $+\infty$, so
$\Phi^*(\xi) = \sup_t \left\{ \iprod{t}{\xi} - \Phi(t) \right\} = + \infty$,
thereby completing the proof.
$\Box$

\medskip
We shall discuss in more detail the dimension of $\dom(\Phi^*)$,
and the directions of linearity of $\Phi$, in Section \ref{sec:equivrel}.
 
One of the main claims
originating from Barthe's arguments and reproduced here, is that
$\Phi^*$ has the domain specified in the following lemma.

\propPhiSfiniterestated*
\noindent{\bf Proof.}
\noindent {\bf First direction: $K_X \subset \dom(\Phi^*)$}

Let $c\in K_X$. The property that we wish to establish, that $c$ belongs to $\dom (\Phi^*)$,
is by definition (see Lemma~\ref{lem:cb})
equivalent to the boundedness of the function of $t$ given by
\[
t \mapsto \iprod{c}{t} -  \log\det\left(Q(t) \right)
= \iprod{c}{t} -  \log\det\left(\sum_{i=1}^ne^{t_i}x_i\otimes x_i\right) .
\]
Assume now that $c = \sum_S \lambda_S {\bf 1}_S$, where the sum is over some set
${\cal S}$ of $d$-tuples $S$ with $\Delta_S> 0$, where $\lambda_S > 0$ for each $S\in {\cal S}$,
and $\sum_{S\in {\cal S}} \lambda_S = 1$. Then, by Lemma~\ref{lem:cb}, we have
\begin{eqnarray*}
    \det\left(Q(t)\right)
    &\ge& \sum_{S\in {\cal S}} \lambda_S \frac{e^{\sum_{i\in S} t_i}}{\lambda_S} \Delta_S
    \ge  \prod_{S\in {\cal S}}  \left(\frac{e^{\sum_{i\in S} t_i}\Delta_S}{\lambda_S}\right)^{\lambda_S} \\
    & = & \prod_{S\in {\cal S}}  \left(\frac{\Delta_S}{\lambda_S}\right)^{\lambda_S} \cdot
    \exp\left(\sum_{S\in {\cal S}}\sum_{i\in S} \lambda_S t_i\right)\\
    & = &  \exp\left(\sum_{i=1}^n \left( t_i\cdot \sum_{S\in{\cal S},\;i\in S} \lambda_S\right) \right) \cdot
    \prod_{S\in {\cal S}}  \left(\frac{\Delta_S}{\lambda_S}\right)^{\lambda_S}\\
    & = &  \exp\left(\sum_{i=1}^n c_i t_i\right) \prod_{S\in {\cal S}}  \left(\frac{\Delta_S}{\lambda_S}\right)^{\lambda_S} ,
\end{eqnarray*}
where we have used the non-negativity of the $\Delta_S$'s and a weighted version
of the arithmetic-geometric mean inequality. This implies
\[
\log\det\left(Q(t)\right)
\ge \iprod{c}{t} - \sum_{S\in {\cal S}}\lambda_S\log (\lambda_S/\Delta_S) .
\]
As this holds for every $t$, we obtain the following upper bound on $\Phi^*(c)$.
\[
\Phi^*(c) \le \sum_{S\in {\cal S}} \lambda_S\log (\lambda_S/\Delta_S) .
\]
That is, we have shown that $\Phi^*(c) < \infty$, as claimed.


\medskip
\noindent {\bf Second direction: $ K_X \supset \dom(\Phi^*)$}

For the converse direction we need to show that,
for $c\in \left(\RR^+\right)^n$, if $\Phi^*(c)$ is finite then
$c$ can be written as a convex combination of the vectors ${\bf 1}_S$, for $|S|=d$ and $\Delta_S > 0$.
Assume then that for all $t\in \RR^n$,
\[
\iprod{c}{t} - \log\det\left(\sum_{i=1}^ne^{t_i}x_i\otimes x_i\right)  \le  D < \infty .
\]
Taking the negative exponent of this inequality, we get that, for every $t$,
\[
\exp{(-\iprod{c}{t})} \det\left(\sum_{i=1}^ne^{t_i}x_i\otimes x_i\right) \ge e^{-D} > 0 .
\]
Write $t_i = -Na_i$, for arbitrary but fixed parameters $a_i$, $i=1,\ldots,n$, and a
common multiple $N$ which we let tend to $\infty$. Then, using (\ref{detx:cb}),
we can rewrite the above inequality as
\[
e^{N\sum c_i a_i}  { \det \left(\sum_{i=1}^n e^{-Na_i} x_i \otimes x_i\right)}=
e^{N\sum c_i a_i}  \sum_{|S|=d} {e^{-N\sum_{i\in S}a_i} \Delta_S} \ge e^{-D} > 0 .
\]
To have this property as $N$ tends to infinity, we need at least
one of the terms in the sum not to tend to $0$, which is equivalent to
$$
\sum_{i=1}^n a_i c_i \ge
\min \left\{ \sum_{i \in S} a_i  \mid |S|=d \mbox{ and } \Delta_S \neq 0 \right\}.
$$
This is a condition that has to hold for every vector $a$. We rewrite it as
\begin{equation} \label{asep}
\min_{ |S|=d, \, \Delta_S \neq 0} \iprod{a}{ {\bf 1}_S}  \le  \iprod{a}{ c} .
\end{equation}
This condition is easily seen to be equivalent to
$$
c \in \conv \left\{ {\bf 1}_S \mid |S|=d \mbox{ and } \Delta_S \neq 0 \right\}.
$$
Indeed, $c$ is in the convex hull if and only if it cannot be separated from
$$
\left\{ {\bf 1}_S \mid |S|=d \mbox{ and } \Delta_S \neq 0 \right\} ,
$$
which follows from condition (\ref{asep}). To see this, take $a$ to be the vector
normal to such a separating hyperplane (if such a hyperplane exists) and pointing
away from $c$. This yields a contradiction to (\ref{asep}),
showing that no such hyperplane exists, so $c$ lies in the desired convex hull.
$\Box$

\subsection{When is $\Phi^*(c)$ attained}
\label{sec:attained}

As we shall see in Section~\ref{sec:isoandphistar}, the question of radial $c$-isotropy
is linked with not only the finiteness of $\Phi^*(c)$, but rather with it actually
being attained, namely with the property that there exists $t\in \RR^n$ such that
\[
\Phi^*(c) = \iprod{t}{c}-\Phi(t) .
\]
Moreover, as we shall see, a solution $t$ of this equation yields the desired
linear transformation that maps $X$ into radial $c$-isotropic position; again, see
Section~\ref{sec:isoandphistar}.

%
%

\lemattainedrestated*

\noindent{\bf Proof.}
The supremum is attained if and only if there exists $t\in \RR^n$ such that
\[
\Phi^*(c)  + \Phi(t)= \iprod{t}{c}.
\]
By the general theory of Legendre transform given in Section \ref{sec:Legduality},
this is equivalent to $c \in \partial \Phi (t)$, which is equivalent to $t \in \partial \Phi^*(c)$.
If $c$ is in the {\em relative interior} of $\dom(\Phi^*)$
then there will be some non-vertical supporting hyperplane for $\Phi^*$ at $c$,
namely some $t$ with  $c = \nabla \Phi (t)$  (since we saw in Lemma \ref{lem:cb} that the function $\Phi$ is differentiable),
and then, as observed above, $\Phi^*(c)$ is attained (at $t$).

For the converse direction, if $\Phi^*(c)$ is attained at a point $t$
then $c = \nabla \Phi (t)$, which means, using that
\begin{equation*}
\Phi(t) = \log \det (Q(t)) = \log  \left(  \sum_{|S|=d} e^{\sum_{i\in S} t_i} \Delta_S\right),
\end{equation*}
that
\begin{equation} \label{eq:ci}
c_i = \frac{\partial}{\partial t_i}\Phi(t) =
\frac{1}{\det(Q(t))} \left( \sum_{|S|=d, i\in S} e^{\sum_{j\in S} t_j} \Delta_S\right) .
\end{equation}
Define
\begin{equation} \label{eq:ci:con}
\lambda_S = \frac{ e^{\sum_{i\in S}t_i}\Delta_S}{\sum_{|S|=d  }e^{\sum_{i\in S}t_i}\Delta_S} =
\frac{ e^{\sum_{i\in S}t_i}\Delta_S}{\det(Q(t))},
\end{equation}
and restrict the definition to the set $\cal S$ of those $d$-tuples $S$ for which $\Delta_S > 0$.
Then (\ref{eq:ci}) means that $c= \sum_{S\in {\cal S}} \lambda_S 1_S$, and as $\lambda_S>0$
for $S\in {\cal S}$, we have written $c$ as a convex combination, with all non-zero coefficients,
of the extremal vectors of $K_X$, meaning in particular that $c$ is in its relative interior.

We remark that all the statements used above are equivalent. In particular, if we find $t$
for which $c$ can be written as the convex combination in (\ref{eq:ci}), that is, find $t$
for which the coefficients $\lambda_S$ given in (\ref{eq:ci:con}) yield a convex combination
of the ${\bf 1}_S$'s that is equal to $c$, then $t$ is the value that attains $\Phi^*(c)$, and,
as we shall see in Lemma~\ref{prop:PhiSfinite-iff-rad-i-pos-exists}, is also the value needed
to transform $X$ into radial isotropic position.
$\Box$

\subsection{Back to Radial Isotropy} \label{sec:isoandphistar}

The following proposition is the main result in Barthe's analysis~\cite{Bar}.
\propPhiSfiniteiffradiposexistsrestated*
\smallskip
\noindent{\bf Remark.}
As we will show (in Lemma~\ref{lem:unique:rip}),
the radial $c$-isotropic position is, after the normalization, unique up to rotation.

\smallskip

\noindent{\bf Proof. First direction: $\Phi^*(c)$ is (finite and) attained implies radial $c$-isotropy.}
Assume that $\Phi^*(c)<+\infty$ and that there exists some $t\in \RR^n$ for which
\[
\Phi^*(c) = \iprod{t}{c}- \Phi(t) =
\iprod{t}{c}- \log \det \left( \sum_{i=1}^n e^{t_i} x_i \otimes x_i \right).
\]
Write the right-hand side of this equality as $F(t)$ (keeping $c$ fixed).
As already observed, $F$ is defined and differentiable (with respect to $t$) everywhere.
Since our $t$ maximizes $F$, we have
$$
0 = \nabla F(t) = c - \nabla \Phi(t) ,
$$
so we get
\[
\nabla \Phi(t) = c.
\]
Equivalently, using Lemma \ref{lem:diff-of-logdetsum}, we have
\begin{equation} \label{eqcj}
c_j=  e^{t_j}|Q^{-1/2}(t)x_j|^2,\quad\quad\text{for $j=1,\ldots,n$} .
\end{equation}
For any vector $t$ we have
\begin{align*}
I_d &= Q^{-1/2}(t)Q(t)Q^{-1/2}(t) = Q^{-1/2}(t)\left(\sum_{i=1}^n
e^{t_i}x_i\otimes x_i\right) Q^{-1/2}(t) \\
& = \sum_{i=1}^n e^{t_i}Q^{-1/2}(t)(x_i\otimes x_i) Q^{-1/2}(t) = \sum_{i=1}^n
e^{t_i}(Q^{-1/2}(t)x_i)\otimes (Q^{-1/2}(t)x_i) .
\end{align*}
Hence, for our special extremizing $t$, we have, using (\ref{eqcj}),
\[
I_d = \sum_{i=1}^n
c_i\frac{1}{|Q^{-1/2}(t)x_i|^2}(Q^{-1/2}(t)x_i)\otimes (Q^{-1/2}(t)x_i) .
\]
In other words, the vectors $x_i$ are sent by the matrix $Q^{-1/2}(t)$ to radial
$c$-isotropic position.

\medskip
\noindent{\bf Second direction: radial $c$-isotropy implies $\Phi^*(c)$ is (finite and) attained.}
For this direction, we shall use the following interesting result.
Roughly, it asserts that if we apply our process to a point set that
is already in radial $c$-isotropic position, we get explicit (entropy-like)
values for the corresponding $\Phi^*(c)$ and for the extremizing vector $t$.
\begin{lem}\label{lem:phiS=entc}
    Let $z_i \in \sph^{d-1}\subset \RR^d$, for $i=1,\ldots,n$, be points that satisfy
    $$
    \sum_{i=1}^n c_i z_i \otimes z_i = I_d .
    $$
    Put $\Phi(t) = \Phi_Z(t) = \log \det \Bigl(\sum_{i=1}^n e^{t_i} z_i \otimes z_i\Bigr)$.
    Then $\Phi^*(c) = \sum_{i=1}^n c_i \log c_i$, and $\Phi^*(c) + \Phi(t) = \iprod{c}{t}$
    for the vector $t$ given by $t_i = \log c_i$, for $i=1,\ldots,n$.
\end{lem}

Before proving the lemma, let us show how it implies the second direction in
Lemma~\ref{prop:PhiSfinite-iff-rad-i-pos-exists}.

\medskip
\noindent{\bf Proof of the second direction in Lemma~\ref{prop:PhiSfinite-iff-rad-i-pos-exists}.}
Assume that a set $X=\left\{x_i\right\}_{i=1}^n$ of vectors can be brought into a set
$Z=\{z_i\}_{i=1}^n$ of corresponding vectors that are in radial $c$-isotropic position,
and let $A$ be the matrix in $GL_d$ that effects this transformation.
Then, by Lemma \ref{lem:phiS=entc}, we have $\Phi_Z^*(c) = \sum c_i \log c_i$,
and it is attained at the point $s$ satisfying $s_i = \log c_i$, for $i=1,\ldots,n$.

We compare the functions $\Phi^*$ associated with the following three sets
of vectors. The original set $\{x_i\}$, the intermediate set $y_i = Ax_i$, and the final one $z_i = y_i /|y_i|$,
for $i=1,\ldots,n$. We write $Y$ and $Z$ for the respective sets
$\{y_i\}_{i=1}^n$ and $\{z_i\}_{i=1}^n$, and put
\begin{align*}
\Phi_X(t) &= \log \det \left( \sum_{i=1}^n e^{t_i} x_i \otimes x_i \right), \\
\Phi_Y(t) &= \log \det \left( \sum_{i=1}^n e^{t_i} y_i \otimes y_i \right),\\
\Phi_Z(t) &= \log \det \left( \sum_{i=1}^n e^{t_i} z_i \otimes z_i \right).
\end{align*}

We have
\begin{align*}
\Phi_Y(t) &= \log \det \left( \sum_{i=1}^n e^{t_i} y_i \otimes y_i \right)
= \log \det \left( \sum_{i=1}^n e^{t_i} Ax_i \otimes Ax_i \right) \\
&= \log \det \left( A \left( \sum_{i=1}^n e^{t_i} x_i \otimes x_i \right) A^T \right) \\
&= \log \det \left( \sum_{i=1}^n e^{t_i} x_i \otimes x_i \right) + 2\log \det (A) = \Phi_X(t) +2\log\det (A) ,
\end{align*}
and
\begin{align*}
\Phi_Y(t) & = \log \det \left( \sum_{i=1}^n e^{t_i} y_i \otimes y_i \right)
= \log \det \left( \sum_{i=1}^n e^{t_i+2\log |y_i|}\cdot \frac{1}{|y_i|^2} y_i \otimes y_i \right) \\
& = \log \det \left( \sum_{i=1}^n e^{t_i+2\log |y_i|} z_i \otimes z_i \right)
= \Phi_Z(s) ,
\end{align*}
where $s\in \RR^n$ is the vector
\begin{equation} \label{eq:si}
s_i = t_i + 2\log |y_i| , \qquad i=1,\ldots,n .
\end{equation}
Therefore,
\begin{align*}
\Phi_Y^*(\xi) & = \Phi_X^*(\xi)-2\log \det (A) ,\quad\text{and} \\
\Phi_Z^*(\xi) & = \sup_s \Bigl\{ \sum_{i=1}^n s_i \xi_i  - \Phi_Z(s) \Bigr\} \\
& = \sup_t \Bigl\{ \sum_{i=1}^n t_i \xi_i - \Phi_Y(t) \Bigr\}
+ 2\sum_{i=1}^n \xi_i \log|y_i| = \Phi_Y^*(\xi)+ 2\sum_{i=1}^n \xi_i \log|y_i|.
\end{align*}
In particular, we see that
$$
\Phi_X^*(c)< \infty \quad\text{if and only if}\quad
\Phi_Y^*(c)< \infty \quad\text{if and only if}\quad
\Phi_Z^*(c)< \infty .
$$
Moreover, if any of these three suprema is attained then so are the other two,
and the points where they are attained are intimately connected.
The first two are attained at the same $t$, and
the third is attained at $s$, where $s$ and $t$ are related as in (\ref{eq:si}).

As we have assumes that  $Z $ is in radial $c$-isotropic position, we have a finite $\Phi_Z^*(c) = \sum c_i \log c_i$,
and it is attained at the point $s$ satisfying $s_i = \log c_i$, for $i=1,\ldots,n$.
The preceding discussion then implies that $\Phi_X^*(c)$ is also finite and attained;
by (\ref{eq:si}), it is attained at the point $t$ satisfying
$t_i = \log c_i -2\log |Ax_i|$, for $i=1,\ldots,n$.
$\Box$

\medskip
\noindent{\bf Proof of Lemma \ref{lem:phiS=entc}.}
Let $z_i \in \sph^{d-1}$, for $i=1,\ldots,n$,
be vectors satisfying $\sum_{i=1}^n c_i z_i \otimes z_i = I_d$. Recall that
\[
\Phi_Z^*(c) = \sup_{t\in \RR^n} \left\{
\iprod{t}{c} - \log \det \left(\sum_{i=1}^n e^{t_i} z_i \otimes z_i\right) \right\} .
\]
One possible choice of $t$ is $t_i = \log c_i$, for $i=1,\ldots,n$, in which case
$\Phi_Z(t) = \log \det (I_d) = 0$. Thus, $\Phi_Z^*(c) \ge \sum_{i=1}^n c_i \log c_i$.
For the other inequality, we need to show that, for any $t\in \RR^n$,
\[
\iprod{t}{c} - \log \det \left(\sum_{i=1}^n e^{t_i} z_i \otimes z_i \right)\le \sum c_i \log c_i ,
\]
which is equivalent to showing that
\begin{equation} \label{eq:detlb}
\det \left(\sum_{i=1}^n e^{t_i} z_i \otimes z_i\right) \ge \frac{ e^{\sum_{i=1}^n t_i c_i} }{\prod c_i^{c_i}}.
\end{equation}
To show this, we use the Cauchy-Binet formula (as introduced in Section~\ref{sec:cb})
for the determinant of $\sum_{i=1}^n c_i z_i \otimes z_i = I_d$. It asserts that
\begin{align*}
1 & = \det(I_d) = \det\Bigl( \sum_{i=1}^n c_i z_i \otimes z_i \Bigr) \\
& = \sum_{|S|=d} \det\left( [c_iz_i]_{i\in S} \right) \cdot
\det\left( [z_i]_{i\in S} \right) \\
& = \sum_{|S|=d} \det\Bigl( \sum_{i\in S} c_iz_i\otimes z_i \Bigr) =
\sum_{|S|=d} \lambda_S ,
\end{align*}
where
$$
\lambda_S = \det \left( \sum_{i\in S} c_i z_i \otimes z_i \right) .
$$
We thus have $\sum \lambda_S = 1$,
and it suffices to consider only those $S$ for which $\lambda_S > 0$, or, equivalently,
$\Delta_S>0$, which are precisely those $S$ participating in the definition of $K_Z$.
Indeed, we have already noted that $\Delta_S>0$ if and only if the set $\{x_i\}_{i\in S}$
is linearly independent. The similar claim for $\lambda_S$ follows since
$$
\lambda_S = \det \left( \sum_{i\in S} (\sqrt{c_i} z_i) \otimes (\sqrt{c_i} z_i) \right) .
$$
Now, for arbitrary values of $t_1,\ldots,t_n$, we have, using again the Cauchy-Binet formula,
$$
\det \left(\sum_{i=1}^n e^{t_i} z_i \otimes z_i\right) =
\det \left(\sum_{i=1}^n \frac{e^{t_i}}{c_i}  c_i z_i \otimes z_i\right) =
\sum_{|S|=d, \Delta_S>0} \left( \prod_{i\in S} \frac{e^{t_i}}{c_i} \right) \lambda_S.
$$
Denote the product
$\prod_{i\in S} \frac{e^{t_i}}{c_i}$ as $\rho_S$, for each $S$. Using a weighted
version of the arithmetic-geometric mean inequality, we get
\begin{align*}
\det \left(\sum_{i=1}^n e^{t_i} z_i \otimes z_i\right) & =
\sum_{|S|=d} \rho_S\lambda_S
\ge \prod_{|S|=d} \rho_S^{\lambda_S}
= \prod_{|S|=d,\;\Delta_S>0} \rho_S^{\lambda_S} \\
& = \prod_{|S|=d,\;\Delta_S>0}\; \prod_{i\in S} \left(\frac{e^{t_i}}{c_i} \right)^{\lambda_S}
= \prod_{i=1}^n  \left(\frac{e^{t_i}}{c_i}\right)^{\sum_{|S|=d,\;\Delta_S>0,\; i\in S} \lambda_S}.
\end{align*}
Comparing this with (\ref{eq:detlb}), what
remains to show is that, under the assumption that
$\sum_{i=1}^n c_i z_i \otimes z_i = I_d$, we have, for each $i=1,\ldots,n$,
$\sum_{|S|=d,\;\Delta_S>0,\; i\in S} \lambda_S = c_i$. Indeed, for each $i$,
\begin{align*}
\sum_{|S|=d,\;\Delta_S>0,\;i\in S} \lambda_S & =
\sum_{|S|=d,\;\Delta_S>0} \lambda_S - \sum_{|S|=d,\;\Delta_S>0,\; i\not\in S} \lambda_S \\
& = 1 - \det (I_d - c_i z_i \otimes z_i) = 1-(1-c_i) = c_i ,
\end{align*}
where, for the one but last equality, one may simply expand the determinant in an orthogonal basis in which the first vector is $z_i$.
$\Box$

\medskip
\noindent{\bf Remark.}
Note that the proof also shows that $\Phi_Z^*(c) = \iprod{c}{t} - \Phi_Z(t)$ if and only if
there is  equality in the arithmetic-geometric mean inequality,
which holds if and only if all terms $\rho_S$ are equal. That is, all products $\prod_{i\in S} \frac{e^{t_i}}{c_i}$
must be equal, for all sets $|S|=d$. In the irreducible case (we discuss this notion in the next section) this implies that 
we have $\frac{e^{t_i}}{c_i} = e^\rho$ for some
$\rho$ and for all $i$, and in the general case where there is a splitting, there can be $k$ different constant $\rho_j$ coming in, so that the
difference between the vector $t$ and the vector $(\ln c_i)_{i=1}^n$ is of the form $\sum \rho_j 1_{\sigma_j}$.

By combining this observation with (\ref{eq:si}), we get the following relationship, which is stated for simplicity in the irreducible case, a notion which will be defined and discussed in Section \ref{sec:equivrel}. 
\begin{lem}
    For a set $X$ of $n$ vectors on $\sph^{d-1}$ which are irreducible and for a vector $c\in(\R^+)^n$, $\sum_{i=1}^n c_i = d$,
    for which $\Phi_X^*(c) < \infty$ and is attained at some vector $t$,
    the extremizing $t$ satisfies, for some constant $\rho$,
    $$
    t_i = \rho + \log c_i - 2\log |Ax_i| ,\quad i=1,\ldots,n ,
    $$
    where $A$ is the matrix that, followed by re-normalization, brings $X$ to radial $c$-isotropic position.
    Consequently,
    $$
    X(t) = \beta \sum_{i=1}^n \frac{c_i}{|Ax_i|^2} x_i\otimes x_i ,
    $$
    for $\beta = e^\rho$.
\end{lem}
We note, though, that this explicit representation of the extremizing $t$ is not so explicit
after all, as it requires knowledge of $A$, which itself (as in Lemma~\ref{prop:PhiSfinite-iff-rad-i-pos-exists})
is a function of $t$.


An interesting corollary arises for the special case where $c$ is the uniform vector $\frac{d}{n}{\bf 1}$
and every $d$-tuple of vectors of $X$ is linearly independent (that is, $X$ is in general position).
In this case a trivial representation of $c$ as a convex combination is
\begin{equation} \label{c:conv:unif}
c = \sum_{|S| = d} \frac{1}{\binom{n}{d}} {\bf 1}_S ,
\end{equation}
and it is easily checked that $c$ also lies in the relative interior of the convex hull.
If $c$ is the uniform vector but not every $d$-tuple is independent, $c$ may fail to be
in the convex hull.

Moreover, it should be emphasized that in general not every convex combination of the ${\bf 1}_S$'s
that represents $c$ gives rise to a transformation that puts $X$ in radial $c$-isotropic position.
That is, verifying that $c$ belongs to the relative interior of the convex hull does not in itself
tell us how to map $X$ into radial $c$-isotropic position. What we need is a ``right'' way of
writing $c$ as a convex combination of the extremal vectors of $K_X$, from which we can find
a mapping of $X$ to radial $c$-isotropic position, as in Lemma~\ref{prop:PhiSfinite-iff-rad-i-pos-exists}.
As it turns out, and detailed in the proof of Lemma~\ref{lem:attained},
the right representation is the one in Equations (\ref{eq:ci}) and (\ref{eq:ci:con}) in that section.
That is, solving these equations for $t$, we obtain the vector $t$ for which $Q^{-1/2}(t)$ brings
$X$ to radial $c$-isotropic position. Moreover, as explained in Section \ref{sec:good-combination},
this representation corresponds to minimizing a certain entropy function (see Lee~\cite{Lee}
for more details). The representation of the uniform vector in (\ref{c:conv:unif})
is most likely not the right representation for this task.


\subsection{The dimension of $K_X$ and irreducibility}\label{sec:equivrel}

In this section we discuss the notion of ``irreducibility'' of a set $X=\{x_i\}_{i=1}^n$ of vectors,
and its connection with both the dimension of $K_X$ and the directions of linearity of $\Phi$.

To illustrate this notion, let us begin with a simple example. Assume that $\RR^d = E_1 \oplus E_2$
with $\dim (E_1)=d_1$, $\dim(E_2) =d_2$ and $d_1 + d_2 = d$, and that each vector $x_i$
falls into one of these two complementary subspaces. In such a case it is clear that any set
$S\subset [n]$ with $\Delta_S>0$ must include $d_1$ elements $x_i \in E_1$ and
$d_2$ elements $x_i \in E_2$. Therefore, the elements ${\bf 1}_S \in K_X$ will satisfy not only
$\iprod{{\bf 1}_S}{{\bf 1}} = d$ but also
$\iprod{{\bf 1}_S}{{\bf 1}_{\sigma_1}} = d_1$ and $\iprod{{\bf 1}_S}{{\bf 1}_{\sigma_2}} = d_2$,
where $\sigma_1 \subset [n]$ denotes the set of indices of those $x_i$ belonging to $E_1$,
and similarly for $\sigma_2$ and $E_2$. This reduces the dimension of $K_X$ by $1$.

In the more general case, consider the basis polytope $K_X$ (Definition \ref{def:basic-poly})
associated with $X$. The dimension of $K_X$ is connected to the number of equivalence classes
of the following relation $\sim$ on $[n]$, considered in Barthe~\cite{Bar}. We say that two indices
$i,j$ satisfy $i\sim j$ if there exists a set $S\subset [n]$ satisfying
$|S| = d-1$ and $S\cup\{i\} \in {\cal S}$, $S\cup\{j\} \in {\cal S}$.
We postpone the task of showing that  this relation is transitive to Section \ref{sec:transitivity}, from which it follows that this is
an equivalence relation.
The equivalence classes of $\sim$ form a partition of $[n]$,
which we denote, as above, as $[n] = \sigma_1 \cup \cdots \cup \sigma_k$.
Let $E_j = \mathrm{span}\{ x_i\mid i\in \sigma_j\}$, for $j=1,\ldots,k$.
As shown in Barthe~\cite{Bar}, and is not hard to verify,
$\RR^d = \bigoplus_{j=1}^k E_j$, and this is the maximal splitting of
$\RR^n$ into a direct sum of subspaces that collectively contain all the vectors of $X$.

\lemequivrelrestated*
\noindent{\bf Proof.}
To understand the dimension of $K_X$, we need to see which linear equations
$\iprod{{\bf 1}_S}{a}= b$ hold for all of the extreme vectors
$\{{\bf 1}_S \mid S\in {\cal S}\}$.

If $\sim$ has $k$ equivalence classes then the codimension of $K_X$ is
at most $k$, since letting $a_i = 1$ if $i \in \sigma_j$ and $0$ otherwise, the
subspace $\sum_{i\in S} a_i = \dim (E_j)$ will include $K_X$, as follows from the
easy fact from linear algebra fact that if $\RR^d = \bigoplus_{j=1}^k E_j$
then every basis of $\RR^d$ picked from these subspaces must include exactly
$\dim (E_j)$ elements from each $E_j$.

Since we are only interested in the case where $K_X\neq \emptyset$, there
is at least one set $S_0 \in {\cal S}$.
Assume that $\iprod{1_S}{a}= b$ holds for all $S\in {\cal S}$.
Given a set $S_1\in {\cal S}$ that differs
from $S_0$ by just one vector, we have that
\[
\sum_{i\in S_0} a_i = b,\,\,\, \sum_{i\in S_1} a_i = b,
\]
and so if $S_0\triangle S_1 = \{ j,i\}$ we get that $a_i = a_j$.
We thus see that for the equation $\iprod{1_S}{a}= b$ to hold for all $S\in {\cal S}$
we must have that if $i\sim j$ then $a_i = a_j$ and thus $a = \sum \alpha_j 1_{\sigma_j}$,
for suitable coefficients $\alpha_j$. Therefore, the affine hull of $K_X$ is the
intersection of $k$ codimension-1 subspaces, as claimed.
$\Box$

\begin{rem}
	When there is only {\em one} equivalence class, we say that $X$ is {\em irreducible};
	in this case the only affine subspace that contains $K_X$ is $\sum z_i = d$.
\end{rem}

We have already shown (Lemma \ref{fact:lineardir}) that $\Phi$ is linear in the direction ${\bf 1}$
and therefore not strictly convex.
The following lemma characterizes all directions on which $\Phi$
is not strictly convex (in fact, linear).

\lemwhenlinrestated*
\noindent{\bf Proof.}
Looking back at the proof of the convexity of $\Phi$
(in Section~\ref{sec:cb}), we see that
the condition for equality is that the vectors $\Bigl(\Delta_S e^{\sum_{i\in S}t_i}\Bigr)_{S\in{\cal S}}$
and $\Bigl(\Delta_S e^{\sum_{i\in S}s_i}\Bigr)_{S\in{\cal S}}$ are proportional. This means that there
is some $\alpha\in \RR$ such that for any $S\in {\cal S}$ we have
\[
\sum_{i\in S}t_i = \sum_{i\in S}s_i + \alpha ,
\]
which in turn means that if $i\sim j$
then $s_i-s_j = t_i - t_j$, or
$s_i-t_i = s_j-t_j$. Therefore, for any equivalence class $\sigma_r$ there is a coefficient
$\alpha_r$ such that $t_i=s_i+\alpha_r$ for any $i\in\sigma_r$, as claimed.

Conversely, since every $S\in {\cal S}$ includes the same number of members from
each equivalence class, the two vectors $s$ and
$t = s + \sum_{i=1}^k \alpha_j {\bf 1}_{\sigma_j}$ produce indeed proportional
vectors $\Bigl(\Delta_S e^{\sum_{i\in S}s_i}\Bigr)$ and $\Bigl(\Delta_S e^{\sum_{i\in S}t_i}\Bigr)$,
implying that $\Phi$ is linear on the line through $s$ and $t$, as needed.
$\Box$


When $\Phi$ is reducible, we get a partition of $\RR^d$ into a nontrivial direct sum
$\bigoplus_{j=1}^k E_j$.
If $E_i$ and $E_j$ are orthogonal for every pair $i\ne j$ then it is easy to check that
\[
\Phi(t) = \sum_{j=1}^k \Phi_j((t_i)_{i\in \sigma_j}) ,
\]
where $\Phi_j$ is the restriction of $\Phi$ to $E_j$.
This  follows from the fact that the determinant decomposes into a product of determinants,
each over the restriction of the vectors to a particular subspace in the sum. Consequently,
everything else factorizes too, so $\Phi^*(c) = \sum_{j=1}^k \Phi_j^*((c_i)_{i\in \sigma_j})$,
and the domain of $\Phi^*$ is the intersection of the domains of the $\Phi^*_j$'s.
This gives another explanation for the codimension $k$ of $K_X$, as its
affine hull is the (transversal) intersection of $k$ subspaces of codimension $1$ in $\RR^n$.
Furthermore, the optimization problem of finding a vector $t$ that maximizes $\iprod{c}{t} - \Phi(t)$,
which we handle using gradient descent (in Section \ref{sec:gd}), decomposes, and we may assume,
in such an algorithm, that $\Phi$ is irreducible.

\medskip
\noindent{\bf Remark.}
For the above decomposition, in the reducible case, to succeed, we must have
\begin{equation} \label{eq:cdecomp}
\sum_{i\in\sigma_j} c_i = \dim (E_j) ,\qquad \text{for each $j=1,\ldots,k$.}
\end{equation}
Put differently, if $X$ is reducible but (\ref{eq:cdecomp}) does not hold,
$X$ cannot be put in radial $c$-isotropic position.
As a concrete example, consider the uniform case $c=\frac{d}{n}{\bf 1}$, and
take $X$ to consist of multiple copies of the standard unit vectors $e_j$,
so that not all multiplicities are equal. In this case the direct sum consists
of one-dimensional spaces (spanned by the $e_j$'s), and (\ref{eq:cdecomp})
clearly does not hold. Thus sets of this kind cannot be put in
radial (uniform) isotropic position.

In case $\Phi$ is reducible but some of the subspaces in the decomposition $\RR^d=\bigoplus_{j=1}^m E_j$
are not orthogonal, we first transform $X$ into isotropic position (not necessarily radial).
After the transformation the decomposition becomes orthogonal and the problem decomposes,
as the following lemma shows.

\orthinisorestated*
\noindent{\bf Proof.}
We argue by contradiction. Let $X_1 = X\cap F_1$ and $X_2 = X\cap F_2$, and let
$\sigma_i = \{k\in [n] \mid x_k \in X_i\}$, for $i=1,2$.
Assume that there exists a unit vector $u$ orthogonal to $F_1$ that does not belong to $F_2$,
and let $u_2$ be a unit vector in the direction of the projection of $u$ on $F_2$. Then
$$
\sum_{k=1}^n c_k \iprod{x_k}{u}^2 =
\sum_{k\in \sigma_1} c_k \iprod{x_k}{u}^2 +
\sum_{k\in \sigma_2} c_k \iprod{x_k}{u}^2
= \sum_{k\in \sigma_2} c_k \iprod{x_k}{u}^2 .
$$
But, since $u\notin F_2$, for any vector $z\in F_2$ we have $|\iprod{z}{u_2}| > |\iprod{z}{u}|$,
and for any vector $z\in F_1$ we have $|\iprod{z}{u_2}| \ge 0$, implying that
$$
\sum_{k=1}^n c_k \iprod{x_k}{u}^2 < \sum_{k=1}^n c_k \iprod{x_k}{u_2}^2 ,
$$
contradicting the $c$-isotropy of $X$.
Hence all vectors orthogonal to $F_1$ lie in $F_2$, and the lemma follows.
$\Box$

\subsection{On transitivity of the relation $\sim$}
\label{sec:transitivity}


For completeness, we show here that the   relation discussed above is indeed transitive. Recall that we are 
given a set of vectors $X$ in $\RR^d$ which span it,
and we define a relation on them in the following way $x\sim y$ if there is a set 
$S\subset X$ with $|S| = d-1$ such that $S\cup \{x\}$ is a basis and so is $S\cup \{y\}$.
Clearly $y\sim \pm y$ for every $y\in X$.

Assume that $x\sim y$ and $y\sim z$, we would like to show that $x\sim z$. 
Assume towards a contradiction that this is not the case, namely $x\not\sim z$. 
By assumption, there is some set $V = \{v_i\}_{i=1}^{d-1}\subset X$ such that 
$\{y\}\cup V$ is a basis as well as $\{x\}\cup V$. Without loss of generality 
we may assume that $z\in V$: Indeed, since the set $V\cup \{z\}$ is not a basis, 
we may write $z$ as a combination of the vectors $v\in V$ and replace one of the 
$v$'s which has non-zero coefficient in this representation of $z$, keeping the 
span of $V$ as it was. So we assume without loss of generality that $v_{d-1} = z$.

Similarly by assumption there is some $\{w_i\}_{i=1}^{d-1} = W\subset X$ with 
$W\cup \{y\}$ and $W\cup \{z\}$ both bases, and again, by the same argument as 
above, we may assume without loss of generality that $w_{d-1} = x$.

Next we notice that as the set $S_1 = \{y\}\cup  \{v_{i} \}_{i=1}^{d-2}$ 
satisfies that $S_1\cup \{z\} = V\cup \{y\}$ is a basis, we know that 
$S_1\cup \{x\}$ cannot be a basis, which means that $x\in {\rm span }(S_1)$. 
However, $x\not \in {\rm span} \{v_i\}_{i=1}^{d-2}$ which means that we may write
\[ 
x = \alpha y + \sum_{i=1}^{d-2} \alpha_i v_i ,
\]
with $\alpha \neq 0$.

Similarly, $S_2 = \{y\}\cup \{w_i\}_{i=1}^{d-2}$ is such that
$S_2 \cup \{x\} = W \cup \{y\}$ is a basis, so we know that $S_2 \cup \{z\}$ 
cannot be a basis, which means that $z\in {\rm span}(S_2)$. As we know that 
$z\not\in {\rm span} \{ w_i\}_{i=1}^{d-2}$, we have here that
\[ 
z = \beta y + \sum_{i=1}^{d-2} \beta_i w_i ,
\]
with $\beta \neq 0$.

We are now in a position to chose a new set $S$ to contradict $x\not\sim z$.
$S$ will be a subset of $\{y\} \cup V\cup W$. Note that this (large) set spans $\RR^d$.

We start with a minimal (in cardinality) set $V_1\subset V$ such that $x$ can be 
written as a linear combination of $y$ and elements $v_i \in V_1$. We add to it 
a subset $W_1$ of $W$ such that $z$ can be written as a combination of $y$, 
elements of $V_1$, and elements of $W_1$. If the set $\{y\} \cup V_1\cup W_1$ 
is not yet spanning $\RR^{d}$, we add to it elements from $V\cup W$ so that it is.

We have thus constructed a set $S\subset X$, which is a basis of $\RR^d$. 
Clearly, writing $x$ as a linear combination of elements in $S$, the ones 
with non-zero coefficients are precisely those in $V_1$ and $\{y\}$. 
In writing $z$ as a linear combination of elements in $S$, we know that the 
nonzero coefficients are those in $W_1$, and at least one of the elements in 
$\{y\}$ or $V_1$ must be non-zero, otherwise it would mean that $z$ may be 
written as a linear combination of elements in $W_1$ alone which we know is 
not the case.

Therefore, there is at least one vector $\xi$ among $\{y\}\cup V_1$ within 
the basis $S$ such that both $x$ and $z$ have non-zero $\xi$ coefficient when 
written as a linear combination of vectors in $S$. This means that omitting 
$\xi$ and replacing it by either one of the two vectors, $x$ and $z$, gives 
a basis on $\RR^d$. In other words, the set $S' = S\setminus \{\xi\}$  is a 
set of $(d-1)$ elements which realizes the relation $x\sim z$, a contradiction 
to our original assumption.

\subsection{Uniqueness}
\label{sec:unique:rip}

A natural question is whether, given some set $X$ of vectors in $\R^d$ and $c\in {\rm relint}(K_X)$,
the radial $c$-isotropic position associated with $X$ is unique, up to an orthonormal transformation.
We show that, up to obvious scaling factors, this is indeed the case.
More formally, we show:

\begin{lemma} \label{lem:unique:rip}
    Let $c$ be a vector in $(\RR^+)^n$ with $\sum_{i=1}^n c_i = d$.
    Assume that $X=\{x_i\}_{i=1}^n$, $Z=\{z_i\}_{i=1}^n$ are sets whose elements satisfy
    $x_i, z_i \in \sph^{d-1}\subset \RR^d$, for $i=1,\ldots,n$, such that
    there exists a matrix $A\in GL_d$ such that $z_i =\frac{Ax_i}{|Ax_i|}$ for each $i$, and
    \[
    \sum_{i=1}^n c_i x_i \otimes x_i = \sum_{i=1}^n c_i z_i \otimes z_i = I_d.
    \]
    Assume further that $X$ (and thus also $Z$) is irreducible.
    Then there is some orthonormal $d\times d$ matrix $U$ and $\lambda>0$ such that $A = \lambda U$, and $z_i = Ux_i$,
    for $i=1,\ldots,n$.
\end{lemma}

\medskip
\noindent{\bf Remark.}
As a corollary we get that if, for some irreducible set $Y$ of vectors
$y_i \in \sph^{d-1}\subset \RR^d$, $i = 1, \ldots, n$, there are two
different linear transformations $A_1, A_2 \in GL_d$, such that the vectors
$\left\{ x_i = \frac{A_1y_i}{|A_1y_i|} \right\}_{i=1}^n$ are in radial $c$-isotropic position,
and so are the vectors $\left\{ z_i = \frac{A_2y_i}{|A_2y_i|} \right\}_{i=1}^n$,
then $A_2^{-1}A_1$ is a scalar multiple of an orthonormal matrix.
Indeed, in this case we have that, for $i=1,\ldots,n$,
$y_i =  \frac{A_1^{-1}x_i}{|A_1^{-1}x_i|}$ (as $y_i$ lies in
direction $A_1^{-1}x_i$ and is of unit length), and so
$z_i = \frac{A_2A_1^{-1} x_i}{|A_2A_1^{-1} x_i|}$ for each $i$. Thus, by
Lemma \ref{lem:unique:rip}, we see that $A_2A_1^{-1} = \lambda U$
for some orthonormal $U$, so $z_i = Ux_i$, for $i=1,\ldots,n$, as claimed.

That is, up to rotation and scaling, the matrix that maps $Y$ to
radial $c$-isotropic position is unique in the irreducible case.

If $Y$ is reducible, we can factor it into its subsets in the components of the
corresponding direct sum $\bigoplus_{j=1}^k E_j$, and then we can associate a unique
nonsingular matrix $A_j$, for each $j=1,\ldots,k$, such that any transformation
that maps $Y$ to radial $c$-isotropic position is a diagonal block matrix, whose
$j$-th block is an (individually) rotated and scaled copy of the corresponding
$A_j$, for $j=1,\ldots,k$.
\medskip

\noindent{\bf Proof of Lemma \ref{lem:unique:rip}.}
For the proof we use the notation of Lemma \ref{lem:phiS=entc}
and the discussion preceding it, within the proof of the second direction in
Lemma~\ref{prop:PhiSfinite-iff-rad-i-pos-exists}. By the lemma, we have
$$
\Phi^*_X(c) = \Phi^*_Z(c) = \sum_{i=1}^n c_i \log c_i .
$$
Inspecting the aforementioned discussion, this implies that
\[
2\sum_{i=1}^n c_i \log |Ax_i| = 2\log \det (A),
\]
namely,
\begin{equation} \label{eqdet}
\prod_{i=1}^n |Ax_i|^{c_i}  = \det(A).
\end{equation}
The assumption that
\[
\sum_{i=1}^n c_i x_i \otimes x_i = I_d
\]
implies that
\[
\sum_{i=1}^n c_i Ax_i \otimes Ax_i = AA^T ,
\]
or, equivalently,
\begin{equation} \label{eq:witha}
\sum_{i=1}^n |Ax_i|^2 c_i z_i \otimes z_i = AA^T.
\end{equation}
Taking the determinant on both sides, using again the Cauchy-Binet formula,
a weighted version of the arithmetic-geometric mean inequality, and the analysis
in the proof of the lemma, we have that (for the parameters $\lambda_S$ defined in that proof)
\begin{align*}
\det(A)^2  &= \det \left(\sum_{i=1}^n |Ax_i|^2 c_i z_i \otimes z_i\right)
= \sum_{|S|=d} \prod_{i\in S} |Ax_i|^2 \lambda_S \\
&\ge \prod_{i=1}^n |Ax_i|^{2\sum_{|S| = d, i\in S} \lambda_S} = \prod_{i=1}^n |Ax_i|^{2c_i} .
\end{align*}
In view of (\ref{eqdet}), we have equality here.
For this to happen, we must have equality in the arithmetic-geometric mean inequality,
which means, using irreducibility, that all the numbers $|Ax_i|$ have to be equal. Denoting their common value by $\rho$,
we get that $y_i = \rho z_i$ for each $i$, or $z_i = Bx_i$ for each $i$, where
$B = \frac{1}{\rho}A$. Also, using (\ref{eq:witha}), we have $AA^T = \rho^2 I_d$; that is,
$B$ is orthonormal, as claimed.
$\Box$

\subsection{Finding a good convex combination for $c$: Another interpretation}\label{sec:good-combination}

Following the technique of Lee~\cite{Lee}, we note here another interpretation of the extremizing
vector $t$ for a given coefficient vector $c$, in terms of minimizing a certain weighted entropy function. 
This interpretation is in close connection with the work of Singh and Vishnoi  \cite{vishnoi1} and  Straszak  and Vishnoi \cite{vishnoi2}, 
where they {\em start} with an entropy minimization problem and reversely engineer an optimization problem with fewer parameters.  

As we have seen throughout the preceding subsections, transforming a set $X=\{x_i\}$ of unit
vectors into radial $c$-isotropic position
is possible if and only if $c$ is in the relative interior of $K_X$, which means that one can write
\begin{equation} \label{eq:conv}
c = \sum_{|S|=d,\; \Delta_S > 0} \lambda_S{\bf 1}_S ,
\end{equation}
for coefficients $\lambda_S > 0$,
with $\sum_{|S|=d,\;\Delta_S > 0} \lambda_S = 1$.
As above, we denote by $\cal S$ the set of all $d$-tuples $S$ with $\Delta_S>0$.

Finding the transformation that sends $X$ into radial $c$-isotropic position is
equivalent to finding $t$ with $\Phi^*(c) + \Phi(t) = \sum c_i t_i$, which in turn
is equivalent to finding $t\in \RR^n$ for which the coefficients
\begin{equation}\label{eq:sconv}
\lambda_S   = \frac{ e^{\sum_{i\in S}t_i}\Delta_S}{\sum_{|S'|=d  }e^{\sum_{i\in S'}t_i}\Delta_{S'}}
\end{equation}
satisfy (\ref{eq:conv}) (see the proof of Lemma \ref{lem:attained}).
Note that the $t_i$'s that satisfy (\ref{eq:sconv}) are not unique; we can add the same constant to all
of them and still satisfy (\ref{eq:sconv}).
In this case, by Lemma \ref{prop:PhiSfinite-iff-rad-i-pos-exists}, the matrix $Q^{-1/2}(t)$
brings $X$ into radial $c$-isotropic position.

Here we focus on the process of finding such a joint solution of (\ref{eq:conv}) and (\ref{eq:sconv}).
Consider sets $\Lambda$ of the coefficients of convex combinations of $\cal S$; that is, sets of
the form $\Lambda = \{ \lambda_S \mid S\in {\cal S} \}$,
with $\lambda_S\in [0,1]$, $\sum_{S\in {\cal S}} \lambda_S = 1$, that satisfy\footnote{%
    We actually want the $\lambda_S$'s to be all positive, but we extend their domain
    in this manner to make it compact.}
\[
c = \sum_{S \in {\cal S}} \lambda_S{\bf 1}_S.
\]
By our assumption $c\in K_X$, so there is at least one feasible convex combination $\Lambda$.
Minimize, on the above domain of feasible linear combinations, the (convex) objective entropy function
\[
H(\Lambda) := \sum_{|S|=d,\; \Delta_S > 0} \lambda_S \log \frac{\lambda_S}{\Delta_S}.
\]

The minimum exists (i.e., it is finite) and is attained, because
we consider only those $S$ with $\Delta_S > 0$, so $H$ is a continuous function over
the compact domain of the feasible values of $\Lambda$.

Using the theory of Lagrange multipliers, we introduce extra variables
$t_1,t_2,\ldots,t_n$ (the Lagrange multipliers), and extremize
\[
\sum_{|S|=d,\; \Delta_S > 0} \lambda_S \log \frac{\lambda_S}{\Delta_S}-
\sum_{i=1}^n t_i \left(\sum_{|S|=d,\; \Delta_S > 0,\; i\in S} \lambda_S-c_i\right).
\]
More precisely, in doing so we seek an extremizing value of $\Lambda$ that lies in
the interior of the feasible domain.
Differentiating with respect to each $\lambda_S$, we get that
\[
\log(\lambda_S/\Delta_S)+1 = \sum_{i\in S} t_i ,
\]
which implies that
\begin{equation} \label{lagrange1}
\lambda_S = \Delta_Se^{\sum_{i\in S} t_i -1} .
\end{equation}

By the condition that $\Lambda$ is a convex combination, this actually yields the expressions in
(\ref{eq:sconv}) where we take the $t_i$'s to be those that make the denominator in (\ref{eq:sconv}) equal to
$e$ (assuming that such parameters $(t_i)$ can be
found, namely that the extremum is indeed attained in the interior of the corresponding domain).
In other words, we get that the desired solution $t$ is the
vector of Lagrange multipliers for the entropy minimization problem.
To find them, we need to solve the system (where the $\lambda_S$'s are given by (\ref{eq:sconv}))
\begin{equation} \label{lambda-and-c}
\sum_{|S|=d,\; \Delta_S > 0,\; i\in S} \lambda_S = c_i , \quad\text{for $i=1,\ldots,n$} .
\end{equation}
This has been addressed in Section \ref{sec:exact}.


\section{Additional material for Section \ref{sec:gd}}
\label{app:gd}

%

\subsection{A construction with a large $|t^*|_\infty$}
\label{sec:large-t}

In view of the upper bounds on $|t^*|_\infty$, given in Lemmas~\ref{R-cll} and \ref{lem:hm-fix},
a natural question is whether $t^*$ can indeed become so large. The following simple example
shows that this indeed is the case, at least in terms of the dependence on $\delta$.

Let $X=\{x_1,x_2,x_3\}$ be the set of the three vectors
\begin{align*}
x_1 & = \Bigl( \cos\theta, \sin\theta \Bigr) \\
x_2 & = \Bigl( \cos\theta, -\sin\theta \Bigr) \\
x_3 & = \Bigl( 0, 1 \Bigr)
\end{align*}
in the plane, and assume that $c$ is the uniform vectore
$\left( \frac{2}{3},\frac{2}{3},\frac{2}{3} \right)$.
We want to replace each $x_j$ by $e^{t_j/2}x_j$, so that
the SVD $U\Sigma V^T$ of the new set of vectors is such that all the rows
of $U$ have the same squared norm $\frac{d}{n} = \frac23$. This is
the condition for the vector $t=(t_1,t_2,t_3)$ to be the extremizer
for $f(t)=\Phi(t) - \iprod{c}{t}$, as defined earlier, with
$c=\left(\frac{2}{3},\frac{2}{3},\frac{2}{3}\right)$.

We write $w_j = e^{t_j/2}$, for $j=1,2,3$, and note that
we can scale these values by any common factor. We take $w_1=w_2=w$
and $w_3=1$. (We are guessing here that $w_1=w_2$, and the guess
is justified because we will get with this assumption the desired
extremizer $t$, as we show next.)

We look for the first singular value $\sigma_1$ and the
right-singular vector $v_1$, which we write as $(\cos\beta,\sin\beta)$,
so we have
\begin{align*}
\sigma_1^2 & = \max_\beta \left\{ \sum_{j=1}^3 w_j^2\iprod{x_j}{v_1}^2 \right\} \\
& = \max_\beta \left\{ w^2 \left( \cos^2(\theta-\beta) + \cos^2(\theta+\beta) \right) + \sin^2\beta \right\} \\
& = \max_\beta \left\{ 2w^2 \left( \cos^2\theta\cos^2\beta + \sin^2\theta\sin^2\beta \right) + \sin^2\beta \right\} \\
& = 2w^2\cos^2\theta + \max_\beta \left\{ \left[ 2w^2 \left( -\cos^2\theta + \sin^2\theta \right) + 1 \right] \sin^2\beta \right\} .
\end{align*}
The maximum is attained at either $\beta=0$ or $\beta=\pi/2$, depending on whether the expression
$$
2w^2 \left( -\cos^2\theta + \sin^2\theta \right) + 1 = 1 - 2w^2\cos 2\theta
$$
is negative or positive, respectively, which in turn depends on whether
$w^2$ is larger or smaller than $\frac{1}{2\cos 2\theta}$, respectively.
(When $\theta>\pi/4$ the expression is always positive, but we consider
here the case where $\theta$ is small.)

In either case, the right singular vectors are the coordinate directions,
and $\sigma_2^2$ is obtained by the same expression, replacing $\beta$
by $\pi/2 - \beta$. Consider the case where $\beta = 0$. Then the singular
values are
\begin{align*}
\sigma_1^2 & = 2w^2\cos^2\theta, \\
\sigma_2^2 & = 2w^2\sin^2\theta + 1 ,
\end{align*}
and the corresponding right-singular vectors are the standard $e_1$ and $e_2$.
$V$ is then the identity matrix, and the rows of $U$ are
\begin{align*}
u_1 & = \left( \frac{1}{\sigma_1} w\cos\theta,\; \frac{1}{\sigma_2} w\sin\theta \right)
= \left( \frac{1}{\sqrt{2}} , \frac{w\sin\theta} {\sqrt{2w^2\sin^2\theta + 1}} \right) \\
u_2 & = \left( \frac{1}{\sigma_1} w\cos\theta,\; -\frac{1}{\sigma_2} w\sin\theta \right)
= \left( \frac{1}{\sqrt{2}} , -\frac{w\sin\theta} {\sqrt{2w^2\sin^2\theta + 1}} \right) \\
u_3 & = \left( 0,\; \frac{1}{\sigma_2} \right)
= \left( 0 , \frac{1} {\sqrt{2w^2\sin^2\theta + 1}} \right) ,
\end{align*}
and their squared  norms are
\begin{align*}
|u_1|^2 = |u_2|^2 & = \frac{1}{2} + \frac{w^2\sin^2\theta} {2w^2\sin^2\theta + 1} \\
|u_3|^2 & = \frac{1} {2w^2\sin^2\theta + 1} .
\end{align*}
Since we want all three squared norms to be equal (to $2/3$), we must have
$$
\frac{1}{2} + \frac{w^2\sin^2\theta} {2w^2\sin^2\theta + 1} =
\frac{1} {2w^2\sin^2\theta + 1} ,
$$
or
$$
\frac{1}{2} = \frac{1-w^2\sin^2\theta} {2w^2\sin^2\theta + 1} ,
$$
or
$$
2w^2\sin^2\theta + 1 = 2 - 2w^2\sin^2\theta ,
$$
or
$$
w = \frac{1}{2\sin\theta} ,
$$
and we note, as a sanity check, that all three squared norms are in this case indeed
equal to $\frac{2}{3}$.

To conclude, we have found an extremizing vector $t$, which satisfies $t_3 = 0$ and
$$
t_1 = t_2 = 2\ln w = 2\ln \frac{1}{2\sin\theta} ,
$$
and they both tend to infinity as $\theta$ tends to $0$.

Note that if we take $\delta < \sin\theta$ and
$\eta=0$ (any  $\eta \le 1/3$ works) then $\left( \frac{2}{3},\frac{2}{3},\frac{2}{3} \right)$ is $(\eta,\delta)$-deep inside
$K_X$, since no two vectors are $\delta$-close to the same direction.
Thus, this example shows that $|t^*|_\infty$ indeed grows proportionally to
$\log \frac{1}{\delta}$. We leave the question of whether the other terms 
in our bounds in Equations (\ref{our-tinf}) and (\ref{our-tinf-uniform}) are essential
for future research.

We also note that for any $\delta$ and $\eta > 1/3$, and for any $\eta$ and $\delta \ge \sin\theta$,
$c=\left( \frac{2}{3},\frac{2}{3},\frac{2}{3} \right)$ is not $(\eta,\delta)$-deep inside
$K_X$.

At the limit, with $\theta=0$, $c=\left( \frac{2}{3},\frac{2}{3},\frac{2}{3} \right)$ gets out of $K_X$, and therefore $X$ cannot be brought
to radial isotropic position.
Indeed, for $J=\{1,2\}$, we have
${\rm dim} ({\rm span} \{x_j\}_{j\in J}) = 1$,
so by Proposition~\ref{cll-version} for $c$ to be inside
$K_X$ we  need that
$\frac23 \cdot |J| \le 1$, or $|J|\le \frac32$, which is impossible since $|J|=2$.

\subsection{Approximate gradient descent} \label{sec:apx:gd}\label{App-Micha-approx-grad}

\def\ax{{\rm apx}}

Consider the gradient descent techniques applied for minimizing a smooth convex 
function $f$ over some convex compact set $K\subset\RR^d$, that is, 
a differentiable convex function $f$ satisfying, for every pair $x,u\in K$,
\begin{equation} \label{smooth0}
\|\nabla f(x) - \nabla f(u)\| \le \beta \|x-u\| ,
\end{equation} 
for some constant (the smoothness parameter) $\beta>0$.
As shown in Bubeck~\cite[Lemma 3.4 and Inequality (3.4)]{Bub}, this,
and the convexity of $f$, imply that, for every $x,u\in K$,
\begin{equation} \label{smooth}
(x-u)^T \nabla f(u) \le 
f(x)-f(u) \le (x-u)^T \nabla f(u) + \frac{\beta}{2} \|x-u\|^2 .
\end{equation}

For such constrained situations, one uses \emph{projected gradient descent}, as reviewed
in Section~\ref{sec:gd}. We recall that this
is an iterative procedure which starts with some $x_1\in K$, and computes,
at each step $s=1,2,\ldots$, 
\begin{equation} \label{proj:step}
x_{s+1} = P_K\Bigl(x_s - \eta\nabla f(x_s)\Bigr) ,
\end{equation}
for $\eta = \frac{1}{\beta}$, where $P_K$ is the projection to $K$.
As shown in \cite[Theorem 3.7]{Bub}, we have
$$
f(x_t)-f(x^*) \le \frac{3\beta \|x_1-x^*\|^2 + f(x_1)-f(x^*)}{t} ,
$$
where $x^*$ is the point at which $f$ attains its minimum.

In practice, computing the gradient $\nabla f(x)$ may suffer from numerical errors,
making the execution of the gradient descent iterations only approximate.
We argue that the convergence of the method and its rate of convergence
are only slightly affected when the gradient is only approximated.

So assume that we are given some error parameter $\eps>0$, and that,
at each step $s$, when we reach a point $x_s\in K$, we get an approximate
gradient at $x_s$, denoted as $\nabla_\ax f(x_s)$, which is sufficiently close
to the exact gradient. Concretely, we assume it to satisfy
\begin{equation} \label{eq:apx}
\| \nabla f(u) - \nabla_\ax f(u) \| \le \eps^2 \|\nabla f(u)\| ,
\end{equation} 
for the prespecified $\eps>0$ and for any $u\in K$. 

We first consider the simpler basic variant of the gradient descent method 
for smooth functions, which is the unconstrained gradient descent (see
\cite[Section 3.2]{Bub} and Section~\ref{sec:gd}). 
Here each step of the iteration, instead of computing
$x_{s+1} = x_s - \frac{1}{\beta}\nabla f(x_s)$, computes 
$$
x^\ax_{s+1} = x^\ax_s - \frac{1}{\beta}\nabla_\ax f(x^\ax_s) ,
$$
starting at the same initial point $x^\ax_1 = x_1$.

We have
\begin{align*}
x^\ax_{s+1} - x_{s+1} & = \Bigl( x^\ax_s - \frac{1}{\beta}\nabla_\ax f(x^\ax_s) \Bigr) 
 - \Bigl( x_s - \frac{1}{\beta}\nabla f(x_s) \Bigr) \\
& = \Bigl( x^\ax_s - x_s \Bigr) - \frac{1}{\beta} \Bigl( \nabla_\ax f(x^\ax_s) - \nabla f(x_s) \Bigr) \\
& = \Bigl( x^\ax_s - x_s \Bigr) - \frac{1}{\beta} \Bigl( \nabla_\ax f(x^\ax_s) - \nabla f(x^\ax_s) \Bigr) 
 - \frac{1}{\beta} \Bigl( \nabla f(x^\ax_s) - \nabla f(x_s) \Bigr) .
\end{align*}
That is, we have,
$$
\| x^\ax_{s+1} - x_{s+1} \| \le 
\| \Bigl( x^\ax_s - \frac{1}{\beta} \nabla f(x^\ax_s) \Bigr) - 
   \Bigl( x_s - \frac{1}{\beta} \nabla f(x_s) \Bigr) \|  
+ \frac{1}{\beta} \| \nabla_\ax f(x^\ax_s) - \nabla f(x^\ax_s) \| .
$$
For the first term in the right-hand side, we apply a variant of the proof of \cite[Lemma 3.5]{Bub}.
Putting $x=x^\ax_s$ and $y=x_s$, we have (this is Equation (3.6) in \cite{Bub})
$$
\iprod{\nabla f(x) - \nabla f(y)}{x-y} \ge \frac{1}{\beta} \|\nabla f(x) - \nabla f(y) \|^2 ,
$$
which implies
\begin{align*}
\| \Bigl( x - \frac{1}{\beta} \nabla f(x) \Bigr) & - 
   \Bigl( y - \frac{1}{\beta} \nabla f(y) \Bigr) \|^2 \\
& = \| x-y\|^2 - \frac{2}{\beta} \iprod{\nabla f(x) - \nabla f(y)}{x-y} + \frac{1}{\beta^2} \|\nabla f(x) - \nabla f(y) \|^2 \\
& \le \|x-y\|^2 - \frac{1}{\beta^2} \|\nabla f(x) - \nabla f(y) \|^2 \\
& \le \|x-y\|^2 .
\end{align*}
For the second term, using (\ref{eq:apx}), we have
$$
\frac{1}{\beta} \| \nabla_\ax f(x^\ax_s) - \nabla f(x^\ax_s) \| \le \frac{\eps^2}{\beta} \|\nabla f(x^\ax_s)\|
\le \frac{M\eps^2}{\beta} , 
$$
where $M= \max_{x\in K} \| \nabla f(x) \|$. Together, we thus get
\begin{equation} \label{eq:dev}
\| x^\ax_{s+1} - x_{s+1} \| \le \| x^\ax_{s} - x_{s} \| + \frac{M\eps^2}{\beta} . 
\end{equation}

\paragraph{Projected gradient descent.}
Consider next the projected gradient descent technique, which computes, 
at each step $s$,
\begin{align*}
x_{s+1} & = P_K\Bigl(x_s - \eta\nabla f(x_s)\Bigr) , \qquad\text{whereas we compute} \\
x^\ax_{s+1} & = P_K\Bigl(x^\ax_s - \eta\nabla_\ax f(x^\ax_s)\Bigr) .
\end{align*}
Rewrite these equations as
\begin{align*}
x_{s+1} & = P_K(y_{s+1}) ,\qquad\text{where}\qquad y_{s+1} = x_s - \eta\nabla f(x_s) \\
& \\
x^\ax_{s+1} & = P_K(y^\ax_{s+1}) ,\qquad\text{where}\qquad y^\ax_{s+1} = x^\ax_s - \eta\nabla_\ax f(x^\ax_s) .
\end{align*}
The preceding analysis has essentially shown (in (\ref{eq:dev})) that
$$
\| y^\ax_{s+1} - y_{s+1} \| \le \| x^\ax_{s} - x_{s} \| + \frac{M\eps^2}{\beta} . 
$$
The fact that the projection onto a convex set is non-expansive asserts that
$$
\| x^\ax_{s+1} - x_{s+1} \| \le \| y^\ax_{s+1} - y_{s+1} \| ,
$$
from which we get the same inequality (\ref{eq:dev}) in this case too.

\paragraph{Wrapping up.}
It follows that, after applying $t$ iterations, we get from (\ref{eq:dev}) that
\begin{equation} \label{apx:works}
\| x^\ax_{t} - x_{t} \| \le \frac{M(t-1)\eps^2}{\beta} . 
\end{equation}
Combining this inequality with either the one in \cite[Theorem 3.3]{Bub} (for the unconstrained case)
or the one in \cite[Theorem 3.7]{Bub} (for the case of projected gradient descent), we get either
\[
f(x^\ax_t) - f(x^*) \le \frac{2\beta\|x_1-x^*\|^2}{t} + \frac{M^2(t-1)\eps^2}{\beta} ,
\]
for the unconstrained case, or
\[
f(x^\ax_t) - f(x^*) \le \frac{3\beta\|x_1-x^*\|^2 + f(x_1)-f(x^*)}{t} + \frac{M^2(t-1)\eps^2}{\beta} ,
\]
for projected gradient descent.
Specifically, we get these bounds by writing the left-hand side as
\begin{equation} \label{eq:final}
f(x^\ax_t) - f(x^*) =
\Bigl( f(x^\ax_t) - f(x_t) \Bigr) + \Bigl( f(x_t) - f(x^*) \Bigr) ,
\end{equation}
and then by bounding the (absolute value of the) first term in (\ref{eq:final})
using the general inequality 
$$
\|f(x)-f(y)\| \le \sup_{z\in K} \|\nabla f(z)\|\cdot \|x-y\| \le M\|x-y\| ,
$$
which, combined with (\ref{eq:dev}), yields the second term in the bounds,
and by bounding the second term using the bound in the respective theorem in \cite{Bub}.

By optimizing $t$ to a suitable multiple of $1/\eps$, we get in both cases
$f(x^\ax_t) - f(x^*) = O(\eps)$. 

\subsection{How to bypass strong convexity} \label{sec:bypass}

The guarantee provided by gradient descent for smooth functions is that after $O(|t^*|^2/\eps)$ steps
 the value of the function $f$ at the current point is  $\eps$-close to its smallest value.
For our purpose here, this is not immediately sufficient, as we want a point in which the gradient of the
function $f$ is close to zero.
In this section we show that for smooth functions we indeed get such a point.

Alternatively, if we can argue that our function is strongly convex, then it would follow
that after $O(\kappa \log |t^*|/\eps)$  steps our point $t$ is close to the minimizer $t^*$ (and by smoothness then the gradient is
close to zero). (Recall that $\kappa$ is the ratio between the largest and the smallest eigenvalues of the Hessian.)
In Section \ref{sec:alpha} we take this alternative approach we give bounds on how strongly convex is our function.

Specifically, here we show that if the algorithm gets us close to a point $t$
for which $f(t)=\Phi(t)-\iprod{c}{t}$ is close to its smallest value,
then at this very point $t$ the gradient of $f$ is close to $0$, so
the gradient of $\Phi$ is close to $c$. This allows us to output $t$,
and use $\nabla \Phi(t)$ as a weight vector $c_{\rm apx}$
that is sufficiently close to the prescribed $c$.

To this end we use the already proven property that
$0\le H \le \frac12 I_n$, as an inequality between positive definite symmetric matrices,
namely $0\le \iprod{H(t)y}{y} \le \frac12|y|^2$ for all $y$.
To introduce the technique in simpler form, let us first consider the
one-dimensional version of the analysis, presented in a more general setting.

\begin{lem}\label{lem:1dgrad}
Let $\varphi:\RR\to \left[0,\infty\right)$ be a twice-differentiable convex function
satisfying $\varphi''(x)\le \rho$ for all $x$, and let $y\in \RR$ be such that
$\varphi(y)\le \eps$. Then $\left|\varphi'(y)\right|\le \sqrt{2\eps \rho}$.
\end{lem}
\medskip
\noindent
{\bf Proof.}
We may clearly assume $\varphi(y)=\eps$. Since $\varphi\ge 0$, and has
value $\eps$ at $y$, we have that $\inf \varphi \in [0,\eps]$.
Let us assume as a first and main case that this infimum is indeed attained,
at some point $z\in \RR$. We may translate $\varphi$, without changing any
of the conditions, to allow us to assume that $z=0$, and, by symmetry, we
may also assume that $y>0$. Moreover, by subtracting
$\varphi(0)\ge 0$ we may further assume that $\varphi(z)=\varphi(0)=0$.

Assume by contradiction that $\varphi'(y)>\sqrt{2\eps \rho}$. Then we must have for all $t<y$ that
\[
\varphi'(t) = \varphi'(y)+(t-y)\varphi''(\xi)\ge \varphi'(y)-\rho(y-t)>\sqrt{2\eps \rho}-\rho(y-t) ,
\]
where $\xi$ is some intermediate value in $[t,y]$ that depends on $t$.
In particular, as $\varphi'(0)=0$, we see that
\[
0>\sqrt{2\eps \rho}-\rho y\qquad{\rm and~so}\qquad y>\sqrt{\frac{2\eps}{\rho}}.
\]
Thus,
\begin{eqnarray*}
\eps &=& \varphi(y)=\int_0^y\varphi'(t)dt \ge \int_{y-\sqrt{\frac{2\eps}{\rho}}}^y \varphi'(t)dt\\
&>& \int_{y-\sqrt{\frac{2\eps}{\rho}}}^y  \left(\sqrt{2\eps \rho}-\rho(y-t)\right)dt= \eps ,
\end{eqnarray*}
which is clearly a contradiction. Thus $\varphi'(y)\le \sqrt{2\eps \rho}$.

In the case where the infimum of $\varphi$ is not attained, we have that
the infimum is in fact a limit of $\varphi$ at either $+\infty$ or
$-\infty$. Again, by symmetry, and without loss of generality,
one may assume the latter, and the preceding argument remains valid if we
replace the lower limit in the integration by $-\infty$, instead of $0$.
$\Box$

We note that the bound is tight, as can be seen by the function
$\varphi (t)= \rho t^2/2$ with $\varphi''(t)=\rho$ and $y = \sqrt{2\eps/\rho}$,
where $\varphi(y)=\eps$ and $\varphi'(y)=\rho y=\sqrt{2\eps \rho}$.

As a consequence we get the following general higher-dimensional property we are after.
\begin{lem} \label{smallf:smallgrad}
Let $\varphi:\RR^n \to [0,\infty)$ be a twice-differentiable convex function whose Hessian satisfies
$H^\varphi \le \rho I_n$, in the same meaning of positive definite symmetric matrices as above,
and assume that $\varphi(y) \le \eps$. Then $|\nabla \varphi(y)|\le \sqrt{2\eps \rho}$.
\end{lem}
\medskip
\noindent
{\bf Proof.}
 Since $\varphi$ is convex so are its restrictions to one-dimensional lines,
 in particular to the line $\ell = y +\RR\nabla \varphi(y)$. Let
 $\psi(t)= \varphi(y+t\nabla \varphi(y))$.
 Clearly $$\psi'(t)= \iprod{\nabla\varphi(y+t\nabla\varphi(y))}{\nabla\varphi(y)}$$ and
 $$
 \psi''(t)= \iprod{H^\varphi(y+t\nabla\varphi(y))\nabla\varphi(y)}{\nabla\varphi(y)}.
 $$
 In particular $\psi'(0)=|\nabla\varphi (y)|^2$.
 On $\ell$ we have that $\psi(t)\ge 0$, $\psi(0)\le \eps$ and, by the assumption on $H^\varphi$,
 we have $\psi''(t) \le \rho|\nabla\varphi(y)|^2=:\rho_0$. Using Lemma~\ref{lem:1dgrad}, we see that
 $\psi'(0)\le \sqrt{2\eps \rho_0}$ which can be rewritten as
 \[
 |\nabla\varphi (y)|^2 \le \sqrt{2\eps \rho|\nabla\varphi(y)|^2}
 \]
 Thus we get $|\nabla \varphi(y)|\le \sqrt{2\eps \rho}$, as claimed.
$\Box$

\begin{cor} \label{cor:smallgrad}
When we terminate either of our gradient descent procedures (projected gradient
descent for Lipschitz functions or unconstrained gradient descent for smooth functions),
at some vector $t_{\rm apx}$ for which $f(t_{\rm apx}) = f(t^*) + \eps$, we have
$$
|\nabla \Phi(t_{\rm apx}) - c | \le \sqrt{\eps} .
$$
\end{cor}
\noindent{\bf Proof.}
Apply Lemma~\ref{smallf:smallgrad} to the function
$f(t) - f(t^*)$, and use the facts that $\nabla f = \nabla \Phi - c$,
and that the constant $\rho$ in this case is $1/2$, as we have shown earlier,
in Lemma~\ref{lem:hnorm}.
$\Box$

In other words, if we put $c_{\rm apx} := \nabla \Phi(t_{\rm apx})$,
we have, by Corollary~\ref{cor:smallgrad}, $|c_{\rm apx} - c| \le \sqrt{\eps}$.
That is, the transformation $Q^{-1/2}(t_{\rm apx})$ brings $X$, after normalization,
to a set in radial $c_{\rm apx}$-isotropic position,
so we obtain the desired approximation to our radial isotropy problem.
To get $f(t_{\rm apx})$ $\eps$-close to $f(t^*)$, we need to apply sufficiently
many iterations, so as to make the right-hand side in either 
 \eqref{eq:sgd} or its projected version smaller than $\eps$.
 
\end{document}